\documentclass[letterpaper,useAMS,usenatbib]{mn2e}
\usepackage{comment}
\usepackage{amssymb}   
\usepackage{amsmath}   
\usepackage{amssymb}  
\usepackage{bm}

\usepackage{xspace}
\usepackage{graphicx}
\usepackage{verbatim}

\usepackage[T1]{fontenc}
\usepackage{aecompl}

\newcommand{\be}{\begin{equation}}
\newcommand{\ee}{\end{equation}}

\title[Narrow-line lensing with the WFC3 grism]{Double dark matter vision: twice the number of compact-source lenses with narrow-line lensing and the WFC3 grism}

\author[Nierenberg et al.]{
A.~M.~Nierenberg$^{1, 2}$\thanks{\tt anna.m.nierenberg@jpl.nasa.gov}, 
D.~Gilman$^{3}$,
T.~Treu$^{3,4}$, 
G.~Brammer$^{5}$,
S.~Birrer$^{3}$,
\newauthor
L.~Moustakas$^{1}$,
A.~Agnello$^6$,
T.~Anguita$^{7,8}$,
C.~D. Fassnacht$^9$,
V.~Motta$^{10}$,
\newauthor
A.~H.~G.~Peter$^{11,12,13}$,
D.~Sluse$^{14}$
\\
\medskip\\
$^1$  Jet Propulsion Laboratory, California Institute of Technology\\
$^2$ NASA Postdoctoral Program Fellow \\ 
$^3$ UCLA Physics \& Astronomy, 475 Portola Plaza, Los Angeles, CA 90095-1547, USA\\
$^4$ Packard Fellow\\
$^5$ The Cosmic Dawn Center, Rockefeller Komplesket, Niels Bohr Institute, Juliane Ma, 2100 K\o benhavn \\
$^6$  DARK, Niels-Bohr Institute, Lyngbyvej 2, 2100 Copenhagen \\
$^{7}$ Departamento de Ciencias Fisicas, Universidad Andres Bello Fernandez Concha 700, Las Condes, Santiago, Chile\\
$^{8}$ Millennium Institute of Astrophysics, Chile\\
$^{9}$ UC Davis Department of Physics, 1 Shields Ave., Davis CA 95616 \\
$^{10}$ Instituto de F\'{\i}sica y Astronom\'{\i}a, Universidad de Valpara\'{\i}so, Avda. Gran Breta\~na 1111, Valpara\'{\i}so, Chile \\
$^{11}$Center for Cosmology and AstroParticle Physics, 191 West Woodruff Avenue, The Ohio State University, Columbus OH 43204, USA \\
$^{12}$Department of Physics, 191 West Woodruff Avenue, The Ohio State University, Columbus OH 43204, USA \\
 $^{13}$Department of Astronomy, The Ohio State University, 4055 McPherson Laboratory, 140 West 18th Avenue, Columbus OH \\
$^{14}$STAR Institute, Quartier Agora - All\'e du six Ao\^ut, 19c B-4000 Li\`ege, Belgium \\
}

\def\be{\begin{equation}}
\def\ee{\end{equation}}

\def\sun{\odot}

\begin{document}
\date{Accepted for publication in MNRAS}
\pagerange{\pageref{firstpage}--\pageref{lastpage}}\pubyear{2016}

\maketitle           

\label{firstpage}

                      
\begin{abstract}

The magnifications of compact-source lenses are extremely sensitive to the presence of low mass dark matter halos along the entire sight line from the source to the observer.  Traditionally, the study of dark matter structure in compact-source strong gravitational lenses has been limited to radio-loud systems, as the radio emission is extended and thus unaffected by microlensing which can mimic the signal of dark matter structure. An alternate approach is to measure quasar nuclear-narrow line emission, which is free from microlensing and present in virtually all quasar lenses. In this paper, we double the number of systems which can be used for gravitational lensing analyses by presenting measurements of narrow-line emission from a sample of 8 quadruply imaged quasar lens systems, WGD J0405-3308, HS 0810+2554, RX J0911+0551, SDSS J1330+1810, PS J1606-2333, WFI 2026-4536, WFI 2033-4723 and WGD J2038-4008. We describe our updated grism spectral modelling pipeline, which we use to measure narrow-line fluxes with uncertainties of 2-10\%, presented here. We fit the lensed image positions with smooth mass models and demonstrate that these models fail to produce the observed distribution of image fluxes over the entire sample of lenses. Furthermore,  typical deviations are larger than those expected from macromodel uncertainties. This discrepancy indicates the presence of perturbations caused by small-scale dark matter structure. The interpretation of this result in terms of dark matter models is presented in a companion paper.
\end{abstract}

\setcounter{footnote}{1}

\section{Introduction}
\label{sec:intro}

The mass function of small scale structure provides crucial insight into the nature of dark matter. In the widely accepted $\Lambda$CDM model, dark matter halos are expected to have a mass function with constant power law slope; forming halos with masses as small as those of planets \citep{Diemand++05, Diemand++08, Springel++08}. 

In some alternate theories of dark matter, the slope of the mass function is substantially different. For instance, in warm dark matter (WDM) models, dark matter particles have a longer free-streaming length which erases structure on small scales \citep[e.g.][]{Colombi++96, Bode++01, Abazajian++05, Abazajian++06, Menci++12, Lovell++12, Lovell++14, Venumadhav++16, Schneider++17}. 
`Fuzzy' cold dark matter made up of ultra-light scalar particles \citep[e.g.][]{Hu++2000, Hui++17} and decaying dark matter models \citep{Peter++10, Wang++12} can also modify or truncate the halo mass function.

Non-gravitational dark matter particle interactions such as self-interaction, interaction with baryonic matter, or `dark photons' can introduce oscillations in the dark matter halo power spectrum, as well as suppression at the low mass end \citep[e.g.][]{Vogelsberger++12, Rocha++13, Cyr-Racine++14, Cyr-Racine++16, Vogelsberger++16, Nadler++19b, Bose++19}. 

The mass-scale and sharpness of turnovers in the mass function are determined by the detailed particle physics of a given dark matter model, thus the characteristics, or absence of such a turnover in the halo mass function can rule out a broad range of dark matter models \citep[see, e.g.,][for a summary of observational consequences for a broad range of models]{Buckley++18}.

Traditional measurements of the halo mass function rely on observations of luminous structure and the assumption that all galaxies are found within dark matter halos. Measurements of the luminosity function and clustering of galaxies in the local Universe indicate that the halo mass function is well fit by a power-law with constant slope down to halo masses of M$_{200}\sim 10^{8}-10^{9} $M$_\odot$ \citep[e.g.,][]{Strigari++07, Tollerud++08,  Behroozi++10, Reddick++12, Jethwa++18, Kim++18, Nadler++19}.  Given the lack of observed turnover in the halo mass function down to this scale, the free streaming length of dark matter must be shorter than that of a $\sim3$ keV sterile neutrino \citep{MaccioFontanot++10, Polisensky++11, Kennedy++14, Jethwa++18, Kim++18, Nadler++19, Drlica-Wagner++19}. 

Pushing the measurement of the luminosity function to very high redshifts can also provide a constraint on the dark matter free streaming length and self-interaction cross section as dark matter halos form later when the free streaming length is longer. Current limits based on observations of the number counts of high redshift galaxies in the Hubble Frontier Fields places a limit of m $>2.4$ keV at the 2$\sigma$ level, competitive with the Local Volume constraints \citep[e.g.][]{Menci++17, Castellano++19, Sameie++19}.

The Ly-$\alpha$ forest provides an alternate window into structure formation. This method uses gas absorption along quasar sight-lines as tracers of structure. The method requires detailed modelling of the temperature variations and evolution of the Inter Galactic Medium (IGM), including the distribution of the UV background, and the redshift of re-ionization, using hydrodynamic simulations. With a weak prior on the evolution of the IGM temperature, the current limit is a particle mass greater than 3.5 keV at 2$\sigma$ confidence for a thermal relic dark matter particle. Imposing a power-law model to the IGM temperature evolution increases the limit to m $>5$ keV for a sterile neutrino \citep{Viel++13, Baur++15, Irsic++17}.

Both the high and low redshift results indicate that if CDM is correct, galaxy formation must become extremely inefficient at present day halo masses of M$_{200}\sim 10^{8}-10^{9}$ M$_\odot$. A variety of mechanisms including re-ionization, supernovae feedback, tidal disruption and stellar winds can preferentially suppress star formation in low mass halos \citep[e.g.][]{Barkana++99, Bullock++2000, Gnedin++00, Benson++02, Somerville++02, Benson++2010,  Hopkins++14, Benitez-Llambay++15, Sawala++16, Fillingham++16, Xu++16, Dawoodbhoy++18, Bose++18, Corlies++18, Garrison-Kimmel++19}. Therefore, measuring the halo mass function and constraining the dark matter free streaming length at halo masses below $\sim 10^9$ M$_\odot$ requires a method which does not use stars as tracers of dark matter halos. 

Strong gravitational lensing provides a powerful probe of the halo mass function at low masses, as it is sensitive to halos even if they do not contain any gas or stars \citep[see e.g.][ and references therein]{Treu++10}. In a strong gravitational lens a distant background source is magnified and multiply-imaged by an intervening massive object. 
The image positions provide a strong constraint on the large-scale (kpc in projection) `macro-model', while the relative magnifications between the images are sensitive to low mass perturbations (pc in projection or smaller depending on source size). Lensed light is sensitive to perturbations along the entire line of sight from the source to the observer and thus provide a constraint on the halo mass function along this entire path \citep{Xu++13, Li++17, McCully++17,  Despali++18, Ritondale++18, Gilman++18}.

In order to be suitable for detecting low mass halos, a gravitational lens must have either four point source images or an extended arc in order to provide a constraint on the large scale mass distribution of the deflector. The source must emit at sufficiently long wavelengths so as to avoid differential dust extinction between lensed images, and it must be extended enough ($\sim$milli-arcseconds in projection) to be unaffected by microlensing \citep[e.g.][]{Anguita++08, Yonehara++08}. 

Gravitational lenses with strongly lensed galaxy sources meet these criteria.  In gravitational imaging, low-mass perturbations to the main deflector mass distribution are observed as astrometric perturbations to lensed arcs \citep{Vegetti++09b, Vegetti++09}. Current imaging with HST/Keck AO yields sensitivity to halo virial masses of M$_{200}\sim 10^9$ M$_\odot$ with this method \citep{Vegetti++12, Vegetti++14, Ritondale++18}. \citet{Ritondale++18} analysed a sample of 17 galaxy-galaxy strong lenses finding one significant detection of a perturbing halo. Constraints from lensed galaxies will improve with the next generation of ground-based telescopes and adaptive optics which will enable the detection of lower mass halos. 

Instead of studying individual detections of perturbers, it is also possible to study the cumulative effect of many low mass perturbers \citep{Hezaveh++14, Cyr-Racine++18, Bayer++18}. \citet{Birrer++17} applied this method to the lens RX J1131-1231 ruling out thermal relic dark matter candidates with masses less than 2 keV at 2$\sigma$ confidence. This result did not include the effects of line of sight structure. 

With fixed spatial resolution, smaller sources are in general more sensitive to lower mass perturbers. Quadruply imaged radio jets are the traditional source for measuring the halo mass function \citep{Dalal++02}. \citet{Hsueh++19} studied a sample of 7 quadruply imaged, radio-loud quasars and included the effects of structure along the line of sight, constraining the WDM particle mass to be greater than 3.8 keV at 2$\sigma$ confidence. This result is competitive with results from Ly-$\alpha$ forest and Local Volume measurements and provides a promising test of this completely independent method. 

In order to make progress it is necessary to increase the sample of lenses which can be used for this analysis. 
\citet{Gilman++19} simulated samples of compact source lenses with realistic line of sight structure populations and varying dark matter halo mass functions. They found that given typical flux ratio measurement precisions of $\sim 4\%$ approximately 10-40 lenses would be needed to rule out 3.3 keV WDM with 2$\sigma$ confidence, depending on the normalization of the subhalo mass function, thereby providing a tighter constraint on the free streaming length of dark matter than conservative Ly-$\alpha$ forest constraints.
Radio loud quasars are rare. There are currently of order ten and it is expected that few of them will be detected in the next decade. 
 Radio weak systems (i.e. systems classified as radio quiet but effectively not radio silent), may extend the existing sample in the radio domain, but the origin of their radio emission is yet debated and generally observed to be significantly extended, making them more suited for gravitational imaging type analyses \citep[e.g.][]{Jackson++15, Hartley++19}. In contrast, wide-field optical imaging surveys have recently discovered large samples of quadruply imaged quasars \citep[e.g.][]{Shajib++19}

Strongly lensed Active Galactic Nucleus (AGN) narrow-line region emission provides an exciting path forward for gravitational lensing studies as it greatly increases the sample of systems which can be used to measure low mass halos without being affected by contamination from microlensing \citep{Moustakas++03}. Nearly all optically detected AGN have significant narrow-emission flux, whereas very few have strong radio emission.  
In addition, the narrow-line region is smooth and extended enough to be unaffected by microlensing \citep{Muller-Sanchez++11}. There are currently a few dozen confirmed quasar lenses, the majority of which have been discovered in the past five years by the Strong Lensing Insights into the Dark Energy Survey (STRIDES) team \citep{Treu++18} and similar efforts in other wide-field surveys \citep[e.g.][]{Anguita++18, Lemon++18, Agnello++18, Lemon++19, Agnello++18b, Ostrovski++18, Rusu++19, Agnello++18c, Schechter++17}. Future surveys such as Euclid and LSST are forecast to contain thousands of such systems \citet{Oguri++10} making narrow-line lensing a promising path forward for strong gravitational lensing studies of dark matter.

\citet{Nierenberg++14} demonstrated that spatially resolved narrow-line flux measurements obtained with Keck OSIRIS provided sensitive enough constraints to be able to detect a M$_{200}\sim 10^7$M$_\odot$ dark matter halo in the plane of B1422+231. Currently, Keck is the only facility with an integral field unit coupled to adaptive optics which gives sufficient wavelength and sky coverage as well as spatial resolution for this experiment. HST offers much more accessibility as it can probe a much larger wavelength range thanks to the lack of atmosphere, and it can also target most of the sky. \citet[][(N17)]{Nierenberg++17} analyzed WFC3 IR grism observations of the gravitational lens HE 0435 (HST-GO-13732, P. I. Nierenberg), finding that the data provided sufficient spatial and spectral resolution to be sensitive to halos with masses M$_{200}\sim 10^{7}$M$_\odot$. 

In this paper we present narrow-line flux ratio measurements for a sample of 8 systems (Section 2). This sample includes the remaining 5 lenses observed from HST-GO-13732, and three of the systems from program HST-GO-15177 (P. I. Nierenberg) which were observed before March 2018.  

This paper is organised as follows. In Section 2 we describe the lens selection. In Section 3 we describe the observing strategy and initial data reduction.   In Section 4 we present our statistical fitting method. In Section 5 we present the resulting integrated emission line fluxes. In Section 6 we compare the measured flux ratios with predictions from smooth model fitting. In Section 7, we discuss the effects of resolved source light on our measurements. In Section 8 we provide a brief summary of the main conclusions of this work.

In order to calculate physical sizes, we assume a flat $\Lambda$CDM cosmology with
$h=0.7$ and $\Omega_{\rm m}=0.3$. All magnitudes are given in the AB
system \citep{Oke++1974}.

 \section{The lens sample}
The lenses in the HST programs GO-13732 and GO-15177 were selected from all known quad quasar lenses at the time of proposing, which had either [OIII] 5007, 4959~\AA~ or [NeIII] 3869, 3968~\AA~observable in a grism filter, and could not be observed from the ground either due to an unsuitable redshift for adaptive optics, or a southern declination, where spatial resolution for integral field spectroscopy at the wavelengths of interest is not yet adequate. As of August 2019, all 6 systems have been observed for program GO-13732 and 7 of 9 systems have been observed for GO-15177. In this work we present results from a new data reduction pipeline which we have applied to all systems observed before March 2018. Table 1 provides a summary of key information about the lenses studied here. 

\begin{table*}
\centering
\begin{tabular}{lllllll}
\hline
\hline
Lens  			  & z source & z deflector &   Narrow Lines$^{a}$   & Obs. Configuration$^b$ & Exposure Time (s)$^c$ & Discovery      
	\\
\hline

WGD J0405-3308     & 1.713   &		 &[OIII]	       & F140W/G141       & 387/2112	           & \citet{Anguita++18} \\
HS 0810+2554	  & 1.506	  & 		 &[OIII]	        & F140W/G141       & 398/2112		   & \citet{Reimers++02} \\
RX J0911+0551	  & 2.763   & 0.769$^{d}$	 &[NeIII]	        & F140W/G141       & 497/5011 		   & \citet{Bade++97}   \\
SDSS J1330+1810    & 1.383  & 0.373	 & [NeIII]	&  F105W/G102	   & 447/5111		   & \citet{Oguri++08}  \\
PS  J1606-2333	  & 1.696   &		  &[OIII]	        & F140W/G141	   & 472/2012		   & \citet{Lemon++18} \\
WFI 2026-4536	  &  2.23    &      	  &[OIII]	        & F140W/G141	   & 472/5312		   & \citet{Morgan++04}\\	
WFI 2033-4723        & 1.66     & 0.661$^{e}$	    &[OIII]	& F140W/G141	    & 447/5312		   &\citet{Morgan++04} \\
WGD 2038-4008     & 0.777	 & 0.230	     &[OIII]	& F105W/G102	   &	497/2062	   &\citet{Agnello++18} \\
\hline \hline
\end{tabular}
\caption{Summary of key lens data and exposure times. $a$~[OIII] refers to the 4960, 5007~\AA~doublet, while [NeIII] refers to the 3870, 3969~\AA~doublet. $b$ WFC3 IR direct/grism filters.$c$~Total time over all dithers. Deflector redshift measurements if different from discovery paper: $d$~\citet{Kneib++2000}, $e$~\citet{Eigenbrod++06}. }
\end{table*}

\section{Observations and Initial Reduction}
\label{sec:data}
The WFC3/IR grism provides slitless dispersed spectra where all of the light within the field of view is dispersed along the x direction of the detector.  
A ``direct image'' in an imaging filter (preferably overlapping with the grism passband) is used to determine a wavelength reference for the dispersed image.
To achieve this, we used the same observing strategy as was used by \citet{Nierenberg++17}.  Spectroscopic and direct imaging observations were split into four point dither patterns with dithers chosen to have quarter pixel offsets to recover sub-pixel information.  The WFC3/IR grism passband was chosen to encompass the narrow-line of interest with necessary exposure times estimated based on observations of optical broad lines and average quasar line ratios of low redshift quasars from \citet{VandenBerk++01}. Each grism exposure was followed or preceded by a short direct exposure at the same location in order to calibrate the wavelength solution, with the direct exposure filter chosen to match the wavelength of the grism exposure. F105W was used in conjunction with G102 (0.8-1.15 $\mu\mathrm{m}$, point source resolution $\sim$20~\AA~per pixel) grism exposures and F140W for G141 (0.8-1.15 $\mu\mathrm{m}$, point source resolution $\sim40$~\AA~ per pixel) grism exposures.  Table 1 provides a summary of the observations.

We use the {\tt grizli} software package to obtain astrometric solutions for the images and to perform flat-fielding, background subtraction, and cosmic ray rejection.\footnote{A full description of the software along with example usage can be found at \tt{https://github.com/gbrammer/grizli}}. 

This software routine differs from the precursor {\tt threedhst} \citep{Brammer++12, Momcheva++16} which was used in \citet{Nierenberg++17} for the reduction of data for the lens HE 0435. The major difference is that {\tt grizli} generates grism wavelength solutions (i.e. how the pixels in the direct frame map onto the grism frame as a function of wavelength \citep{ Kuntschner++10}) in the native frame of detector pixels (i.e., the calibrated FLT files provided by the {\tt wfc3ir} calibration pipeline) rather than in the interlaced frame used by the {\tt threedhst} software. This represents a major improvement in two ways: 

First, it is more robust to large dithers. This is due to the fact that the native pixel frame of WFC3 does not have a uniform collecting area per pixel owing to geometric distortions. For small dithers this fractional variation is small and thus approximately the same fractional pixel information is recovered for each dither, which leads to recovery of sub-pixel information in the interlaced frame. For large dithers, smooth sub-pixel recovery is not obtained. This is particularly important for several of the systems presented here which are extremely bright (i$<$ 17) and thus saturated rapidly in the direct imaging exposure. We used large dithers for these objects to ensure that persistence would not affect our measurements. For these objects the {\tt threedhst} reduction pipeline produced unsatisfactory sub-pixel `interlaced' images.

Second, in the native FLT, the PSF is known to a high degree of sub-pixel accuracy, as it can be obtained directly by imaging star fields. This is not the case for drizzled or interlaced images as the PSF will depend on dither size and exposure times. 
For our analysis we use the empirical PSF models from \citet{Anderson++16} \footnote{These PSF models are available at \tt{http://www.stsci.edu/hst/wfc3/analysis/PSF}}. We found that the effective PSF models provided a consistently much better fit to the data than models based on stars in the images.

\begin{figure*}
\centering
\includegraphics[scale=0.6]{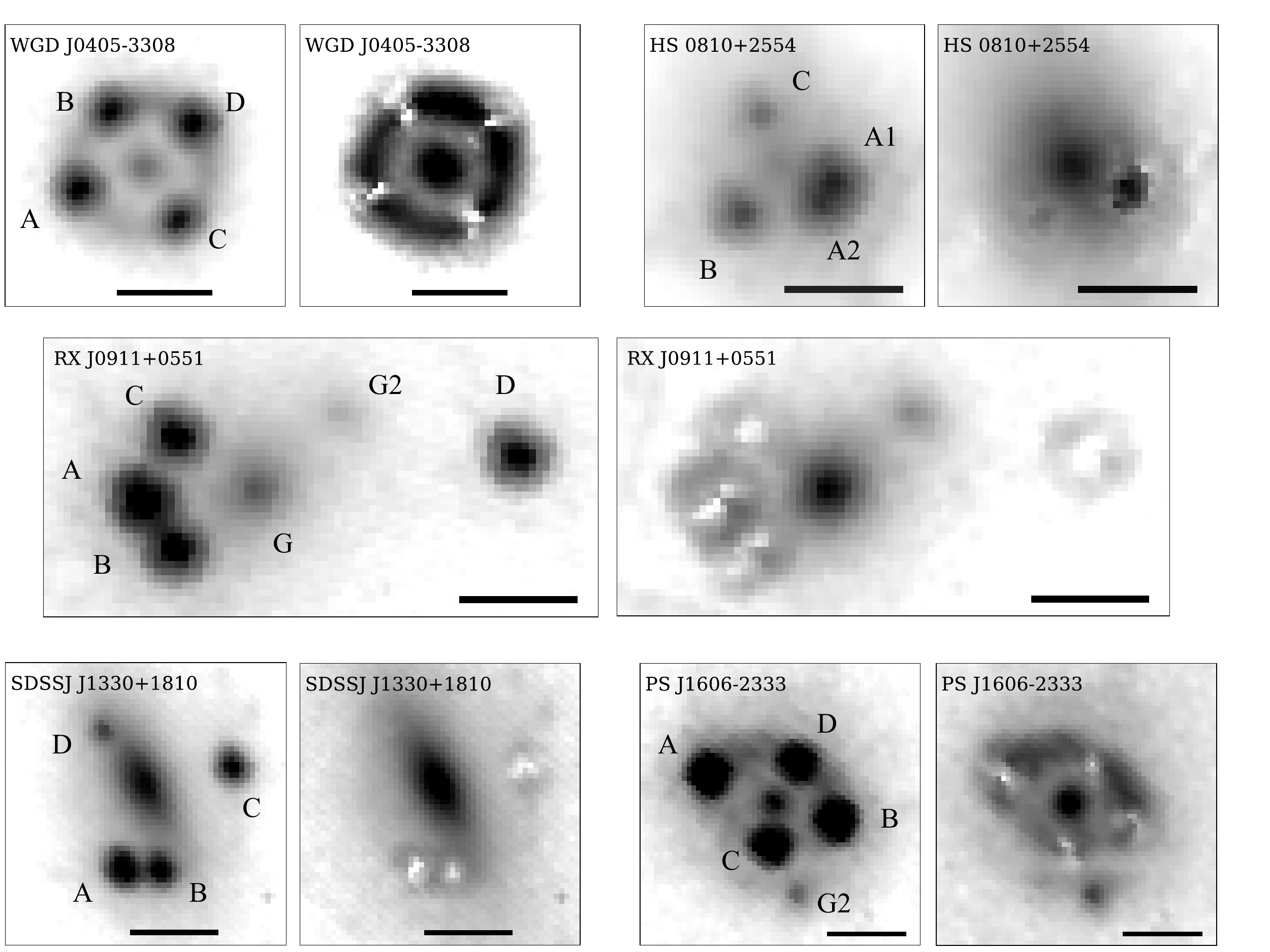}
\hspace{1mm}
\includegraphics[scale=0.6]{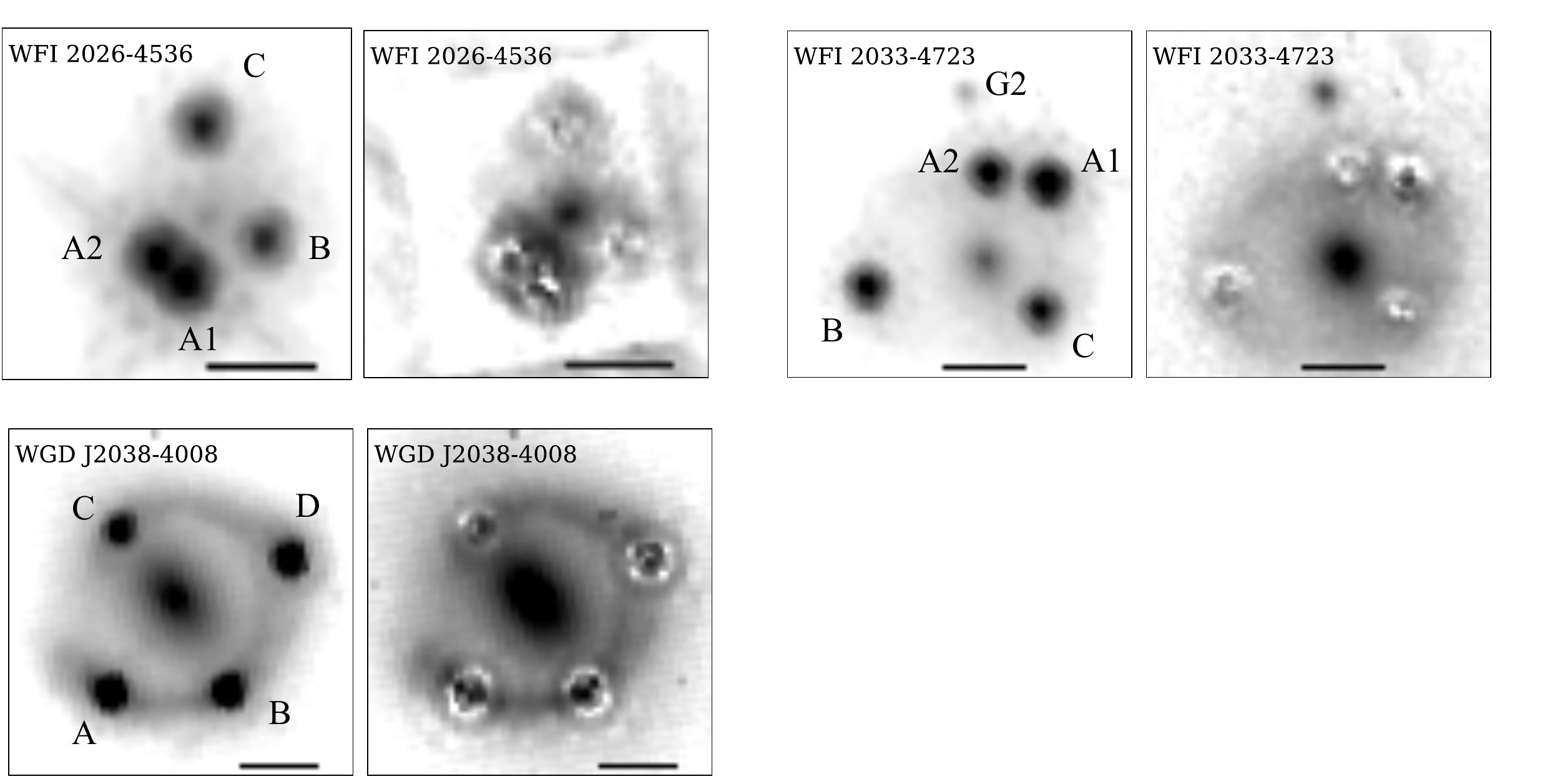}

\caption{Drizzled direct F140W (F105W for SDSS J1330 and WGD J2038) images of the lenses, along with quasar image subtracted residuals. Quasar images are modelled as point sources using the Effective Point Spread Function \citep{Anderson++16} in the native FLT frame (see Section 4.1). All images are rotated relative to the observing frame such that North is up and East left. Bars indicate one arcsecond. With the exception of SDSS J1330, which shows clear evidence for a disk, the deflecting galaxies are smooth, massive ellipticals. The majority of lenses have extended arcs from the strongly lensed quasar host galaxy.}
\label{fig:prmass}
\end{figure*}

\section{Spectral Extraction and fitting}
As described in the Introduction, narrow-line fluxes provide a robust probe of dark matter structure, as the nuclear narrow-line region is $\sim$mas in extent given typical lensing configurations \citep{Muller-Sanchez++11, Moustakas++03}, and thus insensitive to the magnifications induced by stars in the plane of lens galaxies, which have characteristic scales of $\mu$as. Furthermore, narrow-line fluxes are not time-variable on time-scales relevant to galaxy-scale lenses \citep{Peterson++13}.  Here we describe how we extract narrow-line fluxes. 

We account for blending along the y-axis as well as the low grism spectral resolution by adopting a forward modelling procedure. This procedure is nearly identical to that used in \citet{Nierenberg++17},  except that in this work the modelling occurs in the native FLT direct and grism image frames, while in \citet{Nierenberg++17}, the modelling was done in an interlaced frame (see Section 3). Figure 1 of \citet{Nierenberg++17} provides an illustration of this method which can be divided in two steps. First we create a direct-image model for each component for which we wish to infer the spectrum, in the next step we use the {\tt grizli} pipeline to generate simulated 2D grism images based on proposed model spectra and the model direct images.  The full model of the lens plus quasar spectra is then simply the linear sum of these component spectra. The goodness of fit between proposed model spectra and the data is computed in the 2D grism frame to fully account for blending between components, as well as to naturally account for low grism spectral resolution. In the following subsections we discuss the modelling choices in more detail.

\subsection{Direct Image Fitting}

For each direct F140W or F105W image, we generate a separate model for each direct image component that will contribute a spectrum to the grism image. These direct image models include four point sources, one for each quasar, modelled using the effective PSF from \citet{Anderson++16}; a S\'ersic profile \citep{Ser68} for the main deflector and any other nearby galaxies; and an empirical model for the lensed quasar host galaxy if visible. The empirical model for the lensed quasar host light is generated by iteratively subtracting the best fitting quasar and galaxy models which result from the direct image modelling, creating a mask from the residuals, and then refitting the quasar and lens galaxy light with the empirical model for the lens quasar host light masked. {We adopt an empirical model for the lensed quasar host galaxy light, rather than assuming an empirical model for the quasar host galaxy and adopting the best fitting lens model to generate the model lensed arc because the quasar host galaxy light is extended and thus spectrally significantly subdominant to the quasar point sources. }

We fit the direct image components in all FLT frames simultaneously, allowing the sky background and overall normalization of the galaxy, quasar host and quasar images to vary, but keeping other model parameters fixed between the images. We find that the relative offsets we measure between quasar images vary less than 0\farcs005 compared to other results from deep HST imaging \citep{Shajib++19}, while our galaxy light centroid measurement can differ by up to $\sim0\farcs01$. This is likely due to our empirical method of subtracting the ring light, as well as due to our relatively shallow single band imaging. Both of these measurement precisions are more than adequate to construct a robust direct image model which enables accurate extraction of spectra from the grism. The measured quasar image and galaxy positions are listed in Table 2.

Figure 1 shows the drizzled direct images for the lenses, as well as the drizzled residuals after subtracting the best fitting quasar point source models.  The lens galaxies are smooth elliptical galaxies, which are typically expected for lenses, with the exception of SDSS J1330,  
which has a disk which is both evident in this direct imaging data and was also noted by \citet{Rusu++16} who used Subaru imaging.
We discuss the effect of this on the lens modelling in more detail in Section 6.

\subsection{Spectral Fitting}
\label{subsec:specfit}

We model all spectra which may be blended with the quasar spectra. This includes the spectrum of the deflector (and any nearby galaxies or stars close in y-projection), as well as the lensed quasar host galaxy which is visible as a ring or partial ring in all of the systems except HS 0810. We find that the galaxy,  star and lensed quasar host are all adequately fit using straight-line continuum models with the slope and normalization allowed to vary. We find these straight line models are adequate given the short spectral ranges being examined (about 600~\AA~ in the quasar rest frame), and the relative faintness of these spectral components.

We fit quasar spectra in two different rest frame ranges. When made possible by the quasar redshift, we observed the quasar spectra in the range  $\sim 4700-5300$~\AA~ in order to measure [OIII] 4959 and 5007~\AA~lines. Two lenses, RXJ0911 and SDSS J1330 did not fall in an appropriate redshift range. For these, we targeted the fainter [NeIII ] 3870, 3969~\AA~lines which also originate in the quasar narrow-line region \citep[e.g.][]{Osterbrock++06}. 

\subsubsection{[OIII] region spectral fitting}

In the quasar rest frame from  $\sim 4700-5300$~\AA, there can be significant contributions from continuum, broad FeII and broad H$\beta$ in addition to the narrow [OIII] emission which we wish to measure. 

Continuum and broad emission features are all time variable as well as small enough to be differentially magnified by stars in the plane of the lens galaxy. This differential magnification by stars can affect not only the overall amplitude, but also the shapes of the continuum and broad emission, as bluer continuum light and higher velocity broad-line emission 
are emitted from systematically smaller regions which are therefore 
are more susceptible to microlensing \citep{Abajas++02,  Keeton++06, Anguita++08, Mosquera++11, Blackburne++11, Sluse++07, Sluse++11, Sluse++12, Sluse++14, Blackburne++14, Jimenez-Vicente++14, Fian++18, Bate++18}.

To account for these effects we allow the continuum slope and normalization to vary between each quasar image.
The FeII redshift, line broadening and amplitudes are each allowed to vary independently from the H$\beta$ redshift, width, and amplitude for each lensed image.  FeII line broadening is generated by convolving a model spectrum which has no intrinsic velocity dispersion with a constant velocity-space kernel. 

We tested two FeII templates; a purely empirical template based on IzwI \citep{Boroson++92}, as well as a more recent theoretical model based on FeII iron groups \citep{Kovacevic++10}. The \citet{Kovacevic++10} model has more flexibility as the separate iron group amplitudes can vary independently reflecting varying quasar temperature states. In all cases we found that this model provided a fit which is many orders of magnitude better than the purely empirical IzwI model, thus all results here are obtained using the \citet{Kovacevic++10} model.

We tested allowing the FeII line broadening to vary between the lensed images to simulate the effect of microlensing on altering the FeII line shape but found that this did not improve the model fit significantly and had no effect on the target [OIII] flux ratios.

For each lens we tested multiple different models for the H$\beta$ emission in order to test how these model choices affected the inferred flux ratios. These models were \emph{i)} a single Gaussian, \emph{ii)} a fifth order Gauss-Hermite polynomial, \emph{iii)} two Gaussian components with the Gaussian components amplitudes and widths allowed to vary between the images. In the majority of lenses, varying the model choice did not affect the measured [OIII] flux ratios. The exceptions were WGD J0405 and WFI 2033 for which the best fitting model also provided a dramatically improved fit relative to the other models. In all cases including WGD J0405 and WFI 2033,  the flux ratios and models presented here are for the best fitting H$\beta$ model choice.  We note that given the low spectral resolution of the grism we do not attempt to separate narrow and broad H$\beta$ components. Although our models are flexible enough to account for such features, we are agnostic as to our ability to disentangle them accurately, given our goal of measuring [OIII] emission.

Finally, the [OIII] lines are each fit with a single Gaussian, with widths and offsets fixed between the lensed images. The amplitude ratio of the lines is fixed to the quantum mechanical predicted value of 3:1. The model allows for a systemic redshift offset between the [OIII] doublet and broad emission lines.

\subsubsection{[NeIII] region spectral fitting}

We fit two different spectral ranges for RX J0911 and SDSS J1330 owing to the different spectral resolutions of the G141 and G102 grisms.

For RX J0911 for which we used the lower resolution G141 filter, we fit the 2800-5100~\AA~region. We modelled H$\delta$ 4103~\AA ~and H$\gamma$ 4341~\AA~emission as these lines were slightly overlapping with each other and with the [NeIII] 3969~\AA~line.
We allowed the H$\delta$ and H$\gamma$ widths and amplitudes to vary independently between the quasar images. We fit the quasar continuum with a straight-line.

For SDSS J1330, the higher resolution spectrum and narrower broad line widths enabled us to exclude H$\gamma$ from the fit, however the model residuals clearly required H$\epsilon$ at 3971~\AA, which falls under [NeIII] 3969~\AA~line. We required the H$\epsilon$ and H$\delta$ widths to be the same for a given quasar spectrum, but allowed this width to vary between lensed quasar spectra. We held the relative amplitudes of H$\delta$ and H$\epsilon$ fixed between quasar spectra, but allowed the total amplitude of both to vary between lensed spectra. We found that the continuum emission for this system was better fit with a power-law rather than a straight line. Analogously to the case of RX J0911, we allowed the power-law slope and the normalization to vary between the quasar images to account for microlensing. 

The [NeIII] amplitude ratios were held fixed at the quantum mechanical predicted value of 1:3. As in the case of [OIII], we hold the [NeIII] doublet widths fixed between the two lines, and between the different quasar spectra. Also as in the case of [OIII], the model allows a systematic redshift offset between narrow and broad emission lines.

We also model the [OII] 3729~\AA~doublet as a single Gaussian with independent amplitude, width and redshift from the [NeIII] lines. We include this flexibility because [OII] is generally observed to be more strongly excited by star formation than AGN activity \citep[e.g.][]{Ho++05, Davies++14, Maddox++18}. Therefore we assume it is likely to have a complex structure and be more spatially extended than [NeIII] and thus we do not expect it to be lensed in the same way \citep[see e.g.][]{Sluse++07}. We include this region of the spectrum in the fit to ensure that we have a constraint on the continuum emission blue-ward of [NeIII].

\section{Spectral Fitting Results}

Table 2 gives the narrow-line and continuum flux ratios from the spectral fitting with the exception of HS 0810 (see Sections 6.2 and 7). Narrow-line fluxes are integrated over the entire spectral line while continuum fluxes are integrated from 4550 to 5500~\AA~for [OIII] lenses, 2800-5100~\AA~for RX J0911 and 3600-4200~\AA~for SDSS J1330. We do not report broad emission fluxes as our model does not distinguish between broad and narrow emission line components for non-forbidden lines.

 Figures A1 through A8 in the Appendix show detailed results including the model spectra, separate line components, and a comparison between the best fitting model and data PSF-weighted traces in the lower panels. The traces are computed by performing a PSF-weighted extraction in the y- direction on the 2D grism image, and thus include all the effects of blending between the spectra, as well as blurring due to low grism spectral resolution, and the effects of varying grism sensitivity with wavelength.
 
\begin{figure*}
 
\includegraphics[scale=0.35,trim = 60 20 0 20 clip=true]{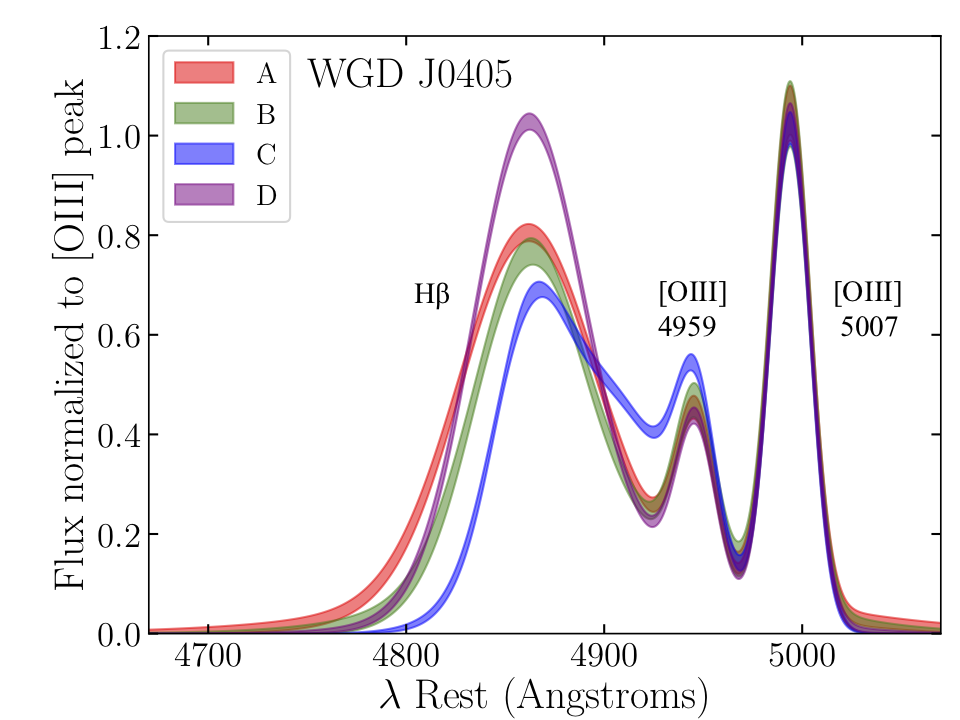} 
\hspace{3mm}
\includegraphics[scale=0.35,trim = 20 20 0 20 clip=true]{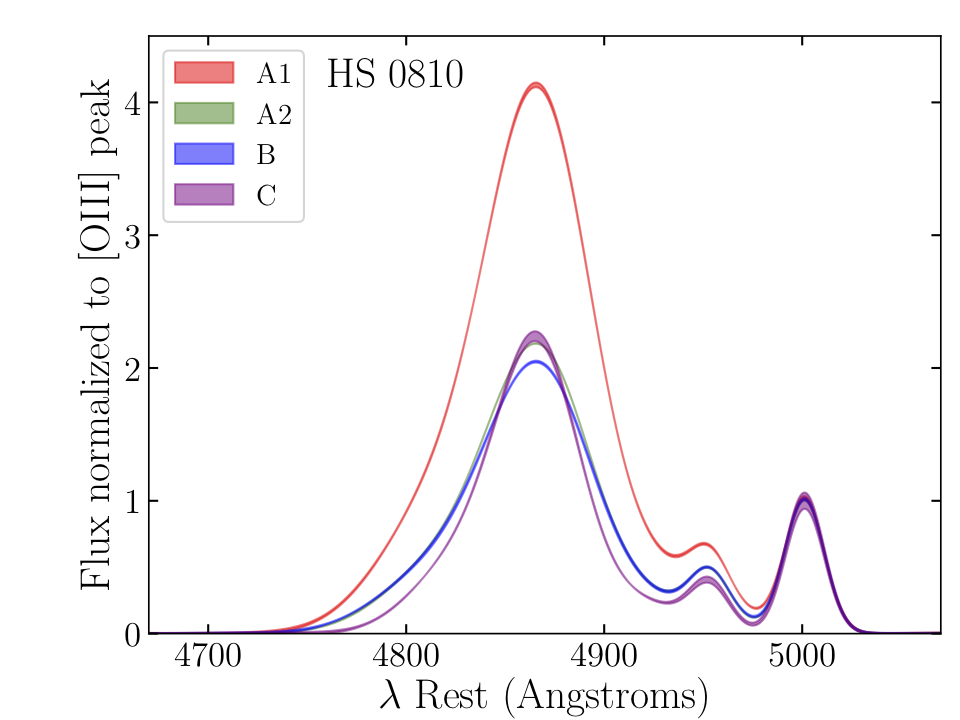} 
\hspace{3mm}
\includegraphics[scale=0.35,trim = 20 20 0 20 clip=true]{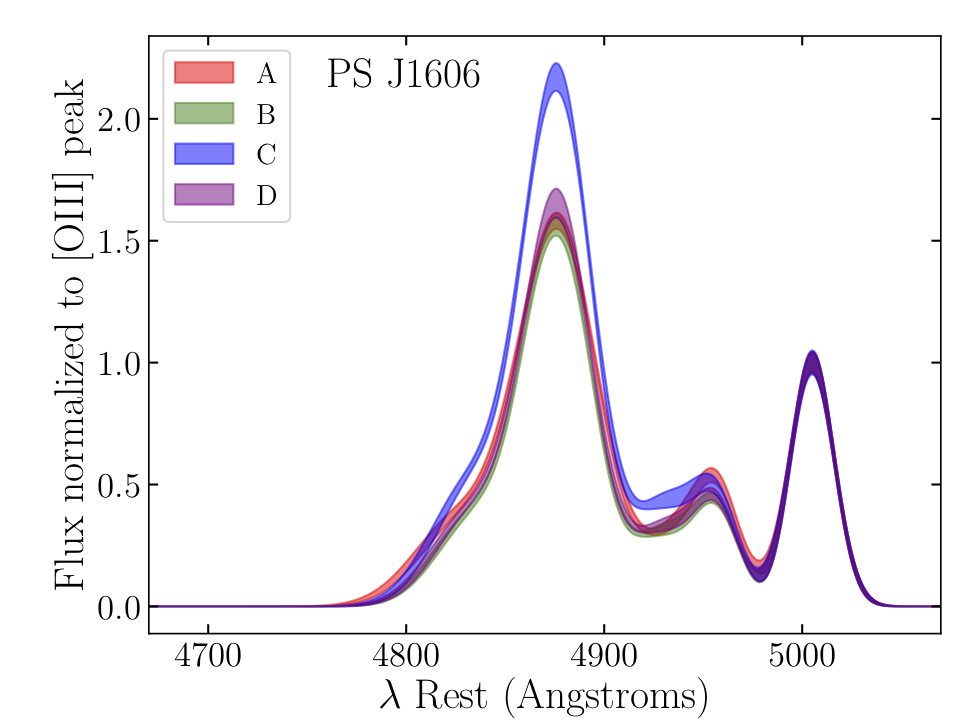} 
\vspace{5mm}

\includegraphics[scale=0.35,trim = 20 20 0 20 clip=true]{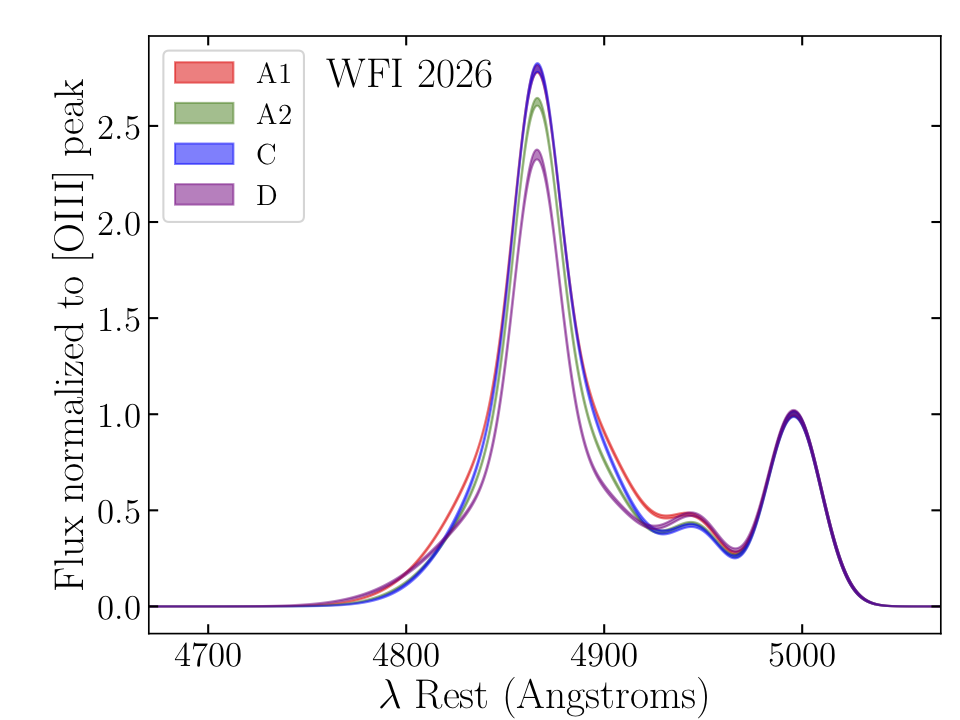} 
\hspace{3mm}
\includegraphics[scale=0.35,trim = 20 20 0 20 clip=true]{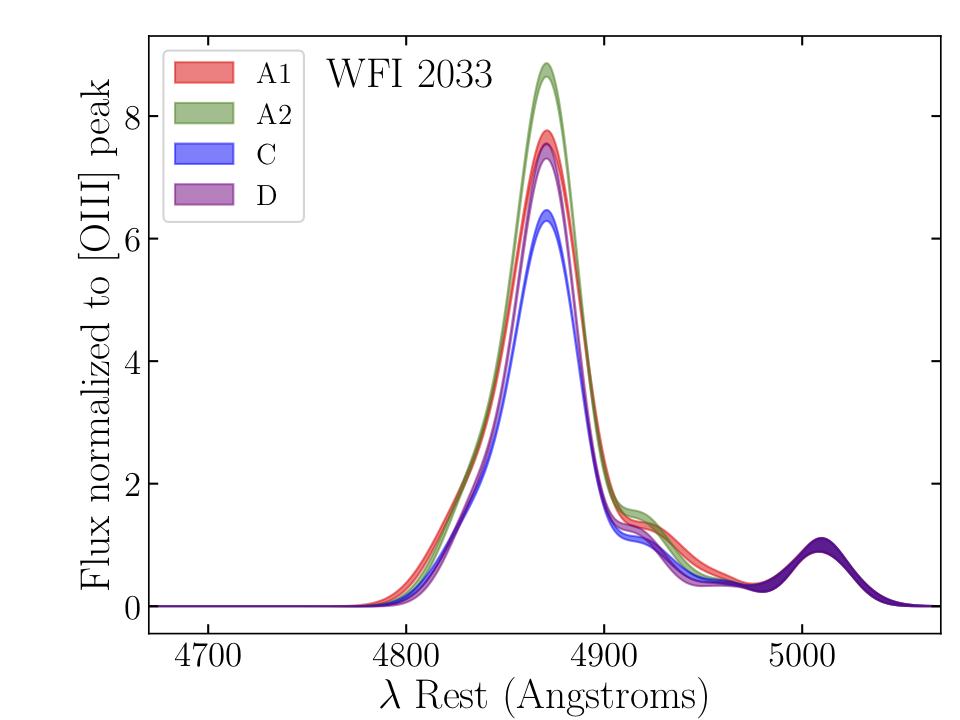} 
\hspace{3mm}
\includegraphics[scale=0.35,trim = 20 20 0 20 clip=true]{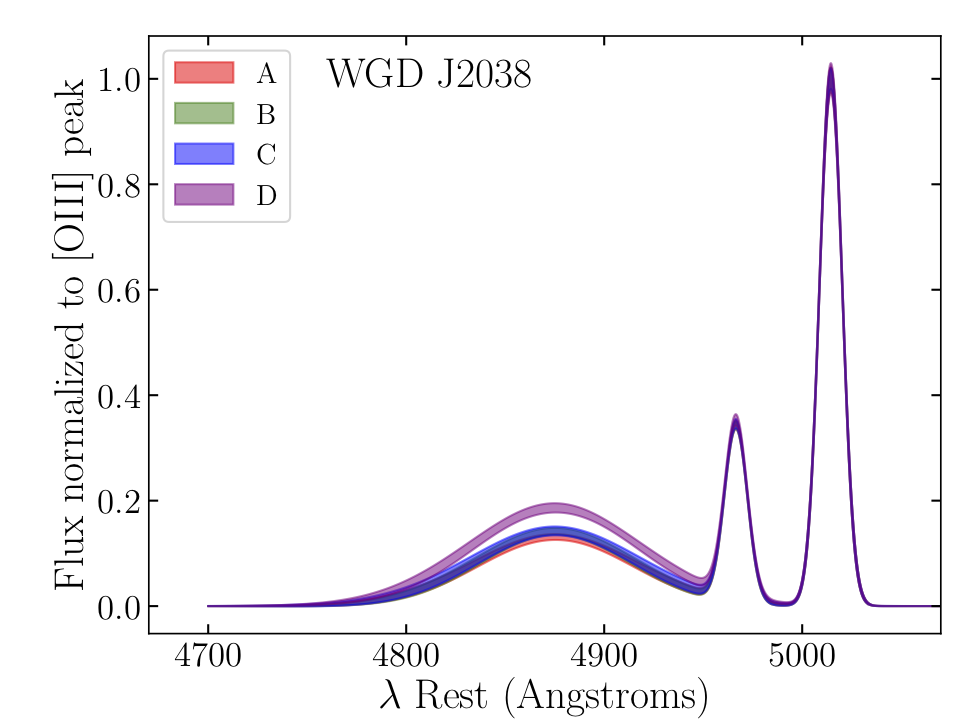}

\caption{Model fit to the broad and narrow emission lines normalized to the peak of the [OIII] flux to highlight differential magnification between the broad H$\beta$ emission and [OIII]. Line widths represent one sigma posterior confidence intervals. H$\beta$ is emitted from a region of $\sim \mu$as in extent making it subject to magnification by stars in the plane of the lens galaxy. WGD J0405, HS 0810 and PS J1606 in particular show significant distortion in images D, A1 and C respectively. The shape of image C in WGD J0405 also shows a relatively magnified red wing compared to the other images, another indication of differential microlensing. We note that in performing spectral fitting we do not separate components of broad and narrow Balmer emission, owing to low grism resolution, thus the signal from microlensing is somewhat under-represented by this comparison, since the narrow component will not be affected by microlensing. }
\label{fig:microlensing}
\end{figure*}

\begin{figure*} 

\includegraphics[scale=0.35,trim = 20 20 0 20 clip=true]{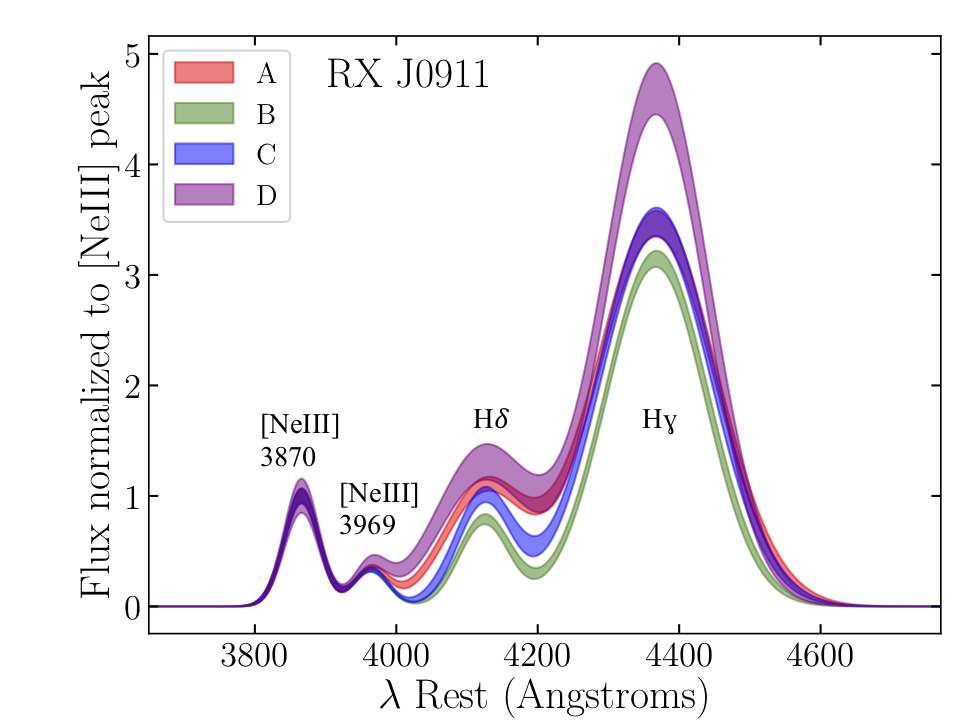} 
\hspace{3mm}
\includegraphics[scale=0.35,trim = 20 20 0 20 clip=true]{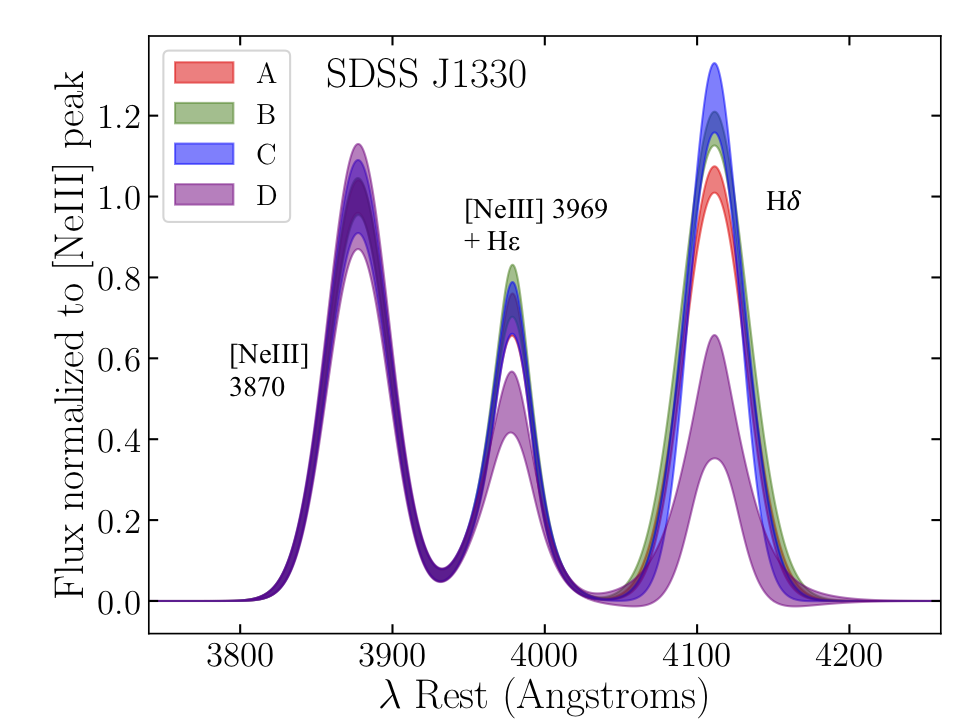} 
\caption{Model fit to the broad and narrow emission lines normalized to the peak of the [NeIII] flux to highlight differential magnification between the broad and forbidden components. Line widths represent one sigma posterior confidence intervals. Broad H$\gamma$, H$\delta$ and H$\epsilon$ are all emitted from a regions of $\sim \mu$as in extent making them subject to magnification by stars in the plane of the lens galaxy. Both lenses show significant differential lensing between the narrow and broad emission features. We note that in performing spectral fitting we do not separate components of broad and narrow Balmer emission, thus the signal from microlensing is somewhat under-represented by this comparison.}
\label{fig:microlensing2}
\end{figure*}

As discussed in Section 4,  for typical quad quasar lens configurations, the quasar broad and continuum emission regions have angular sizes of micro-arcseconds or smaller, making them susceptible to significant magnification by stars in the plane of the lens galaxy. In contrast, the quasar narrow-line region is milli-arcseconds in scale \citep[e.g.][]{Moustakas++03, Muller-Sanchez++11, Sluse++07} and therefore unaffected by microlensing. Comparing the relative magnifications of broad and narrow spectral features for the different lensed images reveals the differential magnification due to stars. We highlight this in Figures \ref{fig:microlensing} and \ref{fig:microlensing2}, where we plot only the Balmer and forbidden narrow-line components from our model fits, normalized to the peak of the narrow forbidden line emission flux.  Note that owing to low grism resolution we do not attempt to differentiate between broad and narrow Balmer components. As the narrow Balmer components are not micro-lensed, this comparison likely underestimates the microlensing signal on the Balmer lines.

The lenses WGD J0405, HS 0810, PS J1606, RX J0911 and SDSS J1330 all show significant deviations in the narrow to broad fluxes for at least one image. As expected, the images with these deviations are also the ones for which the narrow and continuum flux ratios vary the most, as can be seen in Table 2. Continuum fluxes can also vary significantly on the time scales of days which may also contribute to variations between continuum, broad and narrow flux ratios between lensed images.

\begin{table*}
\small
\centering
\begin{tabular}{llllllll}
\hline
\hline
Lens & Image  &M/S$^{a}$ & dRA   & dDec &   Continuum$^b$  &  NL Flux$^{c}$ 	& Model Flux Ratio$^d$     	\\
\hline
 		   & A   & M & 1.066 & 0.323	        & $1.00\pm 0.02$     & $1.00   \pm 0.04$   &     1    \\
                    &B    & S  &    0     &  0           & $0.73\pm 0.01$      & $0.65 \pm 0.04$  &   $0.8^{+0.3}_{-0.2}$            \\
WGD 0405 & C    & S  &  0.721 & 1.159    & $0.87\pm 0.02$       &  $1.25 \pm 0.03$    &    $1.0^{+0.6}_{-0.3}$             \\
		 &	D    & M    &-0.157   & 1.021 & $1.12\pm 0.02$      &  $1.17   \pm 0.04$   & $1.1^{+0.05}_{-0.1}$      \\
		 & G 	&        & 0.358      & 0.567      &				                				&				\\

\hline	
		     & A1  & M      & 0	  &  0 	         &$1.00\pm 0.01$   &   ${e}$      &    1     		 		    \\
		    &  A2  		         & S	    & 0.087    & -0.167& $0.472\pm 0.006$  & 	   &    $0.95^{+0.07}_{-0.08}$    \\
HS 0810      &   B    & M	    & 0.775    & -0.258  &$0.236\pm 0.003$   &        &   $0.25\pm 0.03 $  \\
		    &  C    	& S	    & 0.613    & 0.589                    &$0.078\pm0.002$   &        &    $0.14\pm 0.01$	\\
		    &  G     &            & 0.460    	&	0.150         	&				           &     &		 	\\

\hline

 & A    & S     & 0 	     &  0                           &0.60$\pm$0.02       & $0.56 \pm 0.04$   	&  $0.47^{+0.05}_{-0.07}$   \\
 & B    & M   &   0.258     & 0.405               & 1$\pm0.02$        & 1$\pm 0.05$         &   1 	\\
RX J0911 &  C    &  S   &   -0.016    & 0.959 &  0.47$\pm0.01$      & $ 0.53 \pm 0.04$       &    $0.47^{+0.06}_{-0.03}$  \\
 & D    & M  & -2.971      &  0.791               &0.43$\pm0.01$         & $0.24 \pm 0.04$        &  $0.24\pm 0.04$ \\
 & G    &       & -0.688      & 0.517      &   \\
 & G2  &       & -1.455      & 1.174      & \\

\hline	
& A     &   M	 			 & 0  		&   0            &1 $\pm 0.04$		        &     $ 1.00  \pm 0.05$     &  1   			    \\
& B	&   S	        			&  -0.414    &  -0.012   &0.52 $\pm 0.03$  		& 	$ 0.79   \pm 0.04$ 	   &  $0.94^{+0.05}_{-0.06}$  \\
SDSS J1330 & C	&  M  	 &  -1.249       & 1.167  &0.45$\pm 0.02$	       &   	$ 0.41  \pm 0.04 $     &   $0.42^{+0.04}_{-0.03}$   \\
& D	&  S      				 & 0.237          & 1.582   &0.11$\pm0.005$ 	        & 	$ 0.25 \pm 0.03$       & $0.193^{+0.01}_{-0.009}$   \\
& G	 &        				& -0.226         &  0.978 		  & 			\\

\hline	
 & A   & M	 & 1.622     & 0.589   	  &1$\pm0.02$              & 	$ 1.00   \pm 0.03$ 	   &  1							\\
 & B   & M	 & 0   	  &  0  	   &       1.15$\pm 0.02$		&      $ 1.00  \pm 0.03 $     & $1.24^{+0.05}_{-0.05}$       			  \\
PS J1606 &C   &  S	  & 0.832	  &  -0.316   &0.71$\pm0.01$        &   $ 0.60  \pm 0.02$      &   $0.52^{+0.1}_{-0.09}$			\\
& D   &   S      & 0.495    & 0.739   	   &0.72$\pm0.01$         & 	$ 0.78 \pm 0.02$        & $0.6\pm0.1$ \\
& G1&           & 0.784     & 0.211 		   &		                   & 		\\
& G2 &	   & 0.477    & -0.942 		    &  				             & 		\\

\hline
 & A1   &    M 		&  0.164  & -1.428     & 1$\pm$0.02    	      & $1 \pm 0.02$        & 1 \\
 & A2    &  S   		& 0.417  & -1.213             &  0.65$\pm0.01$      & $0.75 \pm 0.02$   &  $0.72\pm0.06$ \\
WFI 2026  & B      &  M         &  0        & 0     &  0.28$\pm0.01$      & $0.31\pm 0.02$    &   $0.31\pm 0.02$  \\
 & C      &   S 		 & -0.571  & -1.044           &  0.22$\pm 0.01$     & $0.28  \pm 0.01$   & $0.28\pm 0.02$ \\
& G 	&                     & -0.023   & -0.865  & 			       &  	     \\

\hline
 & A1   &   M   &  -2.196  & 1.260                     &    1$\pm 0.02$    &  $1 \pm 0.03$            &     1     \\
& A2   &   S   &  -1.484  & 1.375                      &    0.56$\pm0.01$       & $0.64 \pm 0.03$      &    $0.69^{+0.1}_{-0.09}$         \\
& B     &   M    &  0        & 0                             &     0.55$\pm0.01$    & $0.50\pm 0.02$     &  $0.57^{+0.07}_{-0.08}$		\\
WFI 2033 & C    &    S  & -2.113  & -0.278     &   0.43$\pm 0.01$      & $0.53  \pm 0.02$     & $0.34^{+0.05}_{-0.06}$	\\
  & G 	&     & -1.445   & 0.307    &  	         \\
& G2 	&     & -1.200   & 2.344  &  	         \\

\hline
 & A    & M   & -2.306    & 1.708                     & 1.0$\pm0.08$          & 	$ 1.00 \pm 0.01$	   &  1  	\\
 & B    &  M & 0	  &  0 	  	                  &  0.94$\pm0.09$           &     $ 1.16  \pm 0.02$    & 1.21$\pm 0.01$    	\\
WGD 2038 & C   &   S & -1.518   & 0.029     &  1.09 $\pm$0.08	& 	$ 0.92   \pm 0.02 $    & $0.99\pm0.1$  	\\
 & D    & S   & -0.126    & 2.089                      &  0.45$\pm0.03$   & 	$ 0.46 \pm 0.01$       &	$0.46\pm 0.07$ 	\\
 & G	&	    & -0.832 & 1.220  & 					& 						\\

\hline \hline
\label{tab:fluxRatios}
\end{tabular}
\caption{Summary of measured lens properties. Image and galaxy positions are measured from direct F105W or F140W imaging, while continuum and narrow emission lines are measured from G102/G141 spectra (see Table 1).
Lens data summary: Image and lens galaxy relative positions, as well as lensed image continuum and narrow-line flux ratios. All values assume that the source is unresolved. 
$a$: Whether the image is a maximum or saddle point of the time delay surface based on lens modelling. 
$b$: Measured between 4550-5500~\AA~for all except RX J0911 where it is measured between 2800-5100~\AA~  and SDSS J1330 measured between 3600-4200~\AA~.
$c$: [OIII] 4959, 5007~\AA~ for all lenses except RX J0911 and SDSS J1330, for which [NeIII ] 3870, 3969~\AA~was measured.
$d$: The model fluxes are obtained from fitting the image positions as described in Section \ref{sec:lensmodelling}.
$e$: The merging pair of images in HS 0810 is likely blended owing to the measured finite source size and high magnification (see Sections 6.2 and 7), thus [OIII] fluxes of these images are not well represented by a PSF model. }
\end{table*}

\section{Smooth lens modelling}
\label{sec:lensmodelling}

The narrow-line measurements reported in this paper double the sample of compact-source systems which can be used to measure the low mass end of the halo mass function \citep[e.g.][]{Dalal++02, Gilman++18, Gilman++19, Hsueh++19}\footnote{Although approximately fifteen systems had been previously measured with either radio or mid-IR imaging, only 7 of these are currently useful for detecting dark matter structure owing to a variety of factors including extremely complex deflector morphologies, or uncertainties as to the source of radio emission. See \citet{Hsueh++19} for a detailed description of these systems.}.
Our goal in this paper is primarily to present these new narrow-line measurements.  In this section we aim to provide basic insight into how well our measured narrow-line flux ratios are reproduced by smooth lens models which do not include additional low mass halos. For all lensing calculations we use {\tt lenstronomy} \citep{Birrer++15, Birrer++18}\footnote{https://github.com/sibirrer/lenstronomy} 

\subsection{Lens Model Choices}
As has been traditionally done \citep[recent examples include][]{Gilman++17, Hsueh++18, Hsueh++19}, we model the deflector mass distributions as power-law ellipsoids with an additional contribution from external shear to account for the influence of the group environment of typical gravitational lenses. 

One subset of the power-law ellipsoid model, the Singular Isothermal Ellipsoid (SIE), has projected power-law mass slope of $-2$, and has been shown to generally provide a good fit to the combined stellar and dark matter mass distribution of massive ellipticals \citep{Rusin++03,  Gavazzi++07, Treu++10, Gilman++17, Hsueh++18}. This is traditionally used as a baseline `smooth mass model'. SIE models with external shear are generally observed to fit image flux ratios to better than 10\% in the absence of significant baryonic complexities such as disks, which are readily observable in high resolution data from e.g. Keck with Adaptive Optics or HST. \citet{Hsueh++17, Hsueh++18} demonstrated that such baryonic disks can be incorporated in the lensing model as additional mass components.  Figure 1 shows the lenses with light from the lensed quasars subtracted. The deflectors are all smooth ellipticals with the exception of SDSS J1330, which shows clear evidence for a disk. Thus we do not expect a single component power-law lens model to adequately fit the fluxes for this lens, however exploring the addition of a baryonic disk to the mass profile is beyond the scope of the present work. 

For this work, we adopt a more flexible model for our lenses than SIE. We allow the power-law slope of the projected mass profile to vary between $-1.9$ and $-2.2$. This more flexible range is chosen to encompass the range of mass slopes measured in SLACS gravitational lenses which have been measured with a combination of stellar kinematics, weak and strong lensing \citep{Auger++10, Gavazzi++07}. 
Several of the lenses in this sample including WFI 2033 \citep{Rusu++19}, WGD J0405, SDSS J1330, PS J1606 and WGD J2038 \citep{Shajib++19} have mass models based on the lensed arcs of the quasar host galaxy from deep, multi-band HST imaging. The inferred mass profiles of these systems all fall within our prior range with the exception of WGD J2038 which \citet{Shajib++19} found to have an inferred slope of $-2.35\pm0.04$. For this lens we extend the uniform prior to $-2.4$.  We choose to use relatively uninformative uniform priors in our analysis here rather than tighter priors based on these works as our goal is to provide a preliminary look at a range of lens models that might fit the image positions. 

We note that the framework of \citet{Gilman++18, Gilman++19} has been tested extensively on simulated data sets of flux ratio lenses, and been shown to accurately infer the correct dark matter mass function without incorporating information from lensed arcs. The information from lensed arcs would provide an additional constraint on the macromodel and detection of halos with virial masses $\sim10^9$M$_\sun$ and above, given HST resolution imaging \citep[e.g.][]{Vegetti++12, Vegetti++14, Ritondale++18}, thus future iterations of these pipelines may incorporate such information.

We impose a Gaussian prior on the centroid of the main deflector to be within 0\farcs05 of the light centre measured from direct F105W/F140W imaging. This uncertainty is chosen to incorporate the uncertainty in fitting the centroid of the light profile, as well as to account for imperfections in the mass model which can lead to apparent offsets \citep[e.g.][]{Shajib++19}. We adopt uniform priors for the other parameters of the mass model: $q$, the lens ellipticity, $\phi$ lens orientation, $\theta_E$ the Einstein radius, and $\gamma_{\rm{ext}}$, and $\phi_{\rm{ext}}$ the magnitude and orientation of the external shear.

For lenses RX J0911, PS J1606, and WFI 2033 which each have an additional galaxy close in projection to one of the images (labelled G2 in Figure 1), we include this additional component as a Singular Isothermal Sphere in our mass model, with Gaussian prior of 0\farcs05 on the offset between light and mass centroid as for the main deflector. We adopt broad uniform priors with widths of $\pm$0\farcs3 on the Einstein radii of the G2 galaxies based on results from previous lens models for these systems. For RX J0911, we allow the perturber Einstein radius to vary between 0\farcs$03<\theta_E<$0\farcs6 based on the best fit value of 0\farcs24 from \citet{Blackburne++11}. For 1606 we adopt a range of 0\farcs$03<\theta_E<$0\farcs5 based on the result of \citet{Shajib++19}\footnote{Perturber mass estimate of 0\farcs2 Shajib, Private Communication}.  For 2033 we choose the range 0\farcs$03<\theta_E<$0\farcs3 based on the lens model from \citet{Rusu++19} as they found a mass for the perturber consistent with zero.  

Finally we model the narrow-line source as a Gaussian, with FWHM allowed to vary between 20-50 pc. This range encompasses the range of values measured in nuclear quasar narrow-line emission measured in high resolution, spatially resolved IFU observations of low redshift quasars \citep{Muller-Sanchez++11}. Finite source sizes act to `dampen' the effects of small perturbations from either the macromodel or dark matter structure thus we include this to partially account for these effects, although it is possible that the nuclear narrow-line region for some of these systems may be more or less extended. We discuss this possibility in more detail in Section 7.

\subsection{Flux Ratio Posterior}

Our goal is to explore a comprehensive range of smooth models which can provide a fit to our data. As discussed in the Introduction, the image positions provide strong constraints on the smooth mass distribution, as they are determined by the first derivative of the gravitational potential, while image magnifications are sensitive to the second derivative and thus probe local low-mass fluctuations in the mass distribution. 

The deviation of image fluxes from the fluxes predicted by a fit to the image positions with a smooth mass distribution provide an indication of small scale perturbations to the mass distribution. This can be understood as follows:
 in the absence of perturbations to a smooth mass distribution, we expect that both the image positions and fluxes will be well fit by a smooth mass distribution. The perturbations to the image positions from low-mass halos can typically be absorbed by the smooth macromodel, while the perturbations to image fluxes typically cannot be absorbed by the macromodel and thus reveal the presence of local fluctuations in the lensing potential.
 
To test the extent to which our measured narrow-line fluxes deviate from the range of models which provide adequate fits to the image positions, we use the positions of the measured direct image PSF positions and uncertainties. This assumes that the centroid of the narrow-line region and continuum emission are the same.  This could potentially lead to a mismatch if there is a significant offset between the quasar continuum emission and the centre of the nuclear narrow-line emission, however the position uncertainties of 0\farcs005 correspond to $\sim10$ pc for an image magnification of 10 of a redshift 1.5 quasar. This is significantly larger than typical offsets observed in local Seyfert 1 galaxies \citep{Muller-Sanchez++11}. The \citet{Muller-Sanchez++11} sample, however, is relatively small and it is possible that high redshift or more luminous quasars may exhibit different behaviour.

The lens model we have chosen can provide an arbitrarily good fit to any set of four image positions. For each lens, we draw image positions $1.4\times 10^4$ from Gaussians centred on the measured direct image positions, with widths of 0\farcs005, and use {\tt lenstronomy} \citep{Birrer++15, Birrer++18} to solve for the best-fit macromodel parameters for each draw of a set of image positions. We then calculate the image fluxes predicted by that macro-model based on the source size that was drawn independently for that iteration. This process explores the full parameter space of flux ratios `predicted' by a set of image positions (as well as knowledge of the main deflector and perturber positions), and is similar to that used by \citet{Birrer++19} to assess uncertainties for time-delay cosmography. The last column of Table 2 provides the range of flux ratios predicted by this procedure, while Table 3 provides a summary of the range of macromodel parameters. The bottom row in Figures A1-A8 (with the exception of A2, see Section 7) compare the 2D contours for the model predicted range of flux ratios with the observed narrow-line flux ratios. 

One result of this modelling was that given the image positions of HS 0810, the merging pair of images was predicted to have magnifications of $\sim 120$ each. This is an order of magnitude higher than the magnifications of the images of other lenses in the sample. It also implies that given HST resolution and observed nuclear narrow-line region sizes, the merging pair of images is likely significantly blended and thus not well represented by a `flux ratio'. We discuss this in more detail in Section \ref{sec:resolved}. Given this, we omit HS 0810 from the comparison with `smooth' mass model predictions. We also omit SDSS J1330 from this comparison because it has a disk which requires additional complexity in the macro-model to fit image fluxes and positions adequately \citep[e.g.][]{Gilman++17, Hsueh++17, Hsueh++18}. 

\begin{figure}
\centering
\includegraphics[scale=0.45, trim = 0 0 0 0 clip=true]{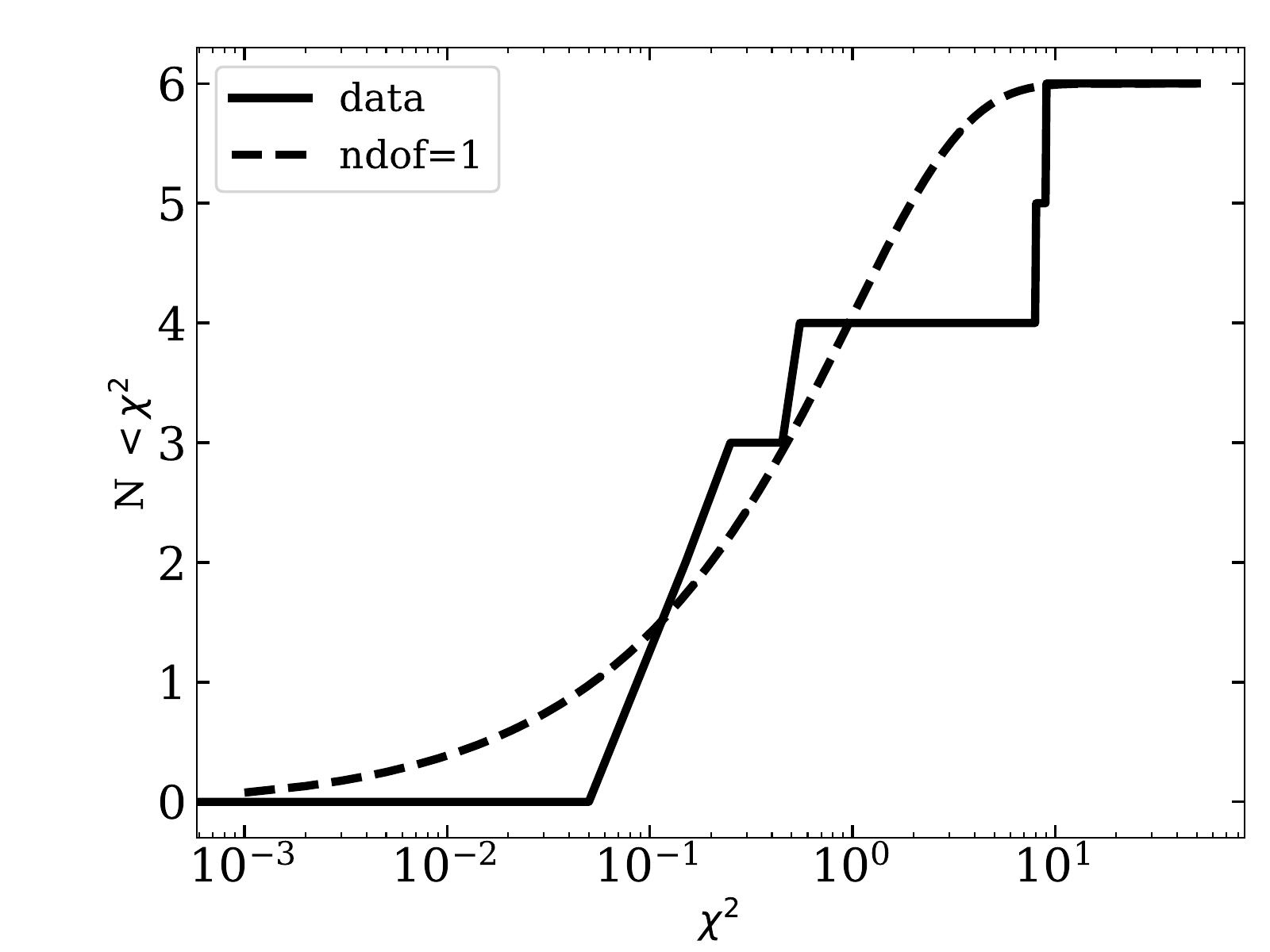} 
\caption{Cumulative distribution of $\chi^2$ values between the measured flux ratios, and the 1D marginalized posterior distributions of smooth gravitational lens model predicted flux ratios for the lenses in this paper excluding SDSS J1330 and HS 0810 (see Sections 6.2 and 7).  Smooth lens model fits are performed using only the lensed image positions, which are fit extremely well. We compare the observed $\chi^2$ distribution for one degree of freedom, which is expected if the image fluxes were drawn from the smooth models. The significant discrepancy in the two distributions indicates the detection of additional required model complexity, such as low mass dark matter halos. These data are used to measure the properties of low mass dark matter halos in \citep{Gilman++19b}.}
\label{fig:xisq}
\end{figure}

We provide a simplified comparison of how well the data is fit by the models by comparing the one dimensional, marginalized posterior distributions for the model predicted flux ratios (given in the last column of Table 2), with the measured flux ratios.  If the measured flux ratios were drawn perfectly from the models, with the only deviations coming from uncorrelated Gaussian flux noise, then the $\chi^2$ values should be Gaussian with one degree of freedom. In Figure \ref{fig:xisq} we show the cumulative $\chi^2$ distribution of model fit to the flux ratios of the lenses compared to the expected distribution for a sample with one degree of freedom. The observed $\chi^2$ values deviate significantly from the one degree of freedom case, indicating that it is unlikely that the measured fluxes were drawn from the smooth model predictions. The probability of the two distributions being the same is p$<0.005$. We note that this comparison ignores the complex, covariant, non-Gaussianity of the posterior distributions for the model flux ratios, and thus significantly under-represents differences between the model predictions and the data. We chose this simplified comparison rather than a more complex comparison as our goal is to simply provide a general sense of the agreement between smooth model and data, rather than to make a definitive statement about the presence of dark matter which is performed in \citet{Gilman++19b}.

\subsection{Sources of `unsmoothness'}
In this subsection we discuss several different factors which might contribute to deviations between observed flux ratios and the values predicted from models with smooth mass distributions.

Differential dust extinction is unlikely to play a significant role in altering the flux ratios. First, because of the high redshift of the source quasars, optical rest-frame quasar light is redshifted into the wavelength range of $\sim8000-10000$~\AA~in the rest-frame of the deflectors. Dust extinction in this wavelength range is typically of order of a few hundredths of a magnitude \citep{Fal++99, Ferrari++99}, which is within the measurement uncertainties for fluxes presented here. SDSS J1330 is clearly a late-type galaxy and thus likely has a higher dust content, which may contribute to differential extinction for images B and D which lie near the disk, however we have not included this lens in our comparison with the distribution of smooth model predictions. 

Elliptical power-law mass distributions provide good descriptions of the combined stellar plus dark matter distributions of early-type galaxies. Deflectors with massive disks such as the one in SDSS J1330 are not well represented by elliptical mass distributions and require explicit modelling of the baryonic component to avoid the spurious detection of dark matter perturbers \citep[e.g.][]{Hsueh++16, Hsueh++17}. 

In the absence of obvious disks, \citet{Gilman++17} studied typical mass model deviations expected from deep imaging of low redshift galaxies in Virgo. This study did not allow the slope of the mass distribution to vary, and they found that maximum flux ratio values of $\sim10\%$ may be observed due to baryonic non-smooth components relative to the smooth model expectation. \citet{Hsueh++17} found a similar result by looking at simulated galaxies. The majority of lenses in our sample have multiple image fluxes which deviate at least this much from the power-law model prediction for multiple images, far above the expectation from elliptical galaxies. 

It is expected that we would observe significant discrepancies relative to smooth model predictions, as we have not included halo mass structure in our models (other than luminous neighbouring galaxies) which lenses are sensitive to. \citet{Gilman++19b} use the statistical inference machinery developed by \citet{Gilman++19, Gilman++18} to simulate populations of dark matter halos along the entire line of sight from the lens to the observers in order to provide a physical interpretation for the discrepancy between the observed and smooth model image fluxes in terms of dark matter models.

\section{Resolved Source}
\label{sec:resolved}
In our spectral fitting, we assumed that the narrow-line emission is unresolved.  In the local universe, \citet{Muller-Sanchez++11} measured narrow emission regions in Seyfert 1 quasars to have typical FWHM of order $\sim 10-60$ pc. 
Given a typical lens source redshift of 1.5 and magnification of a factor of $\sim 5$ this would yield an observed source of $\sim$ mas in scale. In this case, modelling the emission as unresolved would not run the risk of missing flux. Although the quasars in our lens sample are typically 10-100 times brighter than those studied in \citet{Muller-Sanchez++11}, the scaling of the size of the nuclear narrow-line emission with redshift and luminosity is not well known.  \citet{Nierenberg++17} showed that the data strongly disfavoured intrinsic source sizes larger than 100 pc for the narrow emission in HE 0435, and even an intrinsic FWHM of 50 pc was disfavoured relative to the point source model.  When the narrow-line emission was resolved this caused the narrow-line to have characteristic differential widths in the grism spectra of the different images owing to the different axis of shear for each of the images relative to the grism dispersion direction (see Figure 8 of \citet{Nierenberg++17}). 

\citet{Nierenberg++17} also used simulated extended sources of 50 and 100 pc (angular sizes after magnification of 3 and 30 mas) to test the effect of incorrectly using a point source model to measure flux ratios, finding that the error introduced by this was much less than the measurement uncertainties. 

Typical image magnifications for the lenses in this sample are close to $\sim5-20$, similar to magnifications in HE 0435. An important exception is  HS 0810 for which the lens model predicts extremely high magnifications of $\sim 120$ for the two fold images (A1 and A2 respectively). In the case of such high magnifications, a FWHM even as small as $10$ pc would be magnified to an observed size of 90 mas, causing significant blending between the two images which are separated by only 180 mas. Given the orientation of the images in this system, there would be no obvious signature of the differential broadening of the narrow-emission in the 1D spectra since the shear direction is perpendicular to the grism dispersion direction for all but the faintest image.

To test whether we can constrain the intrinsic source size for HS 0810,  we use the best fit lens model based on the image positions to simulate four source sizes; a 1, 40, 100, and 200 pc source. These sizes correspond to observed sizes of 9, 340, 860 and 1700 mas respectively. Simulating resolved sources requires several additional steps relative to simulating unresolved sources. First, we generate a high resolution `true image' based on the best fit macromodel. Next we convolve this pixelized image with a model of the PSF. This PSF model is generated by drizzling EPSF models at the location of the quasar from the four FLT frames.  Finally, after convolving the simulated extended image with the simulated PSF in the drizzled frame, we use the Multidrizzle {\tt blot} function to generate four simulated direct FLT exposures of the extended source. We then  these simulated extended lensed images into {\tt grizli} to represent the light distribution of the [OIII] emission.
 
 Of the four resolved source models, we find that the 40 pc source provides the best fit to the data with a log likelihood improvement of 186 relative to the baseline point source model.  The extended source model has three fewer degrees of freedom because the [OIII] flux ratios are fixed to the lens model predicted values unlike the original model in which they are free to vary independently. The 1, 100 and 200 pc sources are strongly disfavoured with relative log likelihoods of $ -1136$, $-103$ and $-1145$ respectively, thus we conclude that the intrinsic source size is likely between 10 and 100 pc. 

Several of the steps involved in simulating resolved images, in particular the drizzling of the PSF and then blotting into the FLT frames likely introduce noise into the simulated models. The 1 pc source provides a control to measure the extent of degradation of the model relative to the original model based on the fitting the EPSF to the native pixel frame. Given that 9 mas is a tiny fraction of the 120 mas FWHM of the PSF, ideally it should provide a nearly identical model to the original model based on the EPSF. The fact that it does not indicates that significant noise is added by our method of simulating resolved images. Thus we estimate that the source size is likely between 10-100 pc but do not attempt a more precise constraint here. A more robust constraint would require additional simulation using a range of PSF and source size models which we leave to a future work. 

The fact that the HS 0810 narrow-line emission appears to be resolved has exciting implications for studies of the spatial extent of the narrow-line region at relatively high redshifts, however the blending between the images indicates that the narrow emission for this system should not be studied as simple compact-source for this lens.

Given the order of magnitude lower magnifications in the other lenses, as well as the absence of differential narrow-line width between lensed images, we expect the point-source should provide an adequate model for the narrow-line flux in the other lenses.

\begin{table*}
\centering
\begin{tabular}{lllllllllll}
\hline
\hline
Lens  & $\theta_{\rm{E}}^{a}$   & dRa$^{b}$  & dDec$^{b}$ & $\epsilon ^{c}$ & $\phi^{d}$ & $\gamma_{\rm{ext}}^e$ &$\phi_{\rm{ext}}^f$ & $\theta_{\rm{E,2}}^{g}$ & dRa$_{2}^h$ & dDec$_{2}^h$ \\ 
\hline    
WGD J0405 & $0.703^{+0.007}_{-0.004}$ &$0.027^{+0.01}_{-0.009}$ &$-0.010^{+0.009}_{-0.01}$ &$0.030^{+0.02}_{-0.009}$ &$90^{+30}_{-30}$ &$0.13^{+0.1}_{-0.05}$ &$80^{+40}_{-40}$ \\ 

HS 0810 & $0.493^{+0.005}_{-0.004}$ &$0.005^{+0.003}_{-0.003}$ &$0.02^{+0.02}_{-0.01}$ &$0.11^{+0.05}_{-0.04}$ &$21^{+3}_{-4}$ &$0.05^{+0.03}_{-0.02}$ &$20^{+100}_{-10}$ \\ 

RX J0911 & $1.0^{+0.2}_{-0.2}$ &$0.2^{+0.2}_{-0.1}$ &$-0.02^{+0.04}_{-0.04}$ &$0.2^{+0.3}_{-0.1}$ &$160^{+10}_{-70}$ &$0.25^{+0.07}_{-0.1}$ &$100^{+10}_{-5}$ & $0.3^{+0.2}_{-0.2}$ & $0^{+0.05}_{-0.05}$ &$0^{+0.05}_{-0.05}$ \\ 

SDSS J1330 & $0.949^{+0.01}_{-0.008}$ &$-0.02^{+0.02}_{-0.02}$ &$0^{+0.01}_{-0.01}$ &$0.15^{+0.04}_{-0.04}$ &$20^{+2}_{-3}$ &$0.079^{+0.01}_{-0.008}$ &$167^{+9}_{-200}$ \\ 

PS J1606 & $0.67^{+0.03}_{-0.03}$ &$0.01^{+0.02}_{-0.03}$ &$0.02^{+0.02}_{-0.02}$ &$0.2^{+0.1}_{-0.1}$ &$162^{+4}_{-6}$ &$0.11^{+0.04}_{-0.03}$ &$43^{+20}_{-20}$ &$0.3^{+0.2}_{-0.2}$ &$0^{+0.05}_{-0.05}$ &$0^{+0.05}_{-0.05}$ \\ 

WFI 2026 & $0.67^{+0.01}_{-0.01}$ &$-0.059^{+0.008}_{-0.007}$ &$0^{+0.02}_{-0.02}$ &$0.16^{+0.06}_{-0.06}$ &$100^{+4}_{-2}$ &$0.07^{+0.02}_{-0.02}$ &$150^{+10}_{-20}$ \\ 

WFI 2033 & $1.09^{+0.02}_{-0.02}$ &$-0.01^{+0.02}_{-0.03}$ &$0^{+0.02}_{-0.01}$ &$0.08^{+0.03}_{-0.03}$ &$50^{+30}_{-20}$ &$0.09^{+0.08}_{-0.05}$ &$107^{+20}_{-9}$ &$0.2^{+0.1}_{-0.1}$ &$0^{+0.05}_{-0.05}$ &$0^{+0.05}_{-0.05}$ \\ 

WGD J2038 & $1.39^{+0.01}_{-0.01}$ &$0^{+0.01}_{-0.01}$ &$0^{+0.01}_{-0.01}$ &$0.12^{+0.06}_{-0.05}$ &$40.0^{+2}_{-0.9}$ &$0.05^{+0.03}_{-0.03}$ &$122^{+2}_{-5}$ \\ 

\hline \hline
\label{tab:lensModelling}
\end{tabular}
\caption{Lens modelling results based on fitting the quasar continuum emission positions (see Section 6). Columns defined as follows: $(a)$ Einstein radius,$ (b)$: dRA and dDec of the lens mass centroid relative to the lens light centroid, $(c)$ the lens ellipticity defined as $(1-q)/(1+q)$, $(d)$ orientation of the lens major axis counter clockwise from North, $(e)$ magnitude of external shear, $(f)$ orientation of external shear, $(g)$ Einstein radius of G2 if present, $(h)$ dRa and dDec of G2 mass centroid relative to the G2 light centroid. }
\end{table*}

\section{Summary}
Using the WFC3/IR grism, we have measured spatially resolved narrow-line fluxes in 8 new gravitational lenses, doubling the number of compact-source lenses which can be used to measure low mass dark matter halos. 
Our main conclusions are as follows:

\begin{enumerate}
\item We present a new forward modelling pipeline which uses the {\tt grizli} pipeline to extract spectra from grism FLT images using direct {\tt flt} models for the light distributions. This pipeline accounts for blending between neighbouring quasar images as well as with light from the host and quasar host if present. It provides a significant improvement to the fit to quasar point source light, and accommodates large dithers.

\item We present narrow-line fluxes for eight gravitational lenses. We also compare the shape of the narrow emission to the relative height of the broad emission, revealing significant microlensing features in images of five of the eight lenses. 

\item We fit the lensed image positions with a smooth mass distribution for the deflector, with a variable mass profile slope as well as ellipticity, orientation, Einstein radius and external shear. We find that the distribution of flux ratios predicted from the smooth models differs significantly from the observed distribution of narrow-line flux ratios, indicating the need for additional complexity in the models such as low mass dark matter halos.

\end{enumerate}
\section*{Acknowledgments}
We thank the STRIDES team for sharing early information about their lens discoveries which enabled preparation for these observations. 
We thank Matt Malkan and Aaron Barth for extremely helpful conversations regarding AGN physics and spectral signatures.

Based on observations made with the NASA/ESA Hubble Space Telescope, obtained at the Space Telescope Science Institute, which is operated by the Association of Universities for Research in Astronomy, Inc., under NASA contract NAS 5-26555. These observations are associated with programs \#13732 and \#15177.

Support for programs \#13732 and \#15177 was provided by NASA through a grant from the Space Telescope Science Institute, which is operated by the Association of Universities for Research in Astronomy, Inc., under NASA contract NAS 5-26555. AMN, TT and AP acknowledge support from grant \#13732. AMN, TT, SB and DG acknowledge support from \#15177.

AMN acknowledges support from the University of California Irvine, Chancellor's Postdoctoral Fellowship, as well as from the NASA Postdoctoral Program Fellowship, and from the Center for Cosmology and AstroParticle Physics Postdoctoral Fellowship. 

LAM acknowledges support from NASA through grant number \#14305 from the Space Telescope Science Institute, which is operated by AURA, Inc., under NASA contract NAS 5-26555 and from the NASA ROSES program 17-ATP17-0120.

TA acknowledges support from Proyecto FONDECYT N: 1190335.
VM acknowledges partial support from Centro de Astrof\'{\i}sica de Valpara\'{\i}so. 
AA was supported by a grant from VILLUM FONDEN (project number 16599). This project is partially funded by the Danish council for independent research under the project ``Fundamentals of Dark Matter Structures'', DFF--6108-00470. 
Part of this research was carried out at the Jet Propulsion Laboratory, California Institute of Technology, under a contract with the National Aeronautics and Space Administration. 

\bibliographystyle{apj_2}
\bibliography{references}

\appendix
\onecolumn
\section{Appendix}
Here we show detailed results from spectral fitting for each lens, as well as a comparison of the measured flux ratios to the model predicted flux ratios.

\renewcommand{\thefigure}{A\arabic{figure}}

\setcounter{figure}{0}

\begin{figure*}
{\centering
\includegraphics[scale=0.45, trim = 0 0 0 0 clip=true]{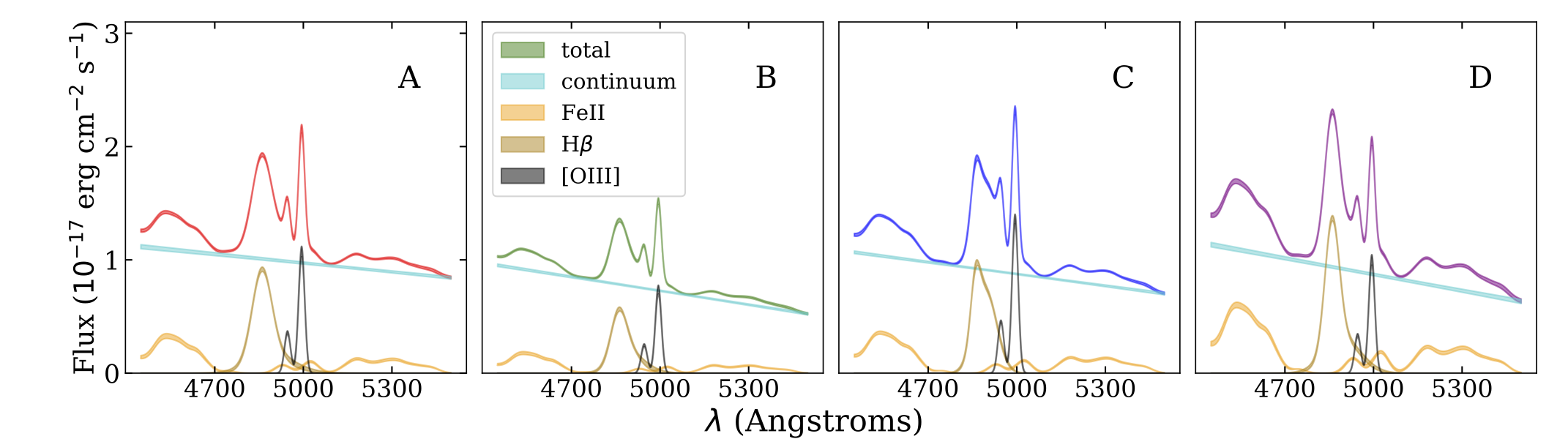}

\includegraphics[scale=0.45, trim = 0 0 0 0 clip=true]{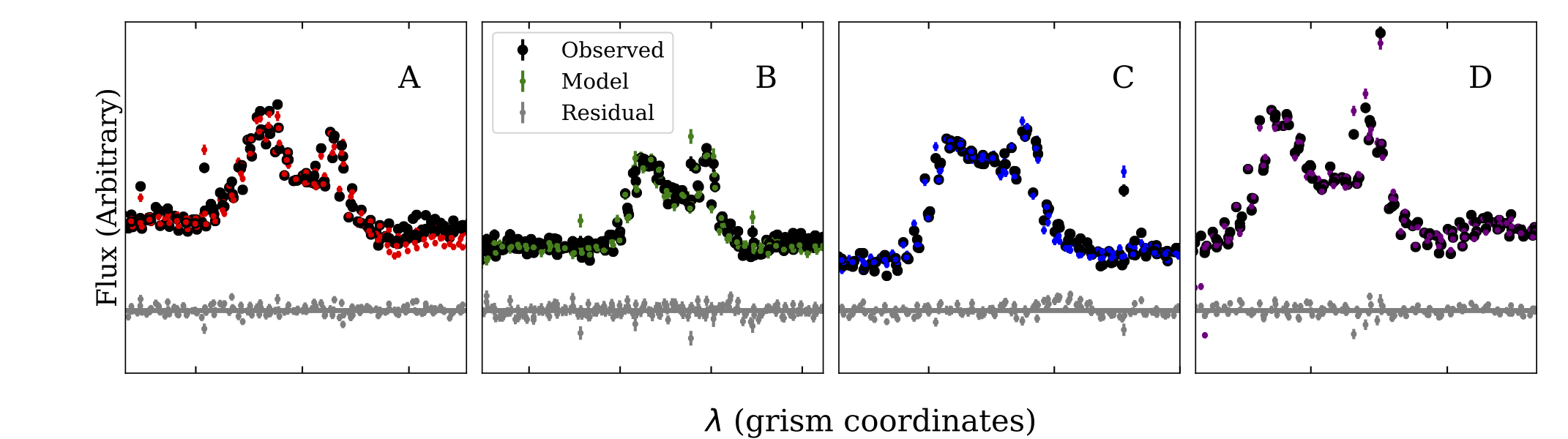}

\vspace{5mm}
\includegraphics[scale=0.35, trim = 20 20 20 20 clip=true]{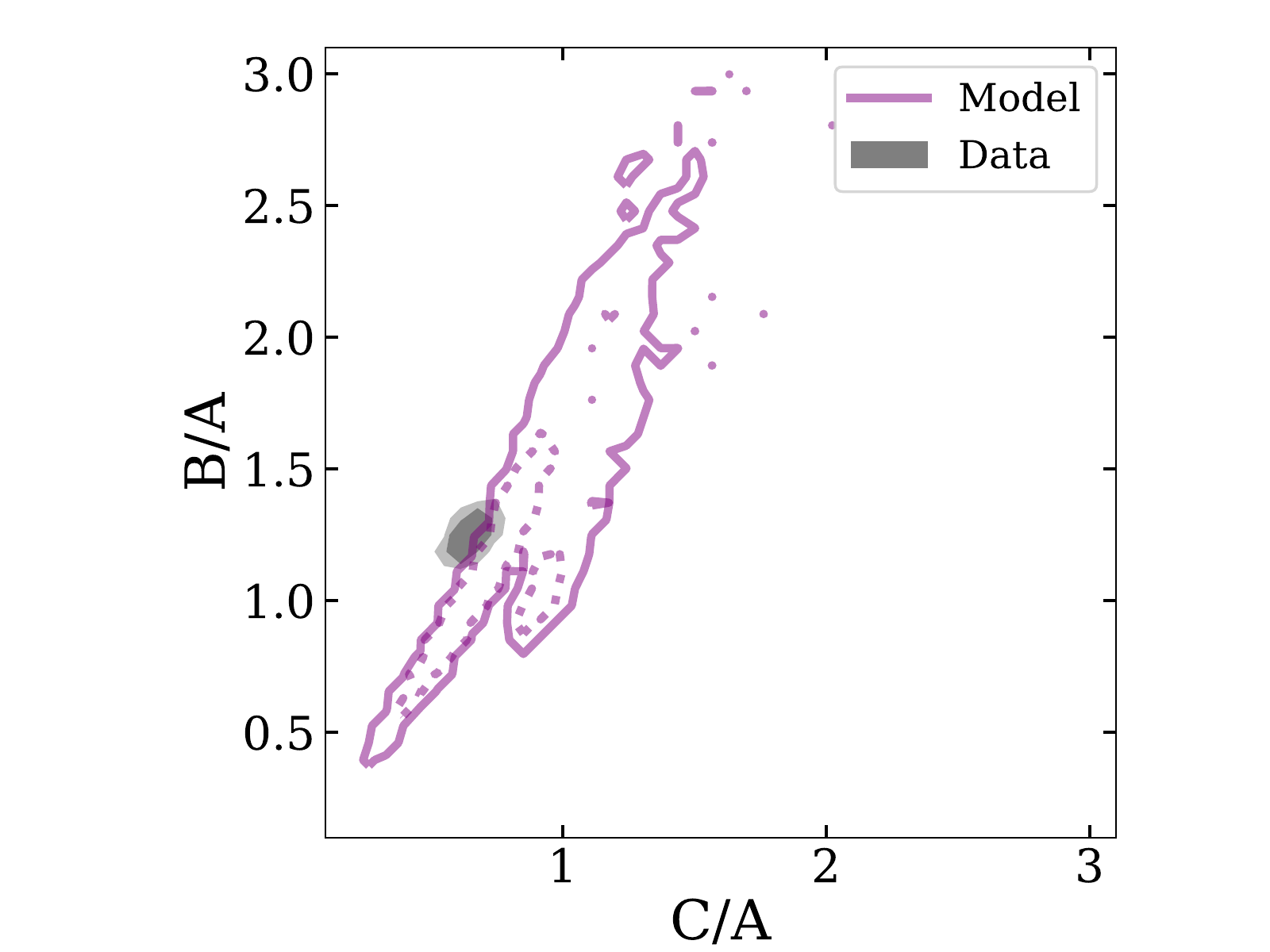}
\includegraphics[scale=0.35, trim = 20 20 20 20 clip=true]{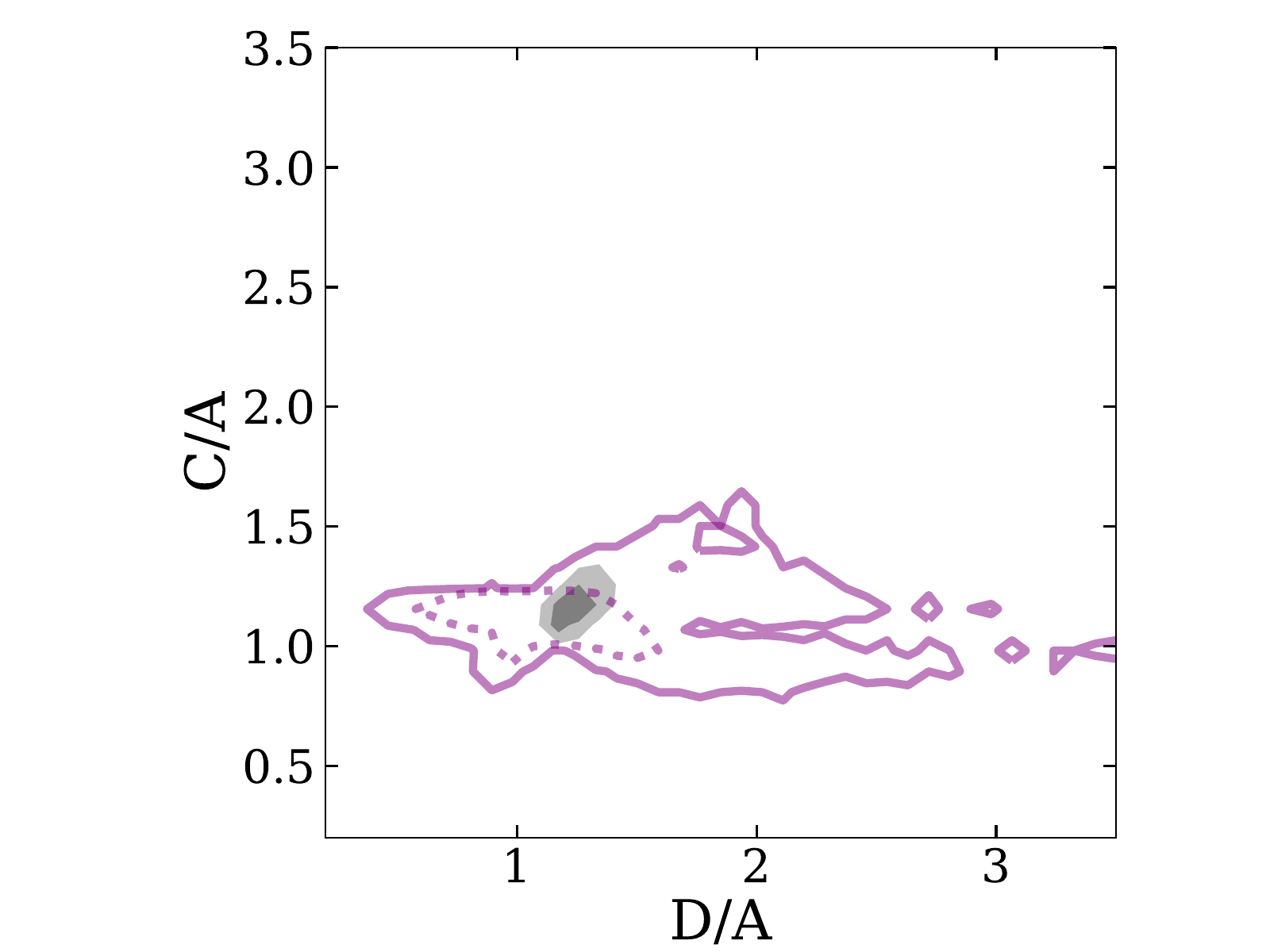}
\includegraphics[scale=0.35, trim = 20 20 20 20 clip=true]{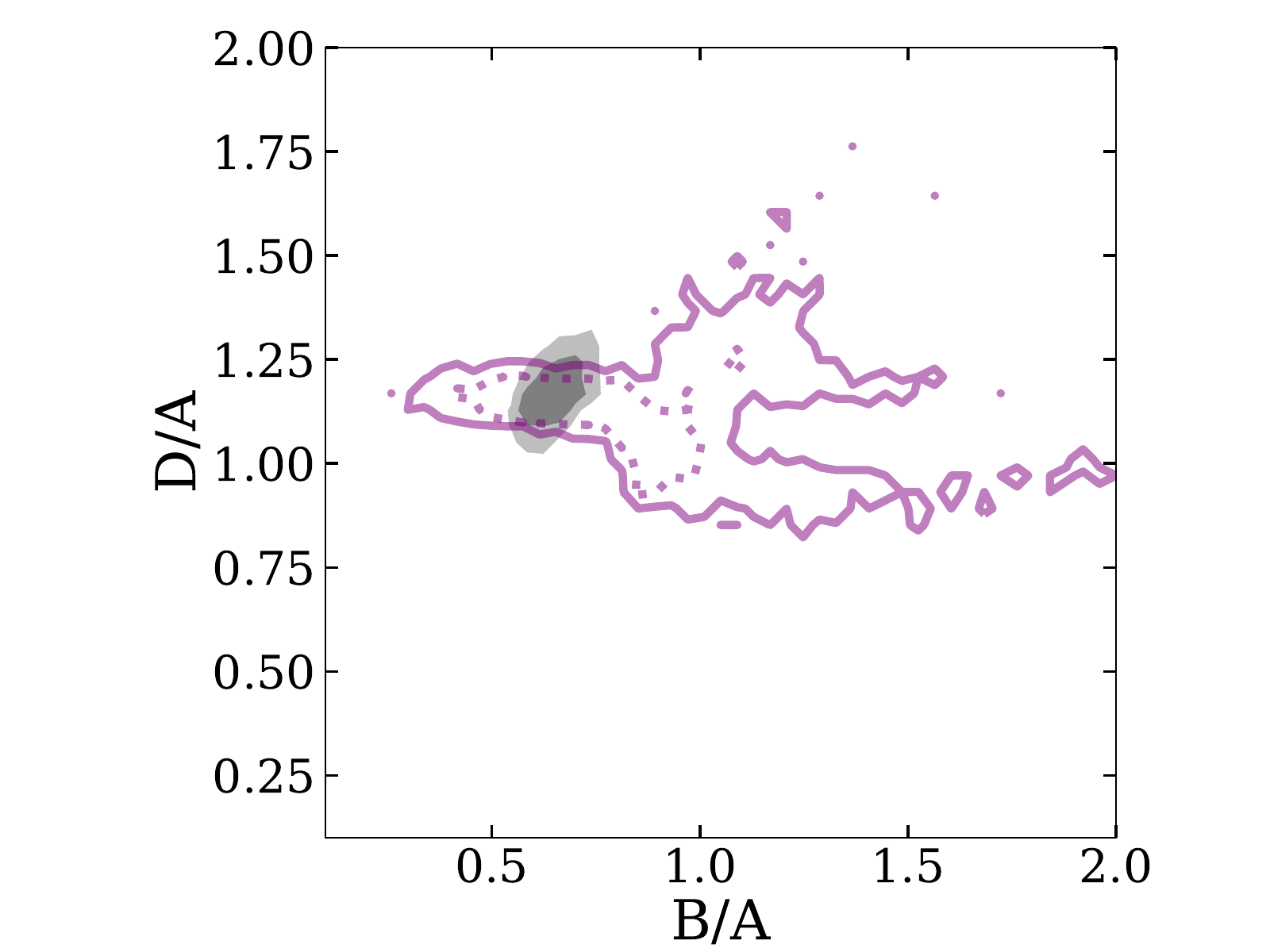}

\caption{Spectral fitting results for WGD J0405. \emph{Top Row}: Model fit with 68\% confidence interval to image spectra showing contributions from all spectral components used in the fit. \emph{Middle Row}: Comparison between model and grism traces in the 2D grism image computed using a PSF-weighted sum along the y axis. Traces are computed after spectra from all images have been added to the 2D image and are thus affected by blending between neighbouring images. \emph{Lower Row}: Comparison between observed narrow-line flux ratios, and predicted flux ratios based on model fit to observed image positions. Dotted lines and dark contours represent one sigma confidence intervals for the model and data respectively, while solid lines and light contours represent two sigma confidence intervals.}
}
\label{fig:prmass}
\end{figure*}

\begin{figure*}
{\centering

\includegraphics[scale=0.45, trim = 0 0 0 0 clip=true]{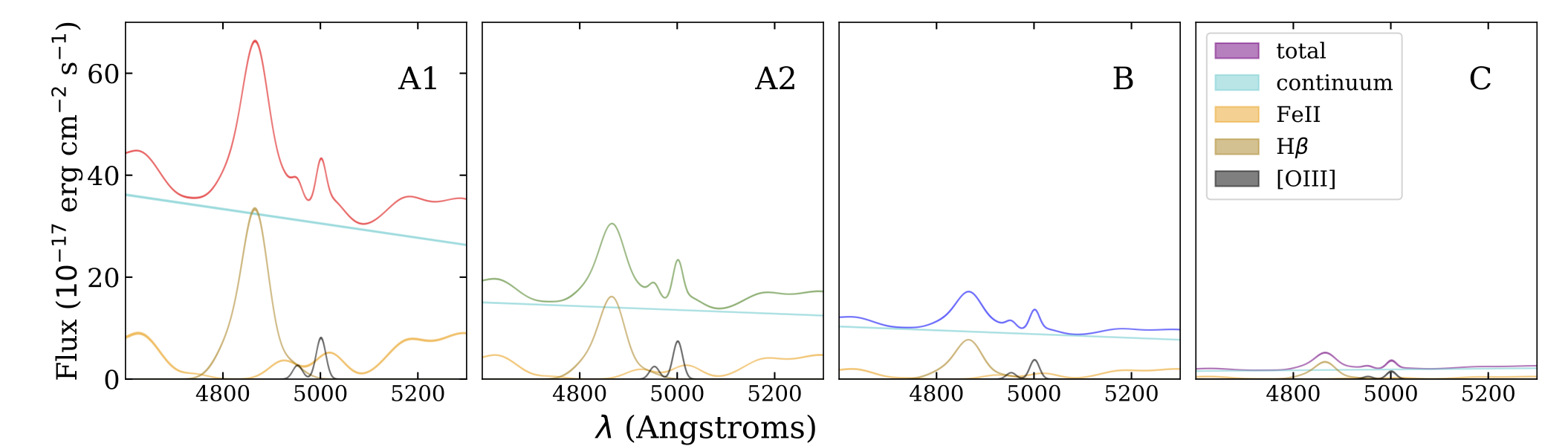}

\includegraphics[scale=0.45, trim = 0 0 0 0 clip=true]{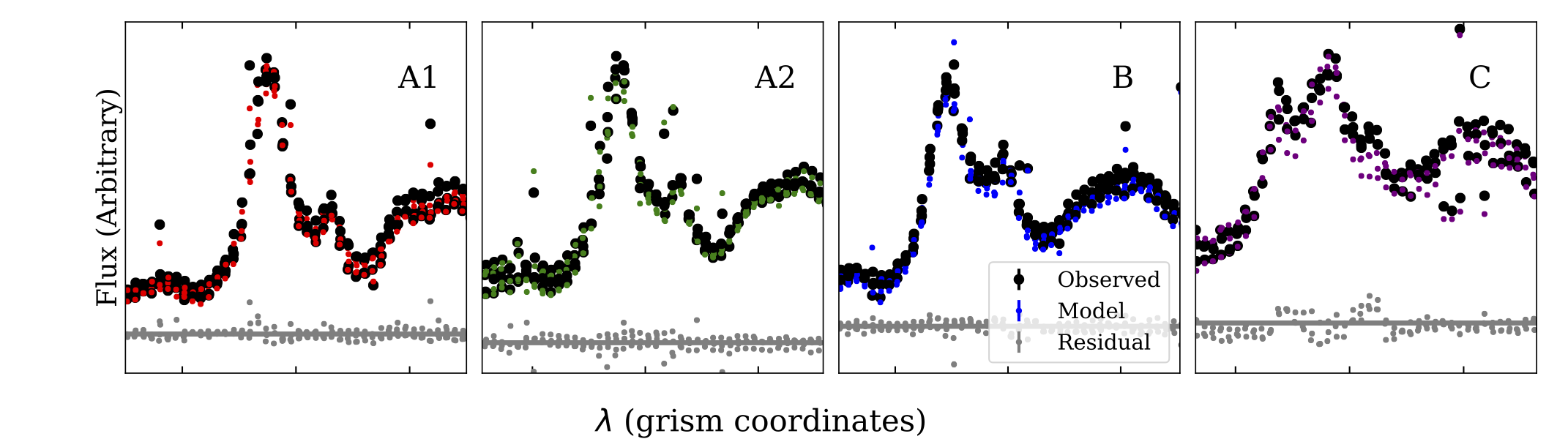}

\caption{Spectral fitting results for 0810. \emph{Top Row}: Model fit with 68\% confidence interval to image spectra showing contributions from all spectral components used in the fit.  \emph{Middle Row}: Comparison between model and grism traces in the 2D grism image computed using a PSF-weighted sum along the y axis. Traces are computed after spectra from all images have been added to the 2D image and are thus affected by blending between neighbouring images. Note that for the bottom row, the y-axis range varies between images. Note:We do not show a comparison between gravitational lens model and measured flux ratios given that we find that narrow emission in images A1 and A2 are likely to be highly blended, owing to their small separation and unusually high magnification of $\sim 120$ in these images. A gravitational lensing comparison would need to account for this blending rather than considering the fluxes separately.}
\label{fig:prmass}
}
\end{figure*}

\begin{figure*}
{\centering

\includegraphics[scale=0.55, trim = 0 0 0 0 clip=true]{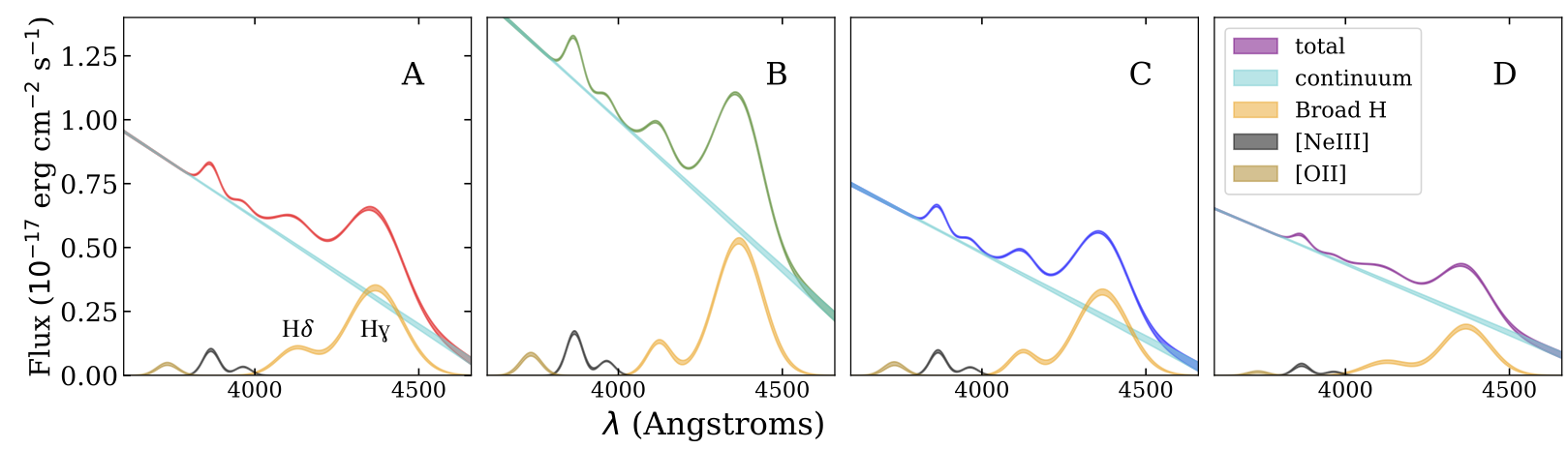}

\includegraphics[scale=0.45, trim = 0 0 0 0 clip=true]{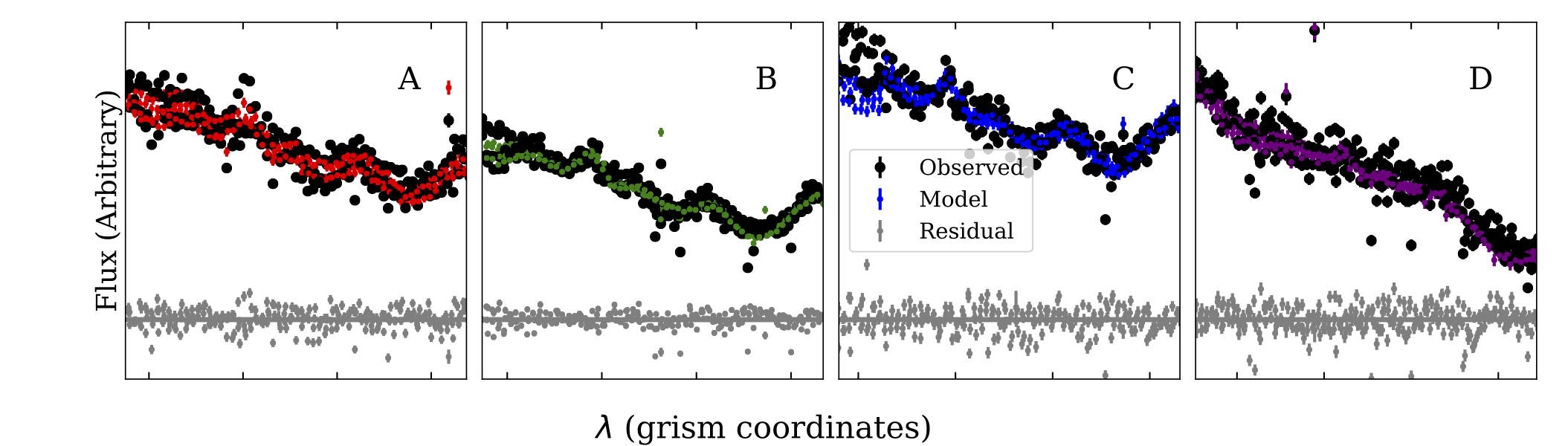}

\vspace{5mm}
\includegraphics[scale=0.35, trim = 20 20 20 20 clip=true]{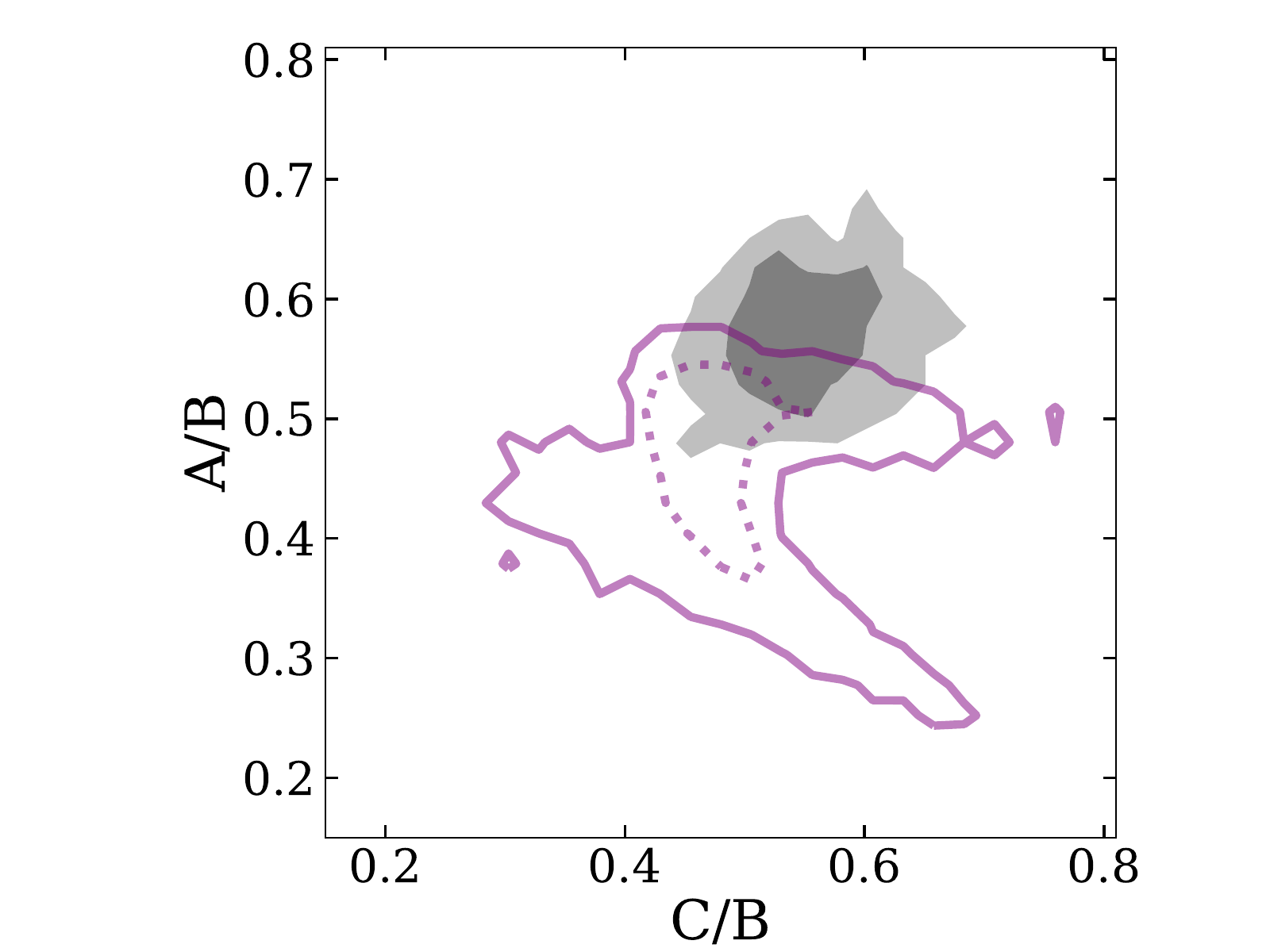}
\includegraphics[scale=0.35, trim = 20 20 20 20 clip=true]{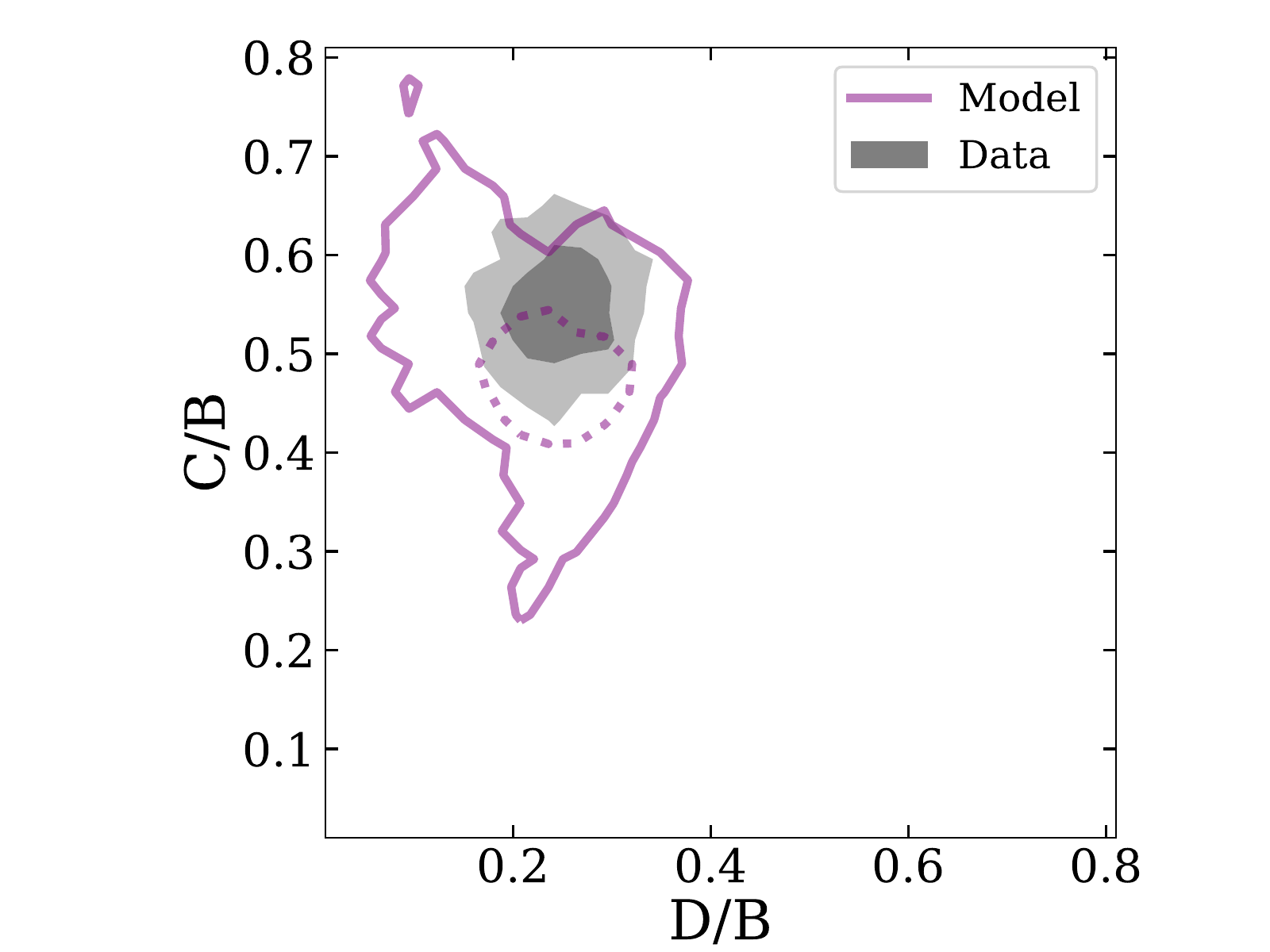}
\includegraphics[scale=0.35, trim = 20 20 20 20 clip=true]{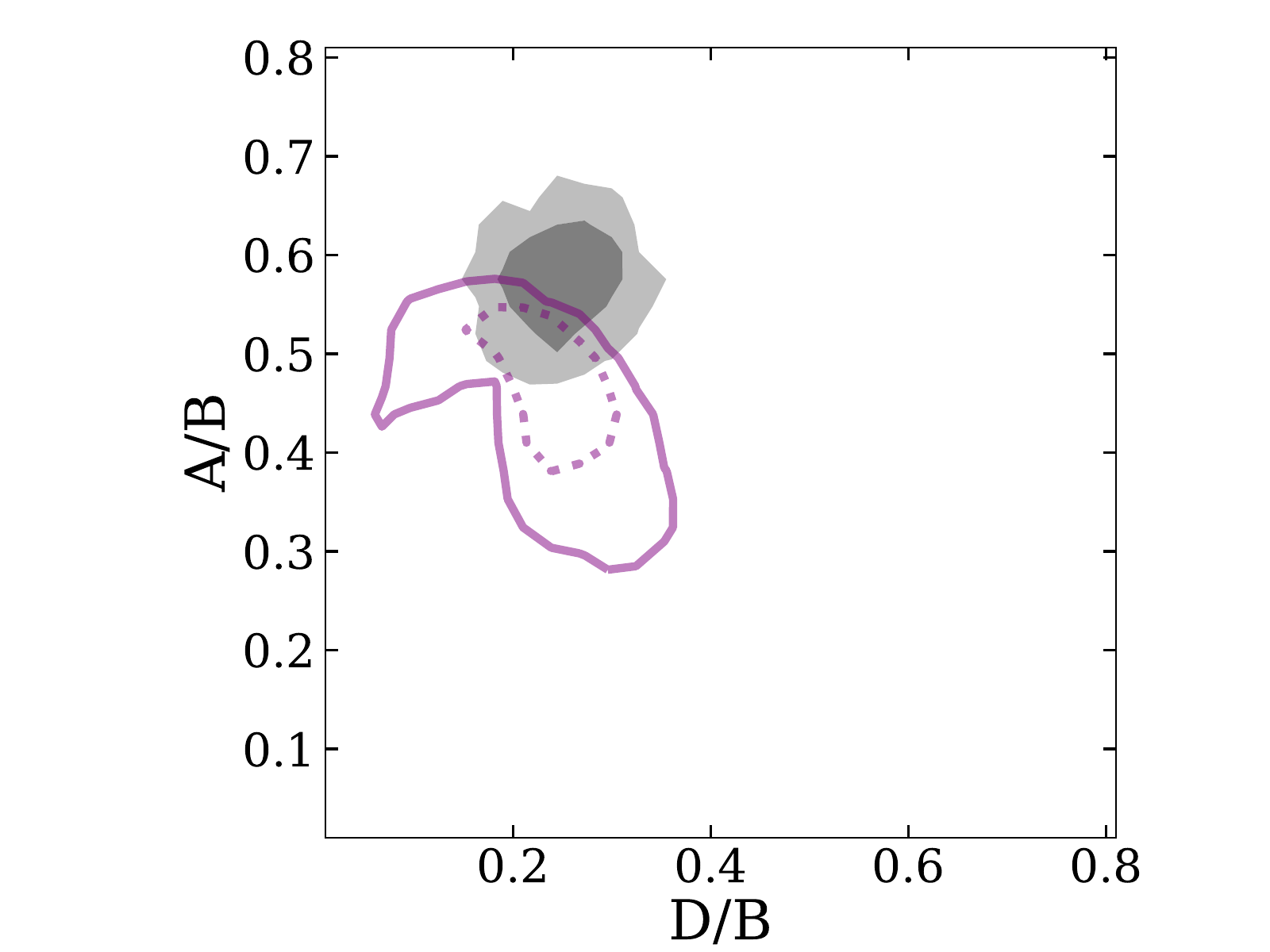}

\caption{Spectral fitting results for RX J0911. \emph{Top Row}: Model fit with 68\% confidence interval to image spectra showing contributions from all spectral components used in the fit. \emph{Bottom Row}: Comparison between model and grism traces in the 2D grism image computed using a PSF-weighted sum along the y axis. Traces are computed after spectra from all images have been added to the 2D image and are thus affected by blending between neighbouring images.  \emph{Lower Row}: Comparison between observed narrow-line flux ratios, and predicted flux ratios based on model fit to observed image positions. Dotted lines and dark contours represent one sigma confidence intervals for the model and data respectively, while solid lines and light contours represent two sigma confidence intervals.}
\label{fig:prmass}
}
\end{figure*}

\begin{figure*}
{\centering

\includegraphics[scale=0.55, trim = 0 0 0 0 clip=true]{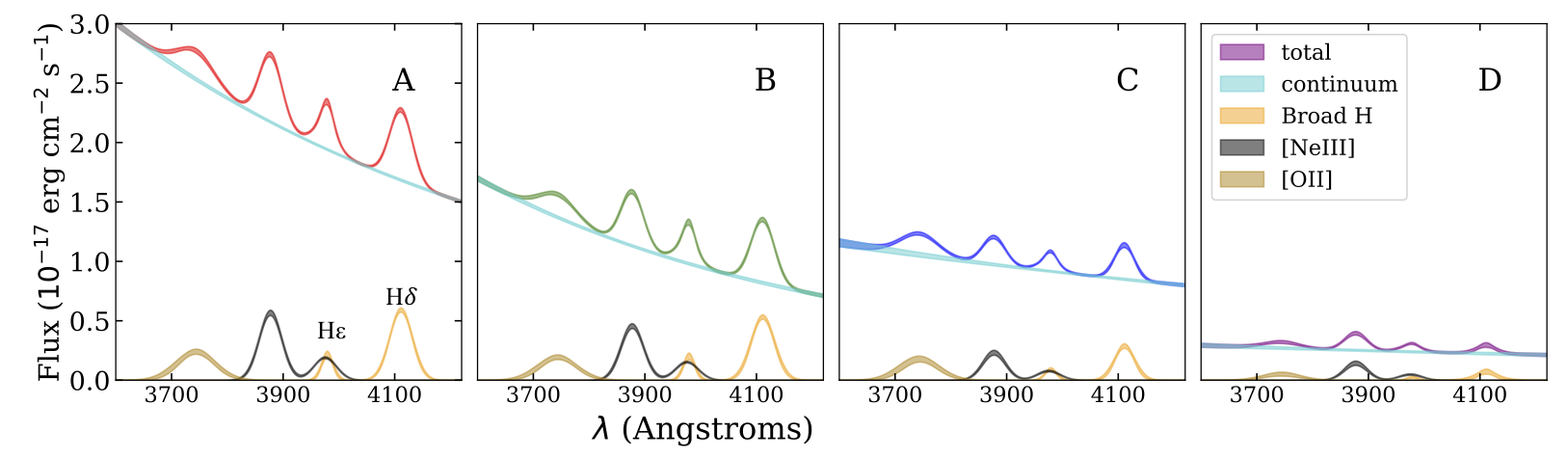}

\includegraphics[scale=0.45, trim = 0 0 0 0 clip=true]{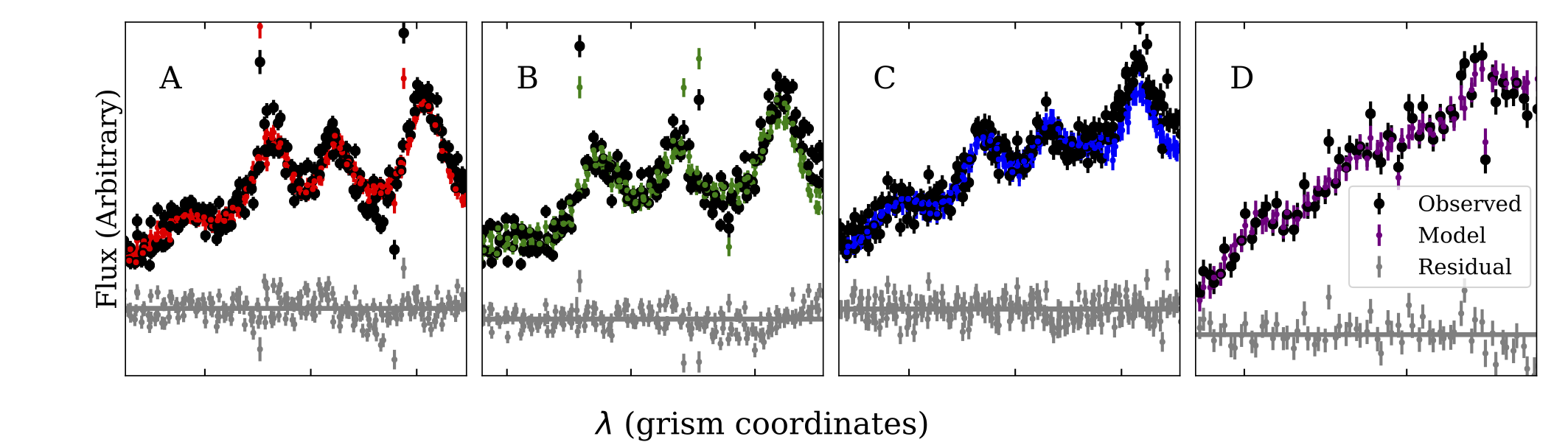}

\vspace{5mm}
\includegraphics[scale=0.35, trim = 20 20 20 20 clip=true]{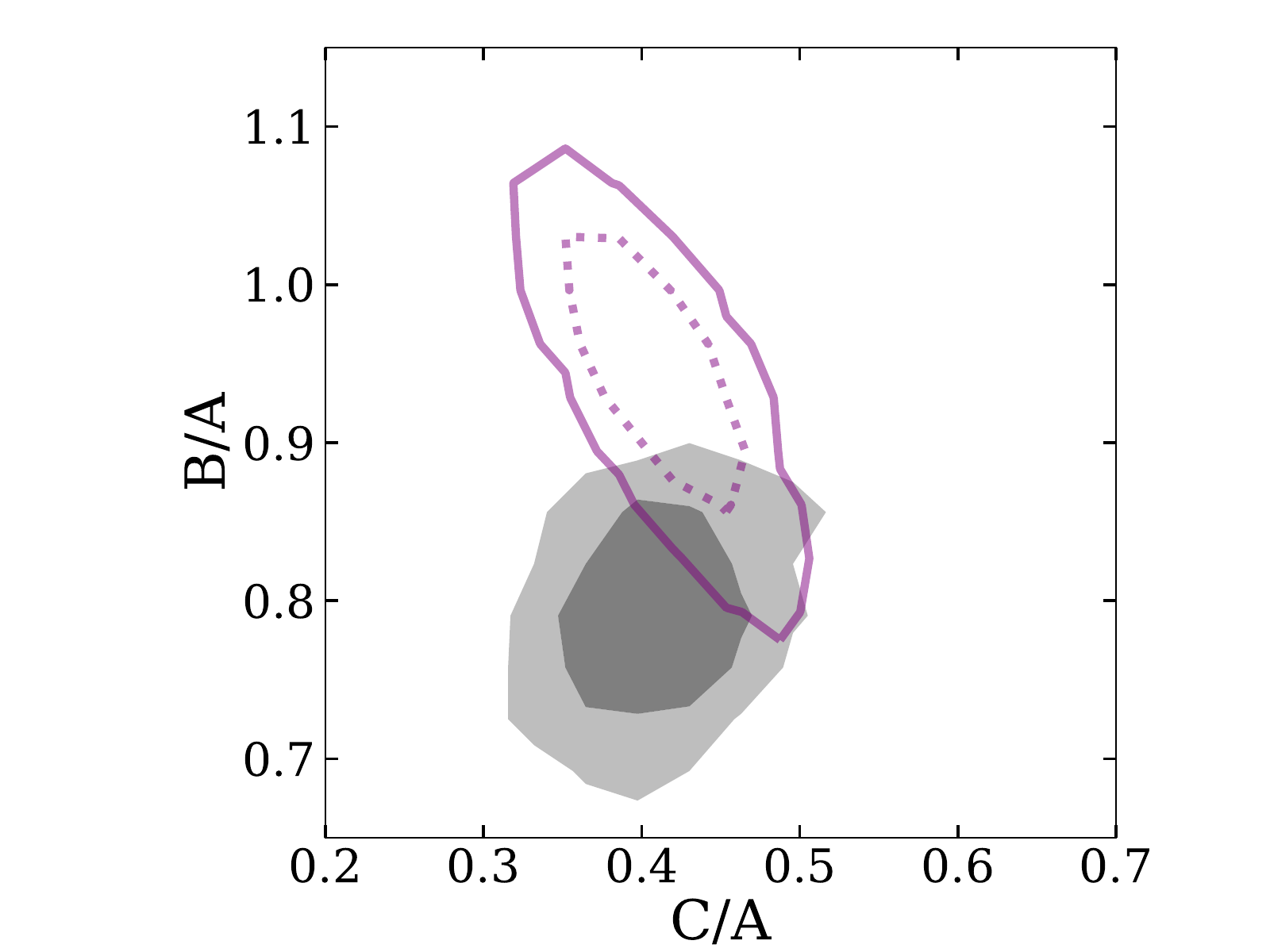}
\includegraphics[scale=0.35, trim = 20 20 20 20 clip=true]{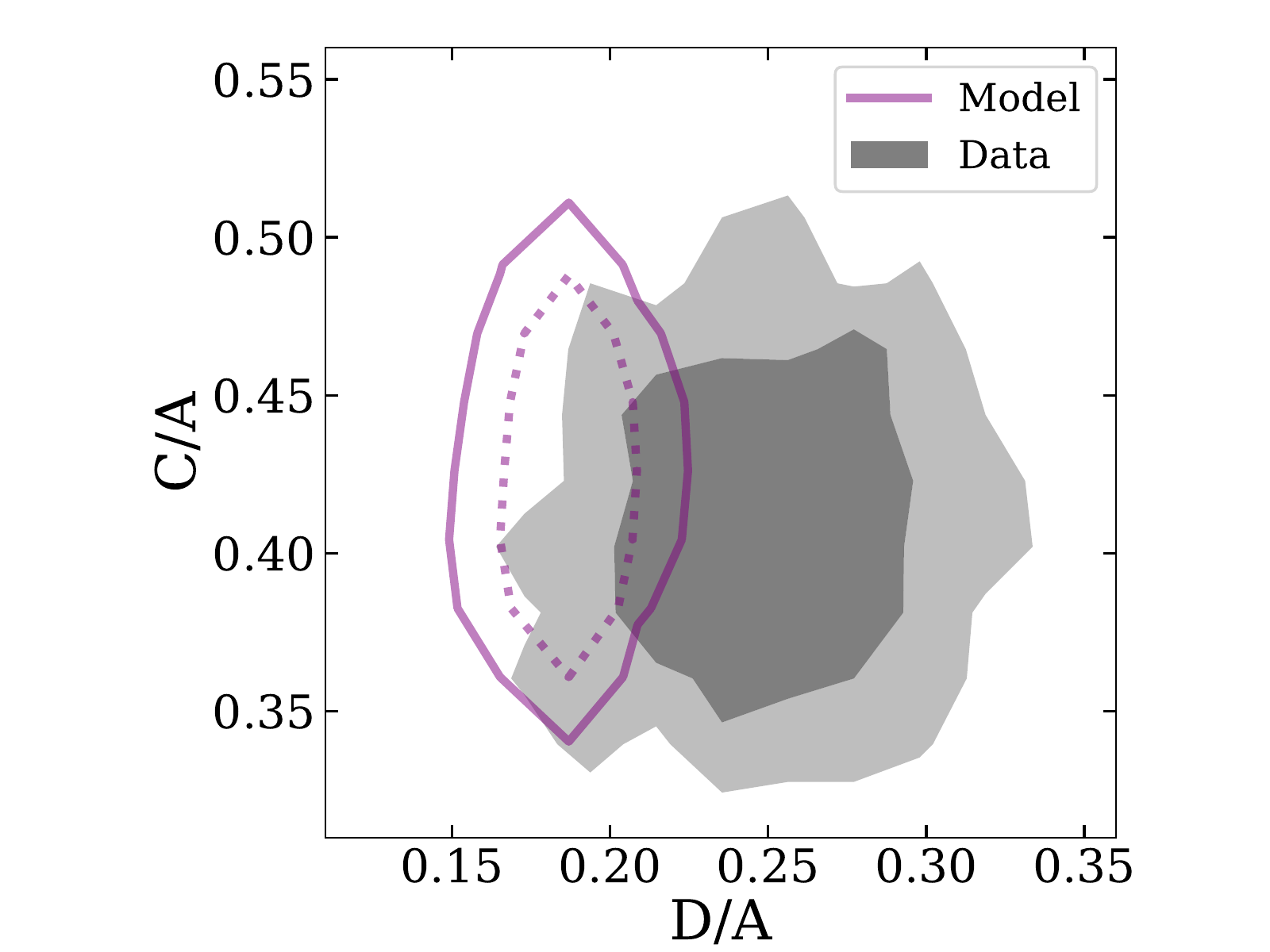}
\includegraphics[scale=0.35, trim = 20 20 20 20 clip=true]{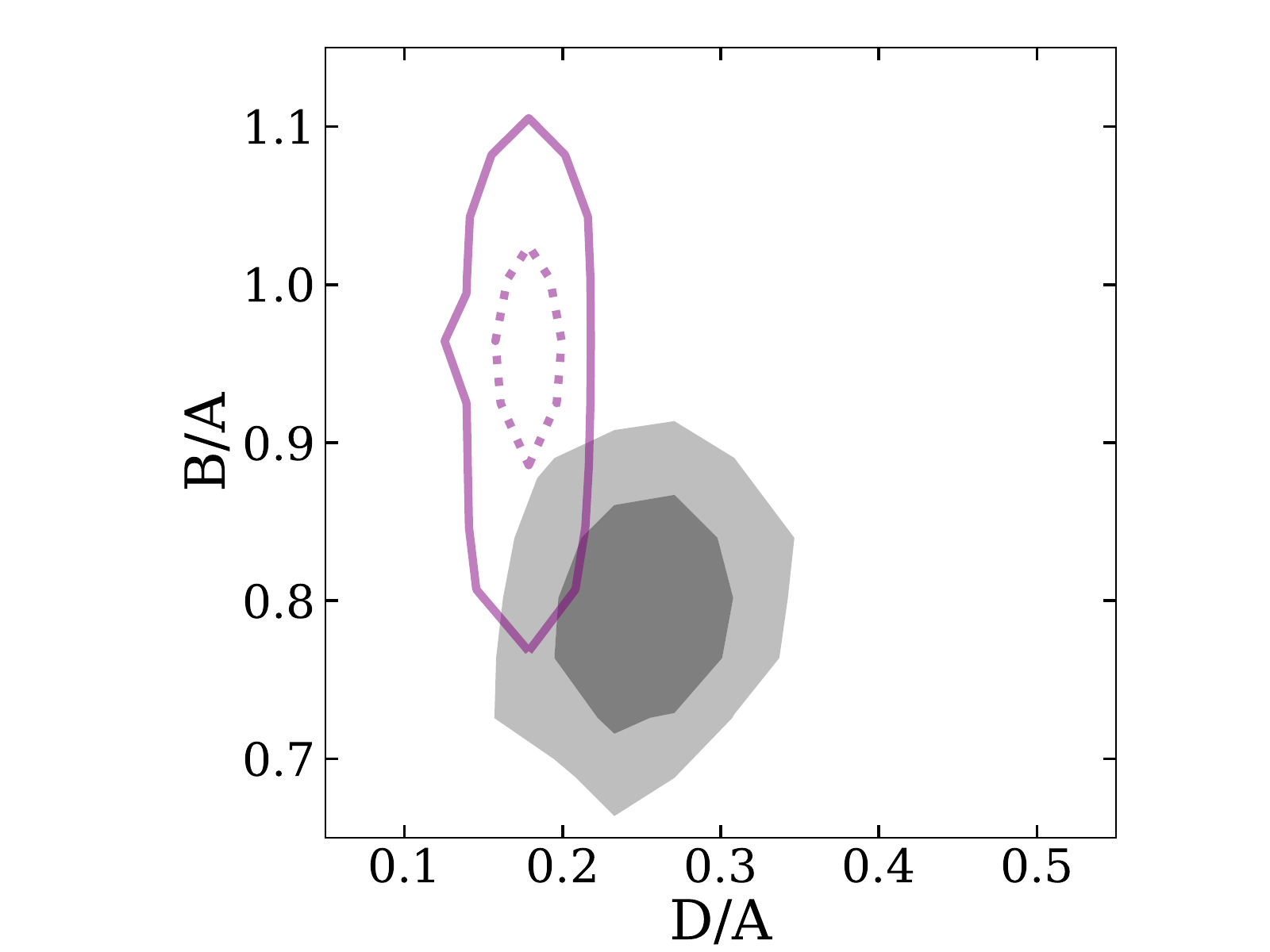}

\caption{Spectral fitting results for SDSS J1330. \emph{Top Row}: Model fit with 68\% confidence interval to image spectra showing contributions from all spectral components used in the fit. \emph{Bottom Row}: Comparison between model and grism traces in the 2D grism image computed using a PSF-weighted sum along the y axis. Traces are computed after spectra from all images have been added to the 2D image and are thus affected by blending between neighbouring images.  \emph{Lower Row}: Comparison between observed narrow-line flux ratios, and predicted flux ratios based on model fit to observed image positions. Dotted lines and dark contours represent one sigma confidence intervals for the model and data respectively, while solid lines and light contours represent two sigma confidence intervals.}
\label{fig:prmass}
}
\end{figure*}

\begin{figure*}
{\centering

\includegraphics[scale=0.4, trim = 0 0 0 0 clip=true]{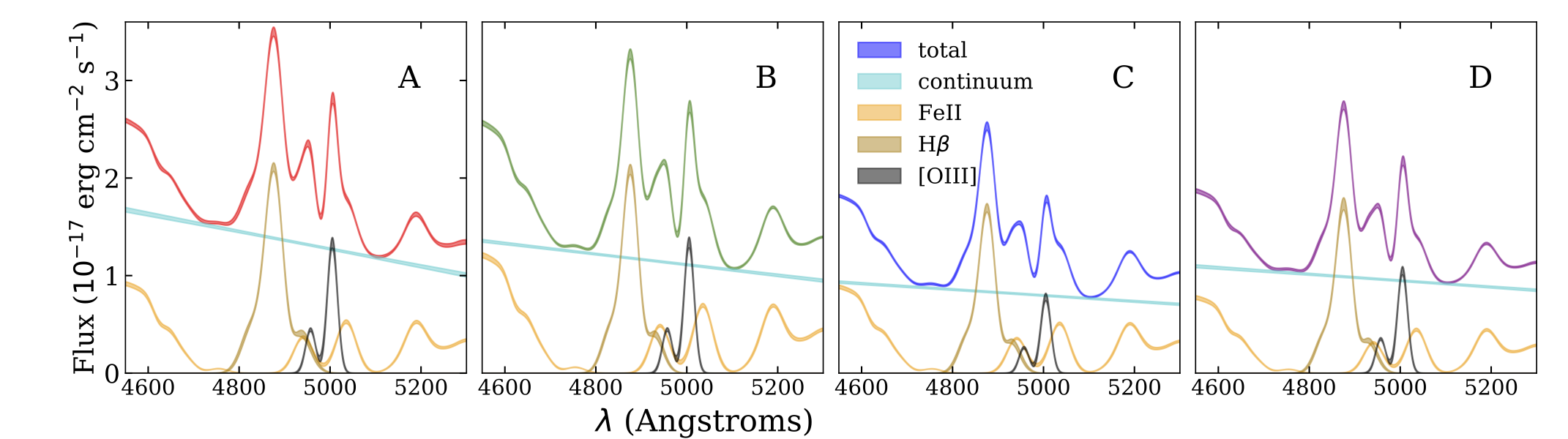}

\includegraphics[scale=0.4, trim = 0 0 0 0 clip=true]{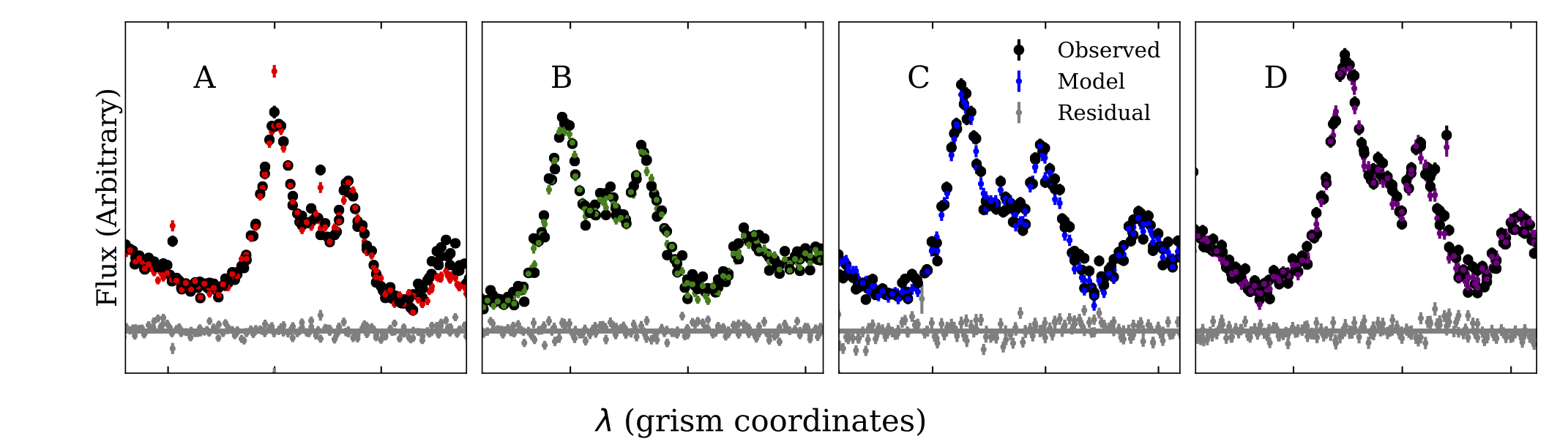}

\vspace{5mm}
\includegraphics[scale=0.4, trim = 0 0 0 0 clip=true]{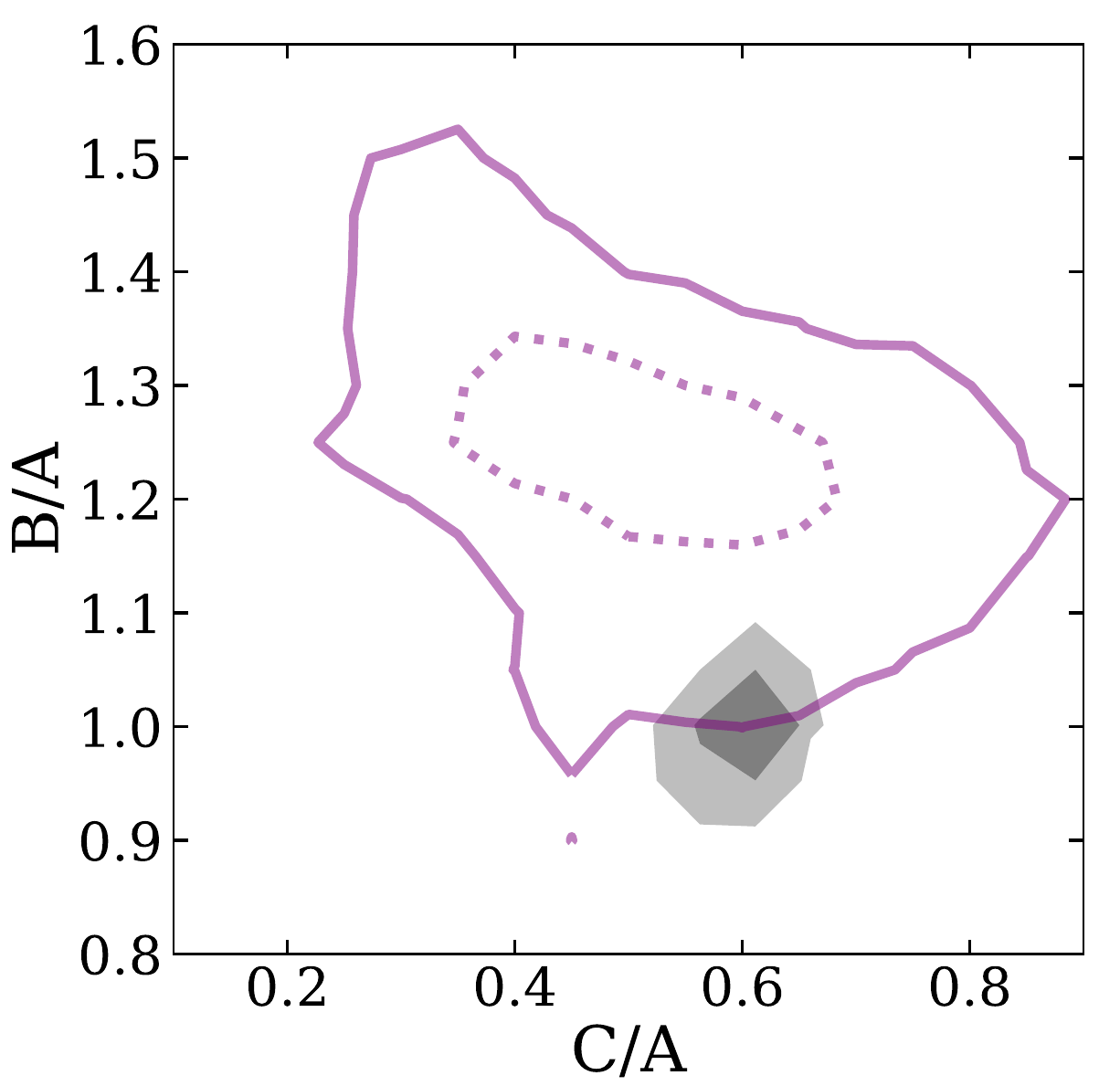}
\includegraphics[scale=0.4, trim = 0 0 0 0 clip=true]{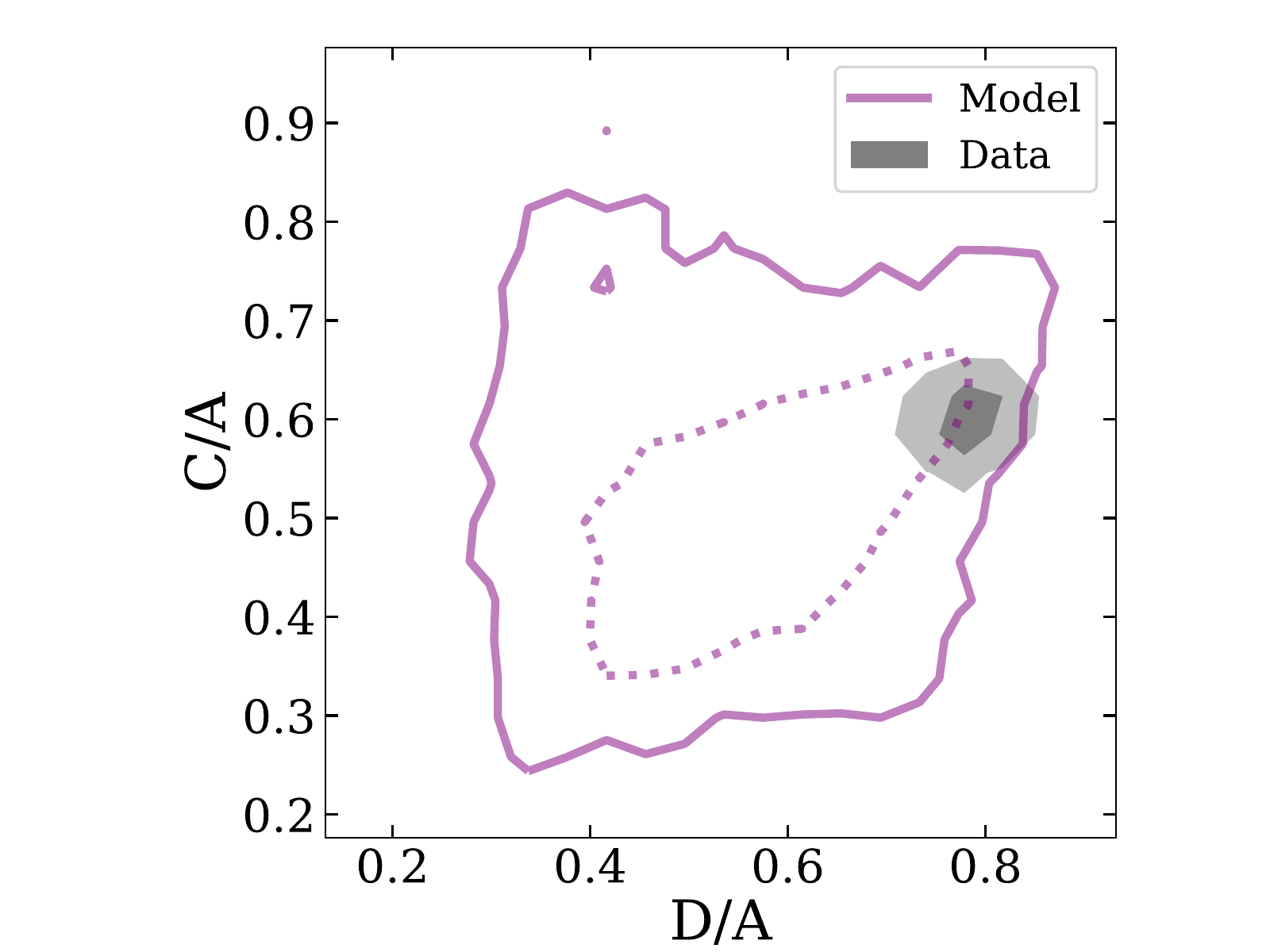}
\includegraphics[scale=0.4, trim = 0 0 0 0 clip=true]{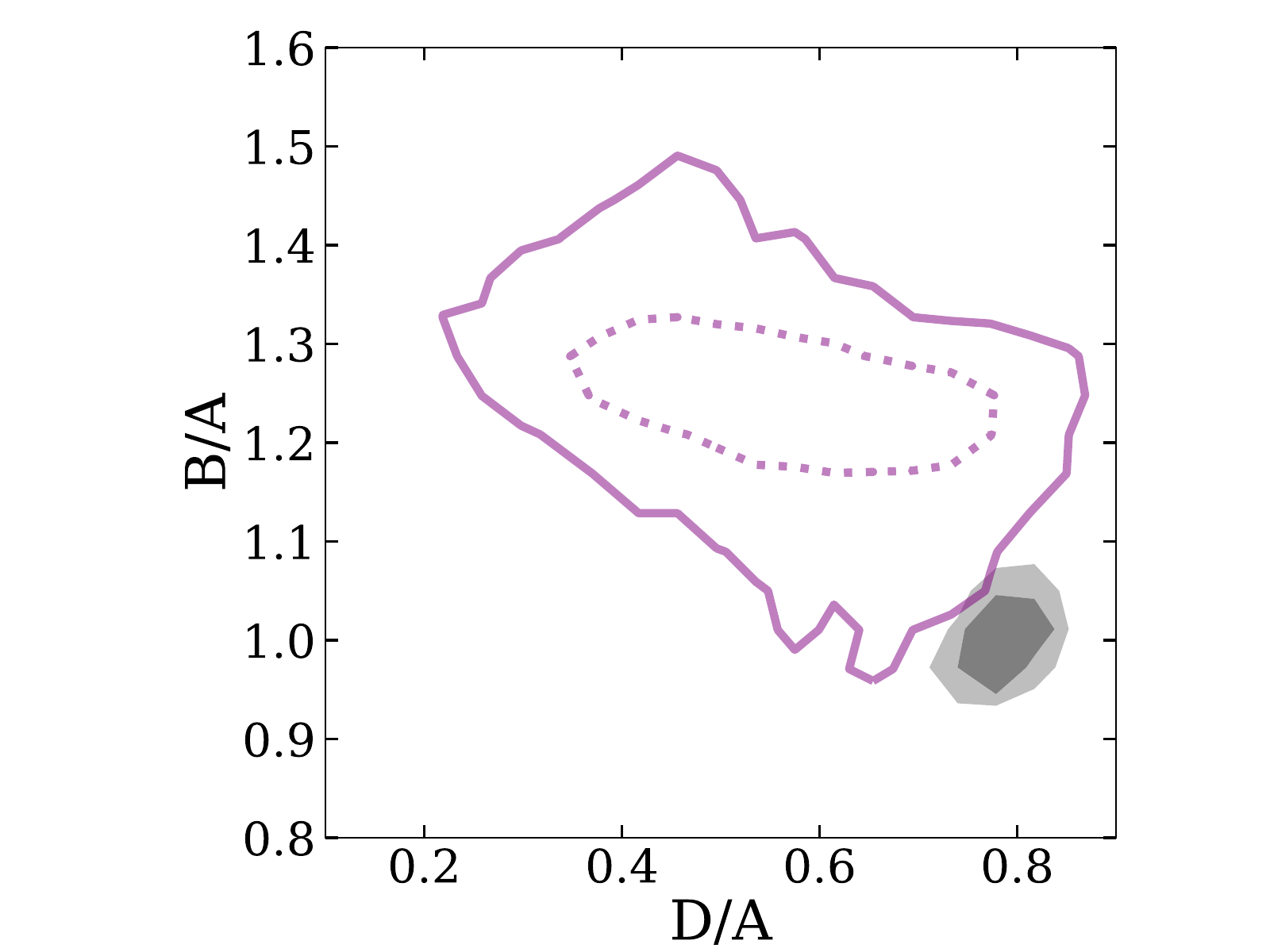}

\caption{Spectral fitting results for PS J1606. \emph{Top Row}: Model fit with 68\% confidence interval to image spectra showing contributions from all spectral components used in the fit.  \emph{Bottom Row}: Comparison between model and grism traces in the 2D grism image computed using a PSF-weighted sum along the y axis. Traces are computed after spectra from all images have been added to the 2D image and are thus affected by blending between neighbouring images.   \emph{Lower Row}: Comparison between observed narrow-line flux ratios, and predicted flux ratios based on model fit to observed image positions. Dotted lines and dark contour represent one sigma confidence intervals for the model and data respectively, while solid lines and light contours represent two sigma confidence intervals.}
\label{fig:prmass}
}
\end{figure*}

\begin{figure*}
{\centering

\includegraphics[scale=0.2, trim = 0 0 0 0 clip=true]{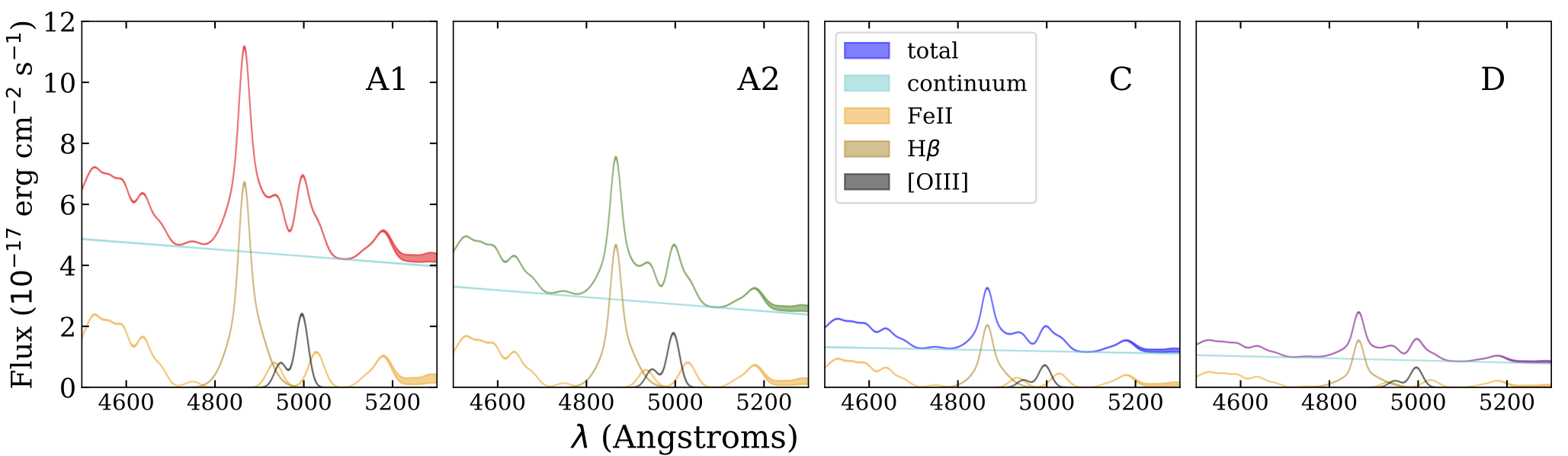}

\includegraphics[scale=0.2, trim = 0 0 0 0 clip=true]{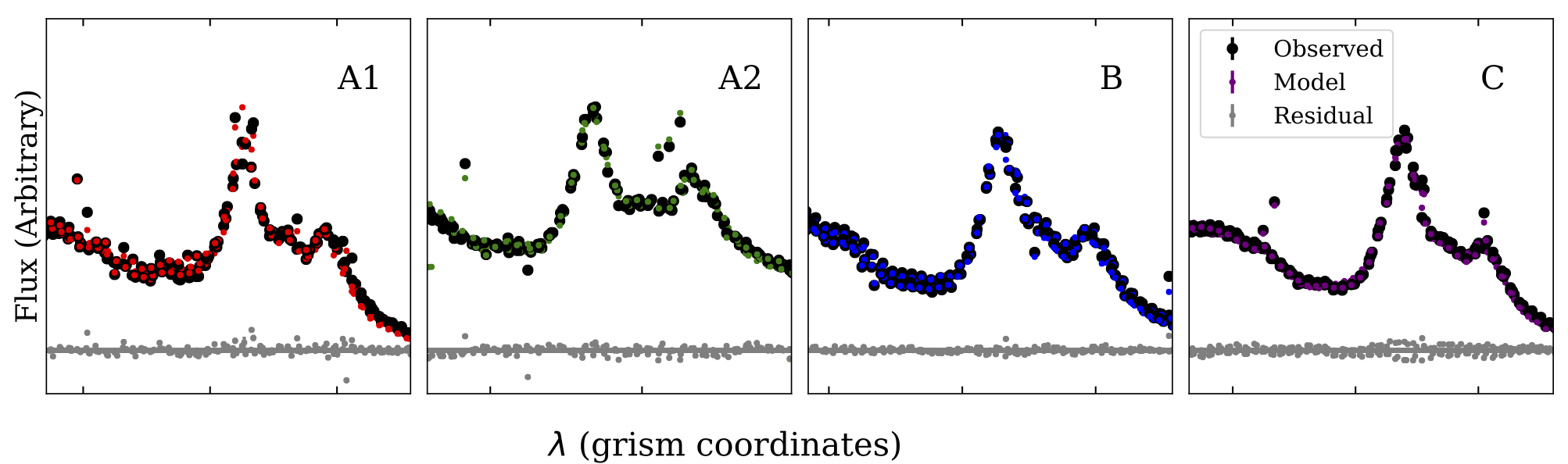}

\vspace{5mm}
\includegraphics[scale=0.4, trim = 0 0 0 0 clip=true]{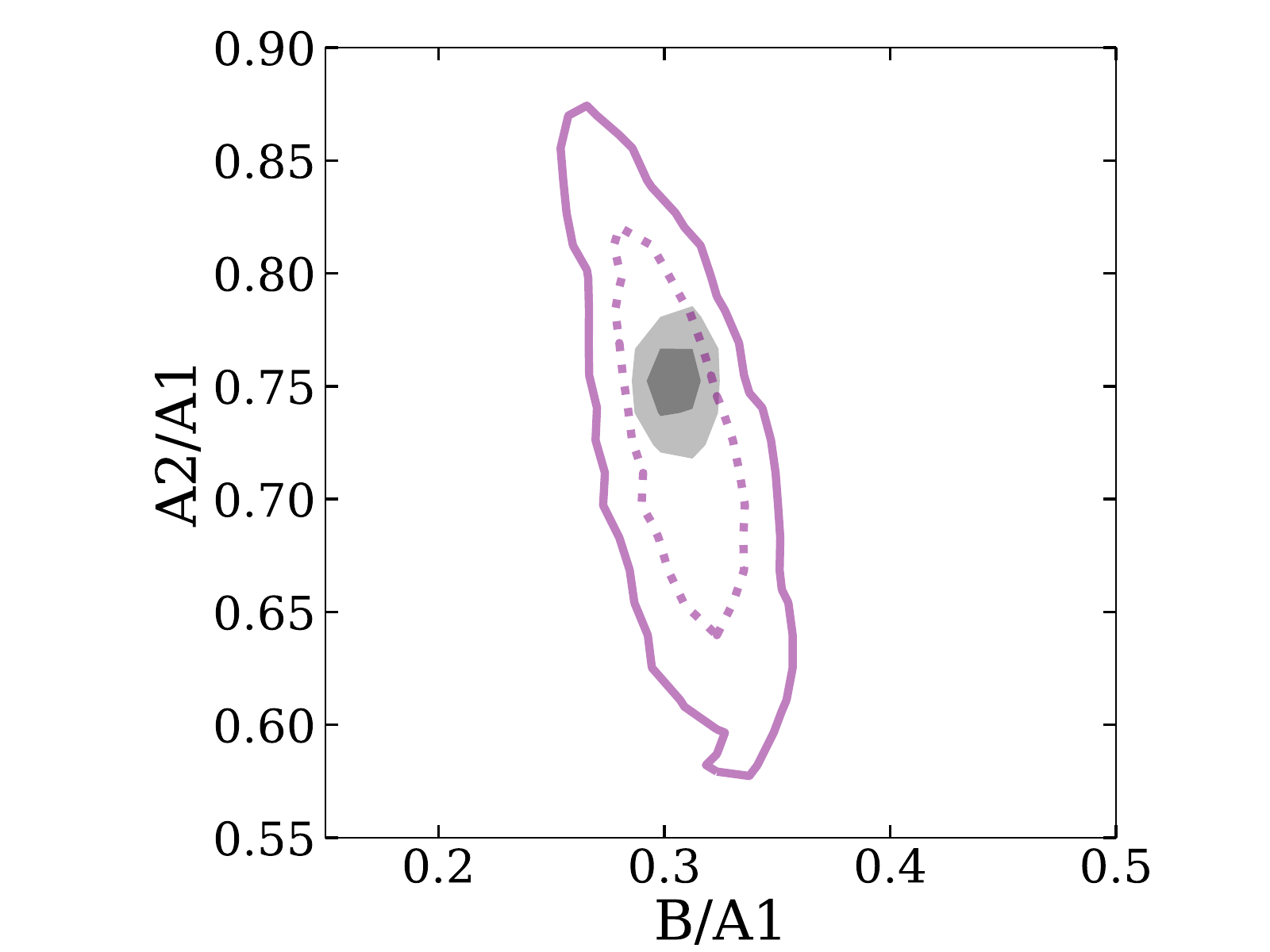}
\includegraphics[scale=0.4, trim = 0 0 0 0 clip=true]{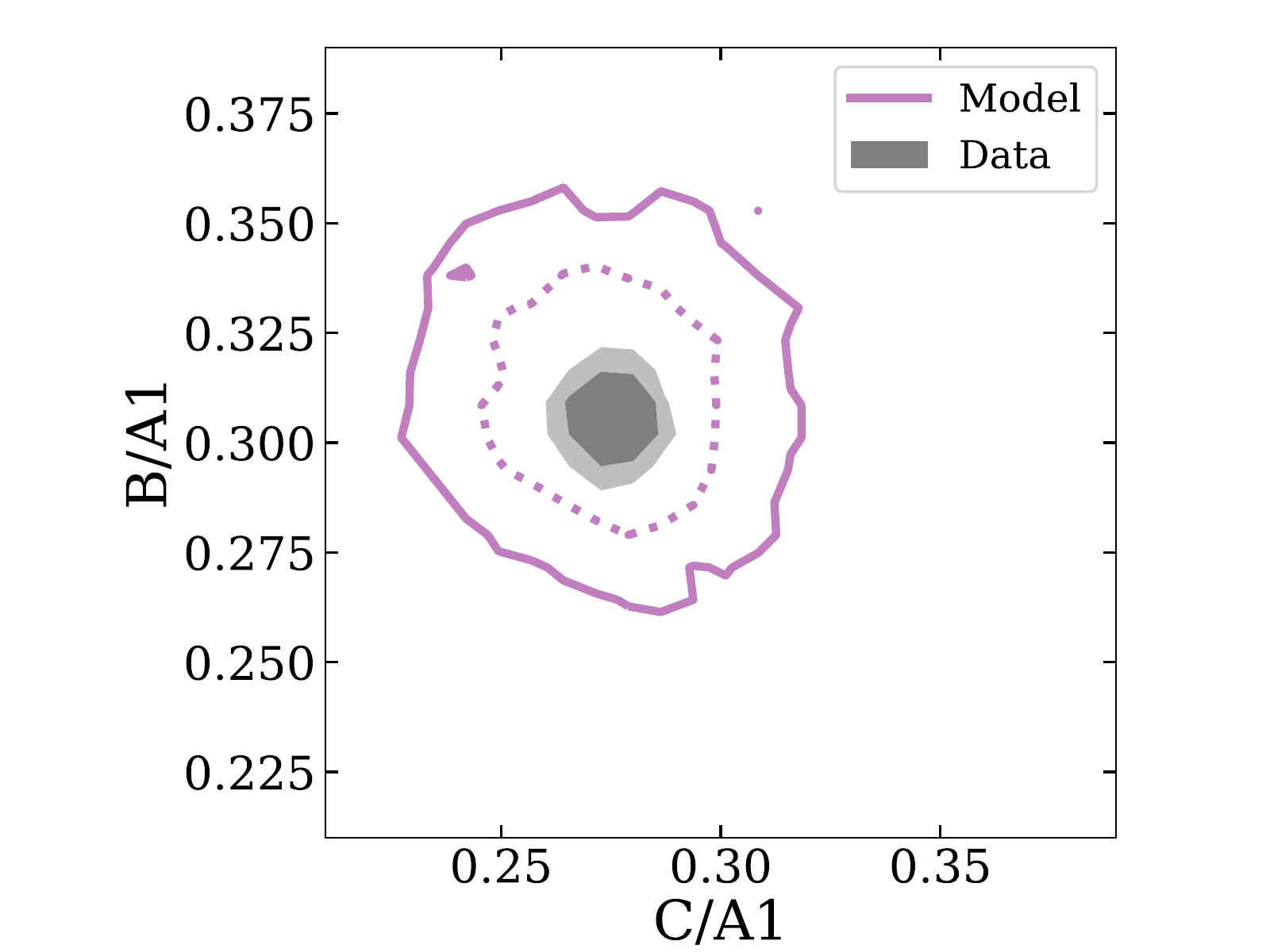}
\includegraphics[scale=0.4, trim = 0 0 0 0 clip=true]{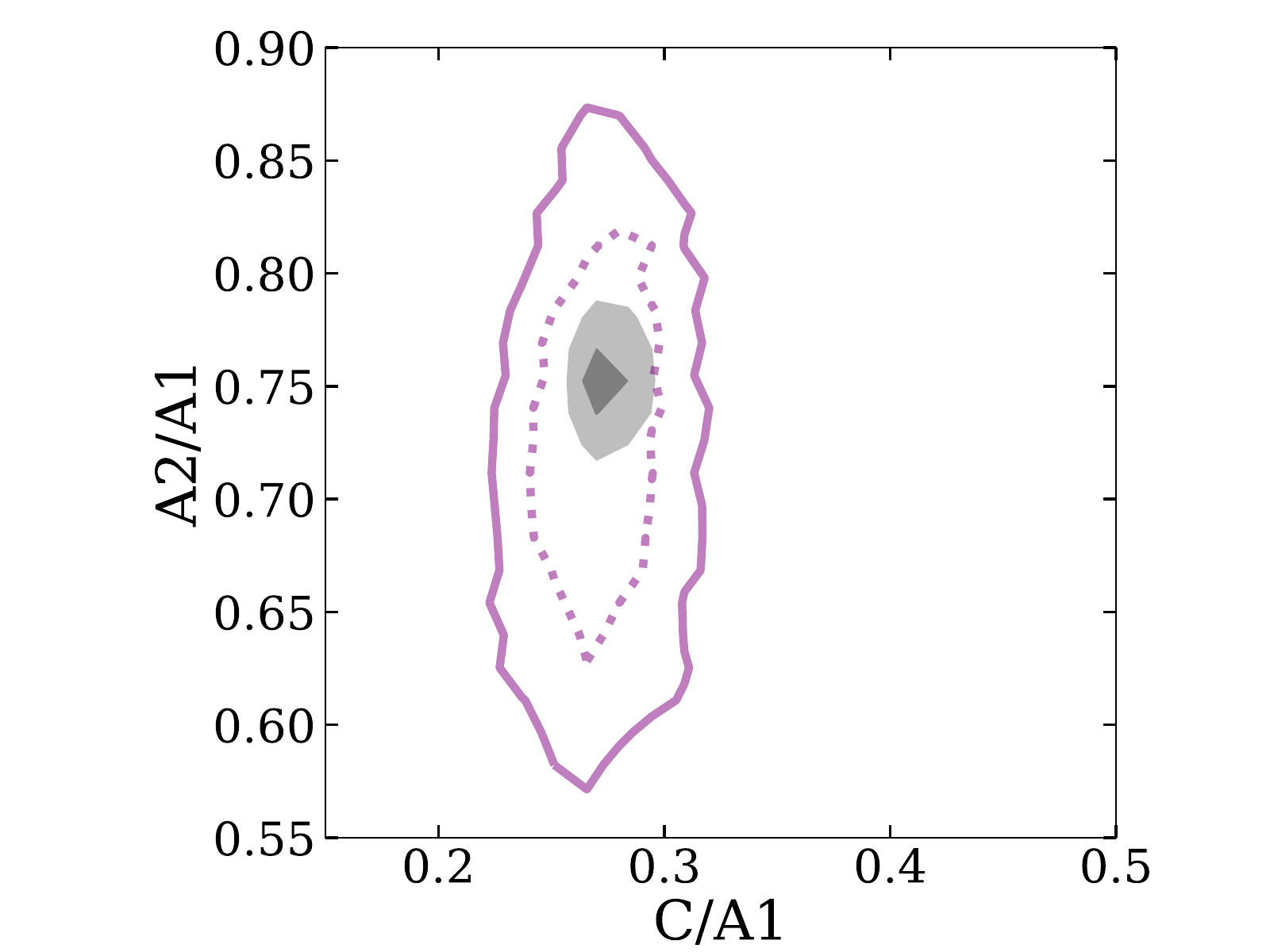}

\caption{Spectral fitting results for 2026. \emph{Top Row}: Model fit with 68\% confidence interval to image spectra showing contributions from all spectral components used in the fit. \emph{Bottom Row}: Comparison between model and grism traces in the 2D grism image computed using a PSF-weighted sum along the y axis. Traces are computed after spectra from all images have been added to the 2D image and are thus affected by blending between neighbouring images.  \emph{Lower Row}: Comparison between observed narrow-line flux ratios, and predicted flux ratios based on model fit to observed image positions. Dotted lines and dark contour represent one sigma confidence intervals for the model and data respectively, while solid lines and light contours represent two sigma confidence intervals.}
\label{fig:prmass}
}
\end{figure*}

\begin{figure*}
{\centering

\includegraphics[scale=0.2, trim = 0 0 0 0 clip=true]{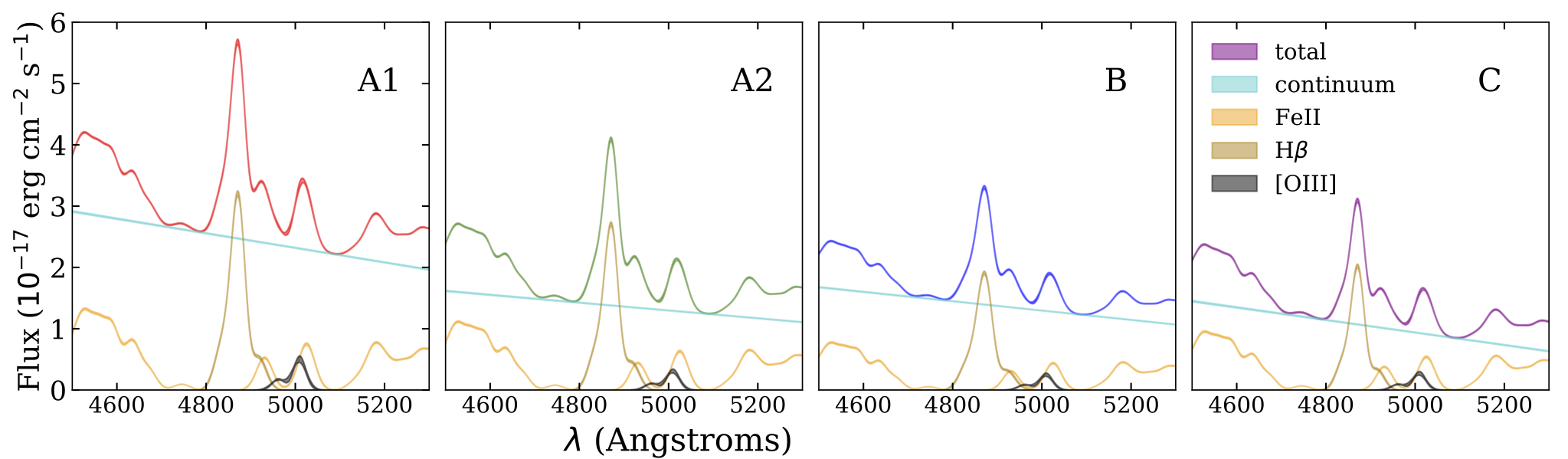}

\includegraphics[scale=0.2, trim = 0 0 0 0 clip=true]{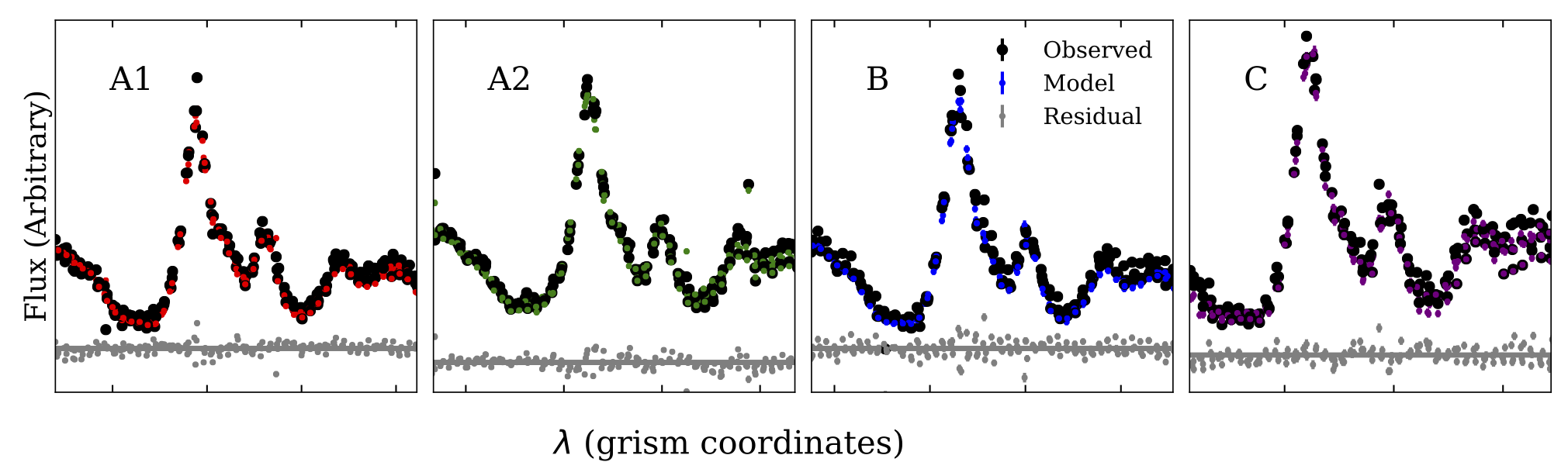}

\vspace{5mm}
\includegraphics[scale=0.35, trim = 0 0 0 0 clip=true]{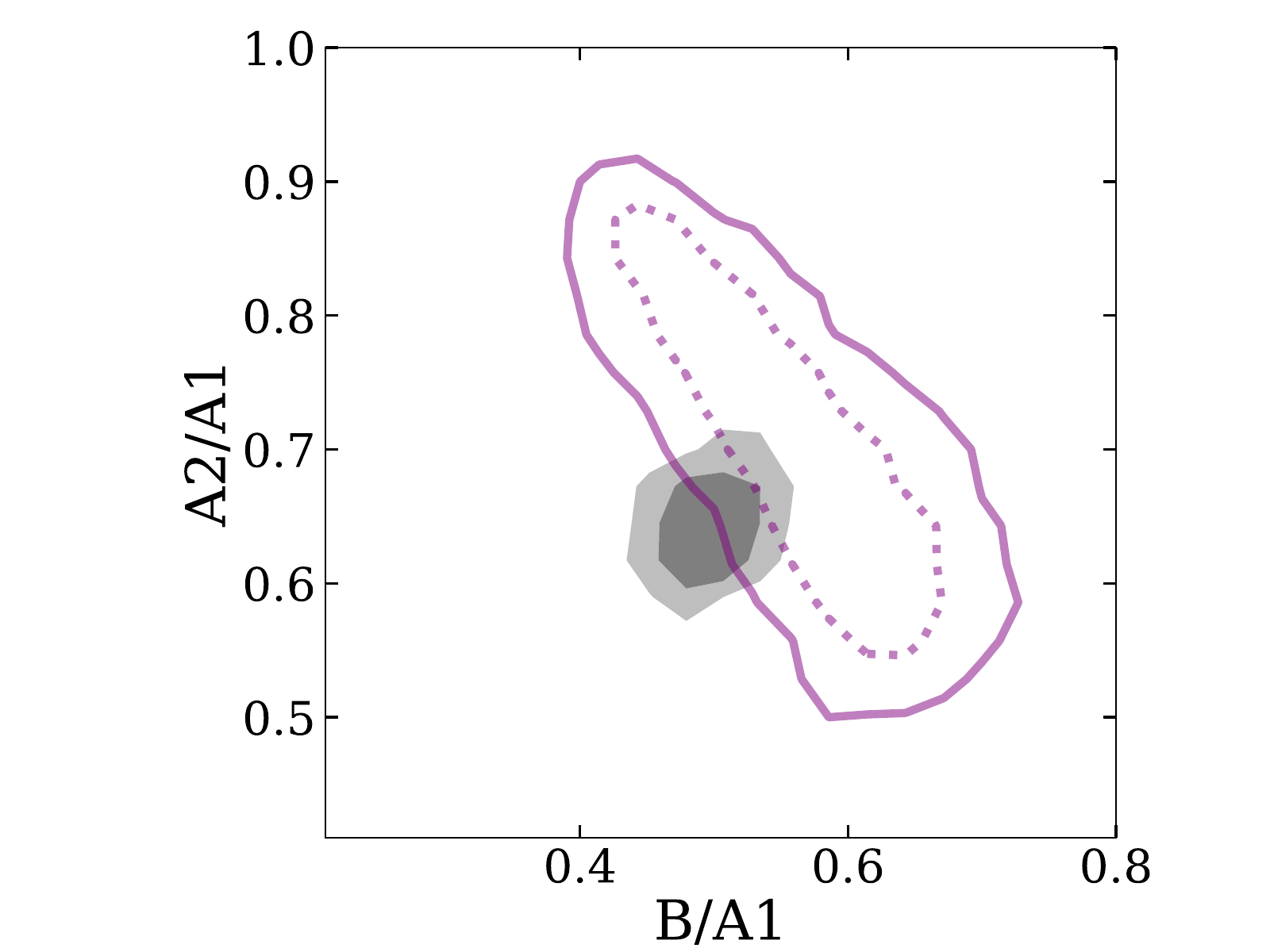}
\includegraphics[scale=0.35, trim = 0 0 0 0 clip=true]{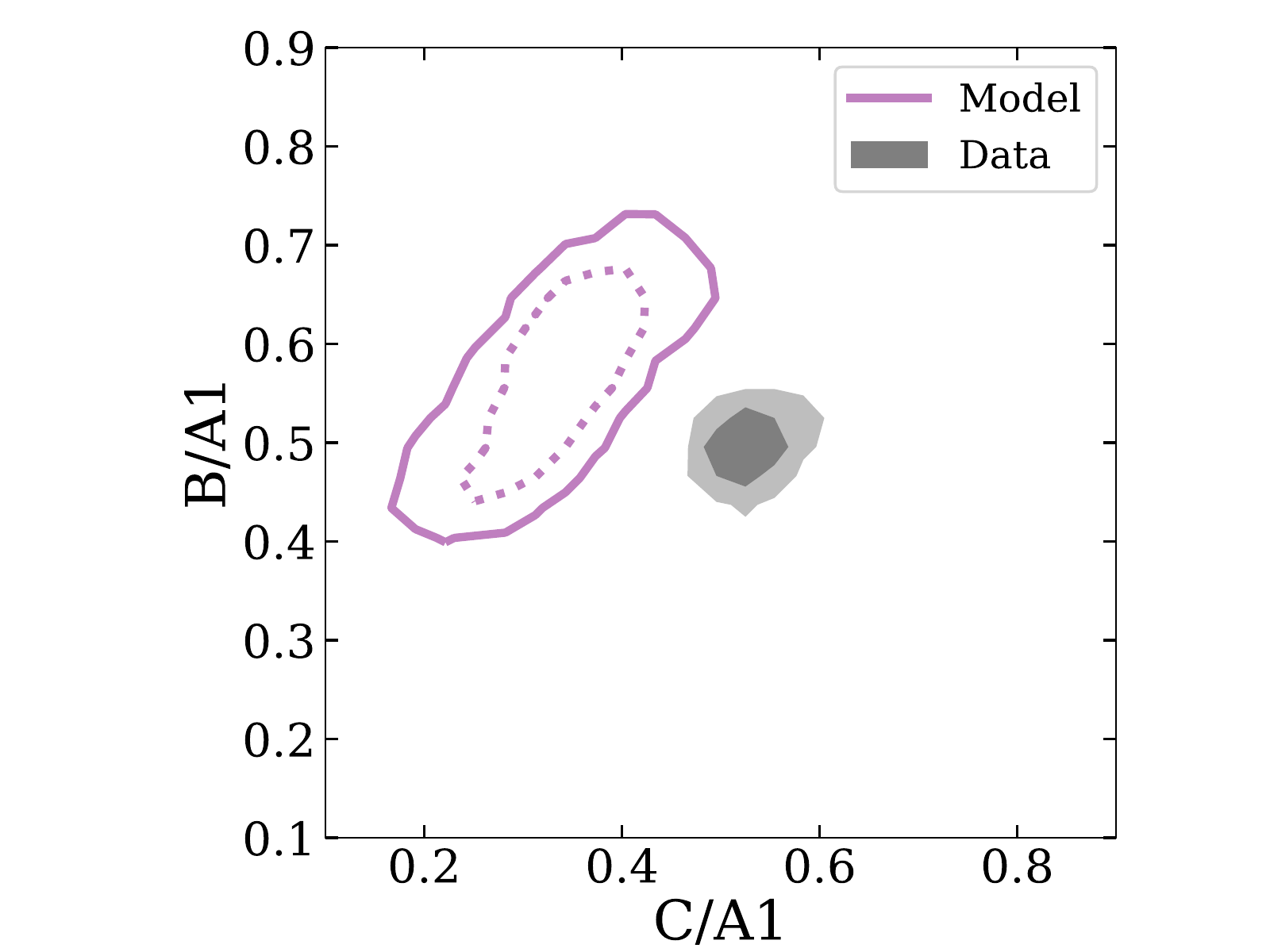}
\includegraphics[scale=0.35, trim = 0 0 0 0 clip=true]{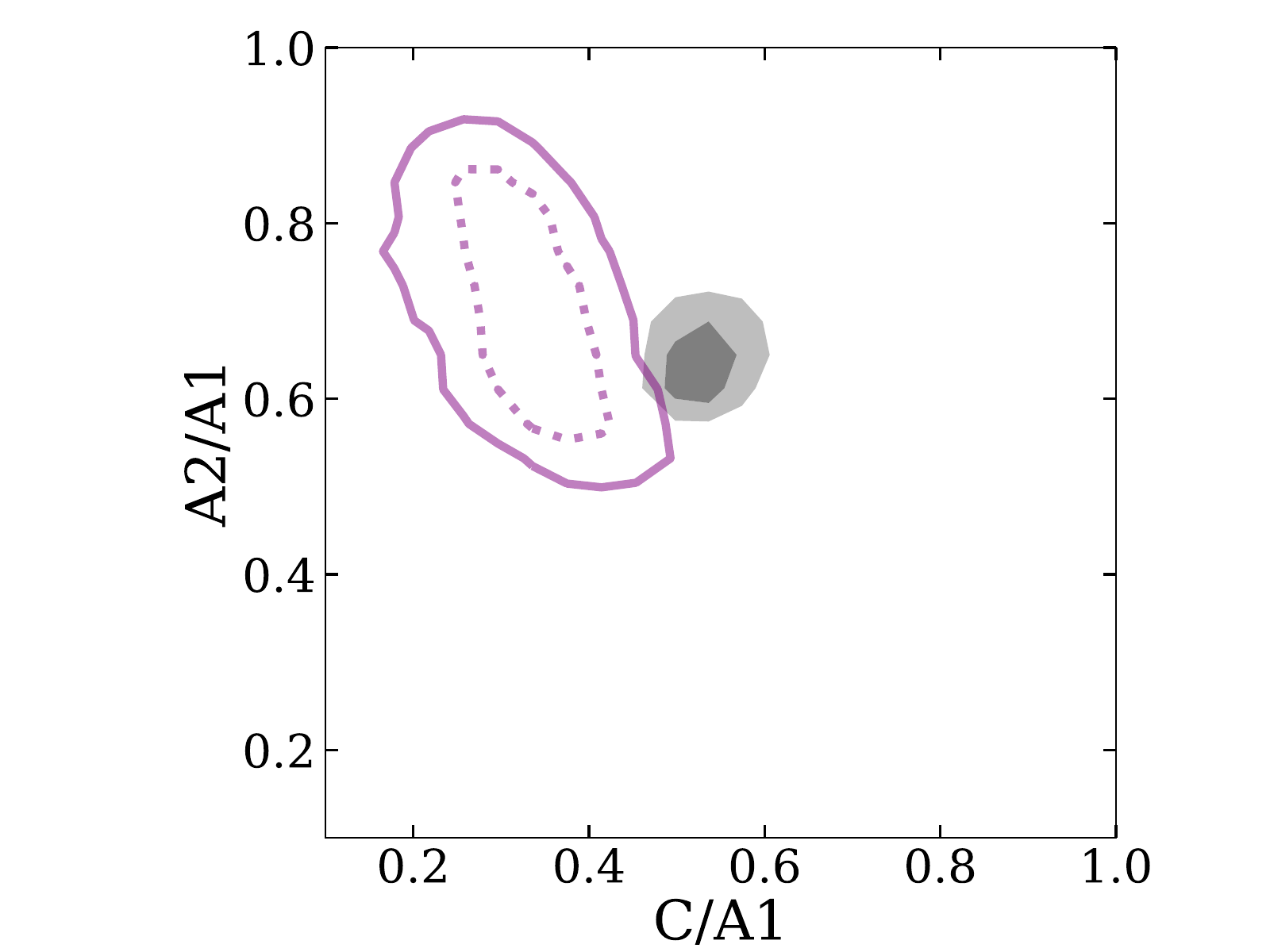}

\caption{Spectral fitting results for WFI 2033. \emph{Top Row}: Model fit with 68\% confidence interval to image spectra showing contributions from all spectral components used in the fit. \emph{Bottom Row}: Comparison between model and grism traces in the 2D grism image computed using a PSF-weighted sum along the y axis. Traces are computed after spectra from all images have been added to the 2D image and are thus affected by blending between neighbouring images.  \emph{Lower Row}: Comparison between observed narrow-line flux ratios, and predicted flux ratios based on model fit to observed image positions. Dotted lines and dark contour represent one sigma confidence intervals for the model and data respectively, while solid lines and light contours represent two sigma confidence intervals.}
}
\end{figure*}

\begin{figure*}
{\centering

\includegraphics[scale=0.45, trim = 0 0 0 0 clip=true]{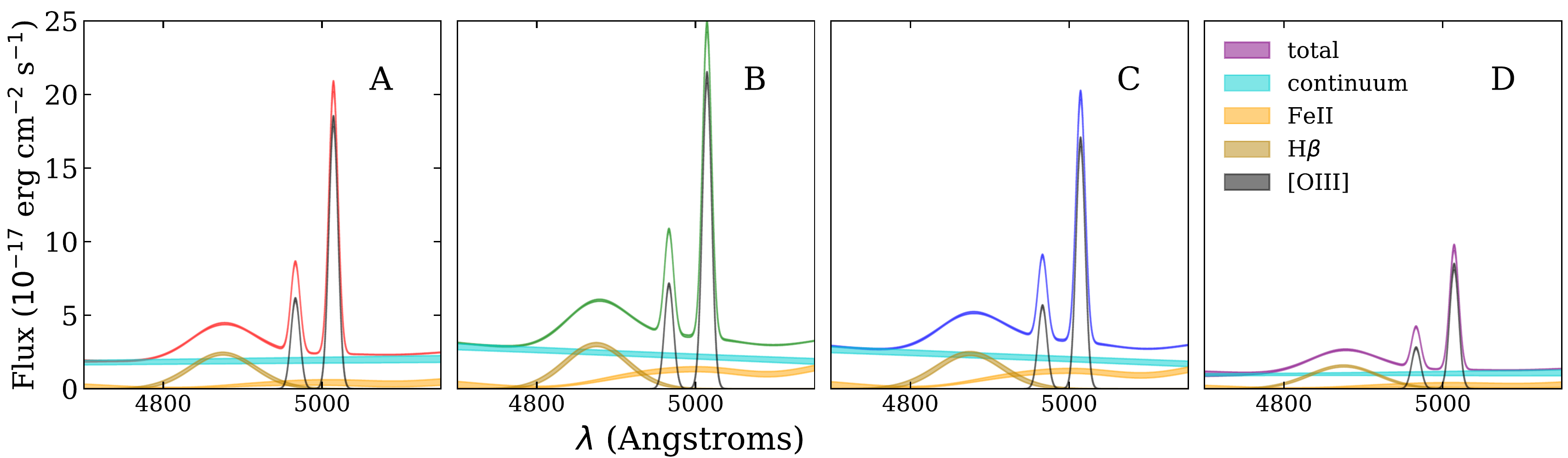}

\includegraphics[scale=0.45, trim = 0 0 0 0 clip=true]{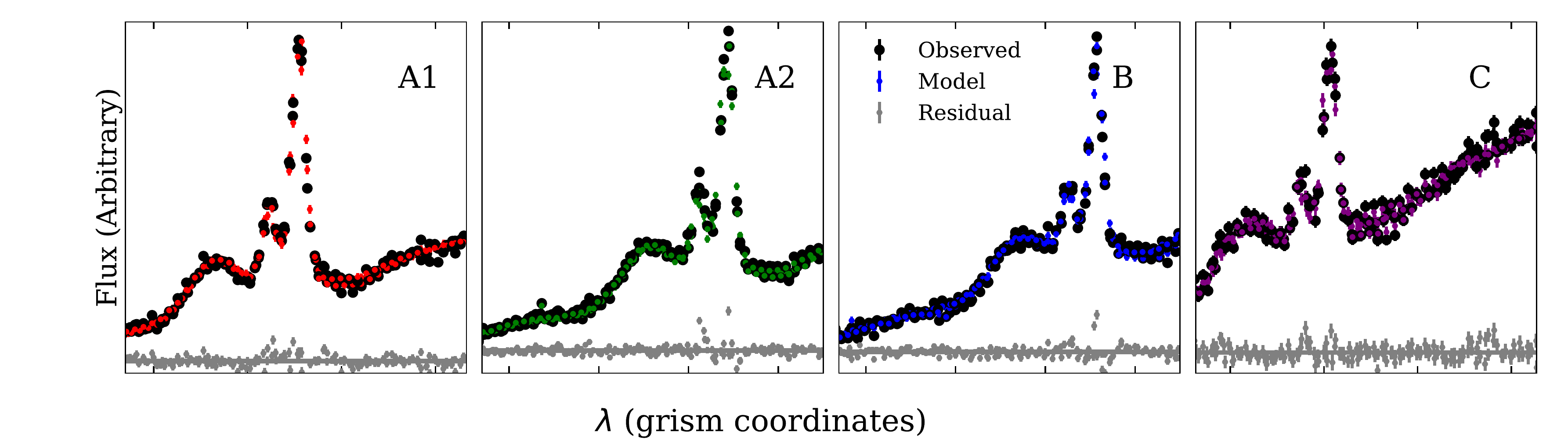}

\vspace{10mm}
\includegraphics[scale=0.4, trim = 0 0 0 0 clip=true]{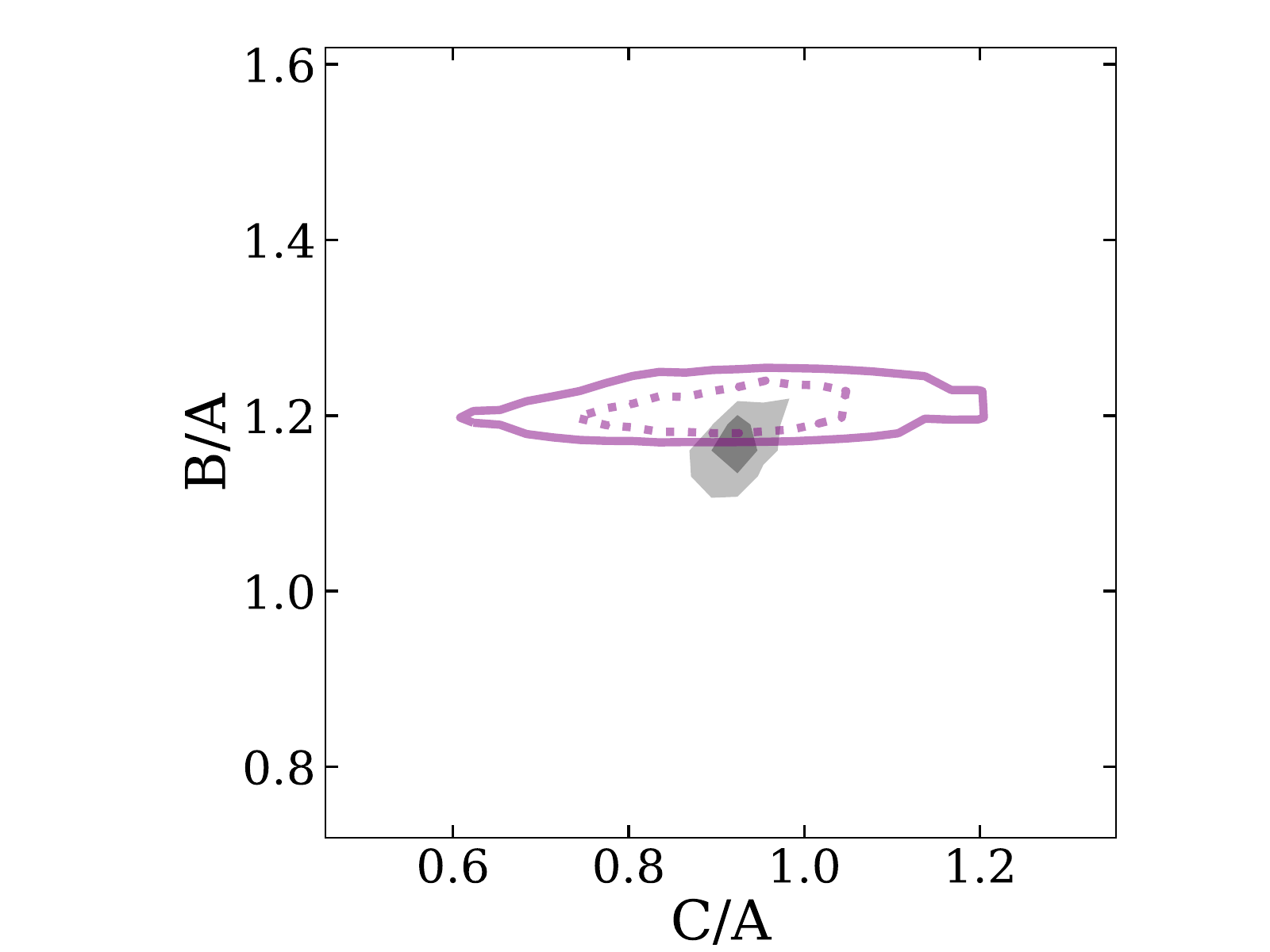}
\includegraphics[scale=0.4, trim = 0 0 0 0 clip=true]{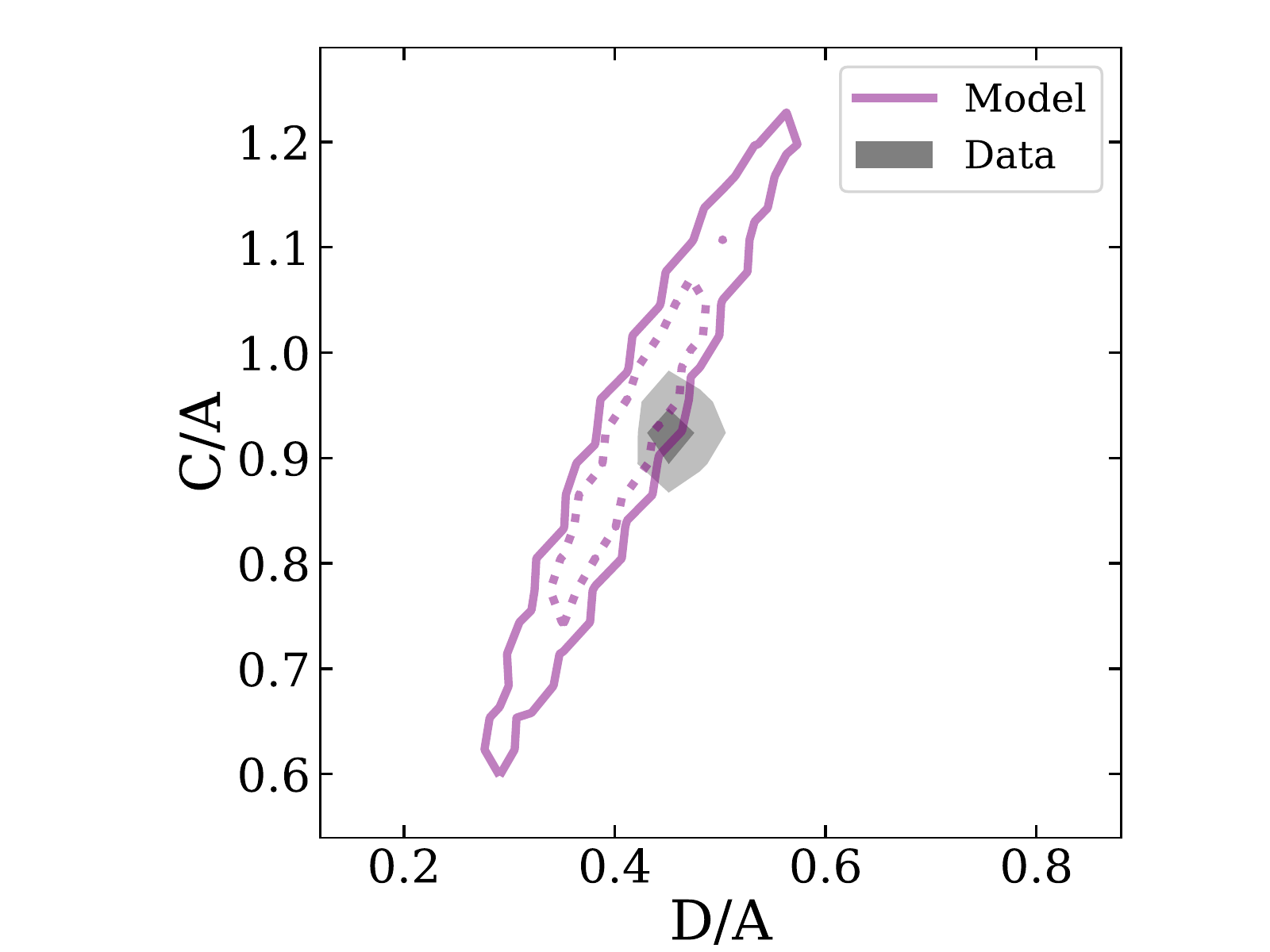}
\includegraphics[scale=0.4, trim = 0 0 0 0 clip=true]{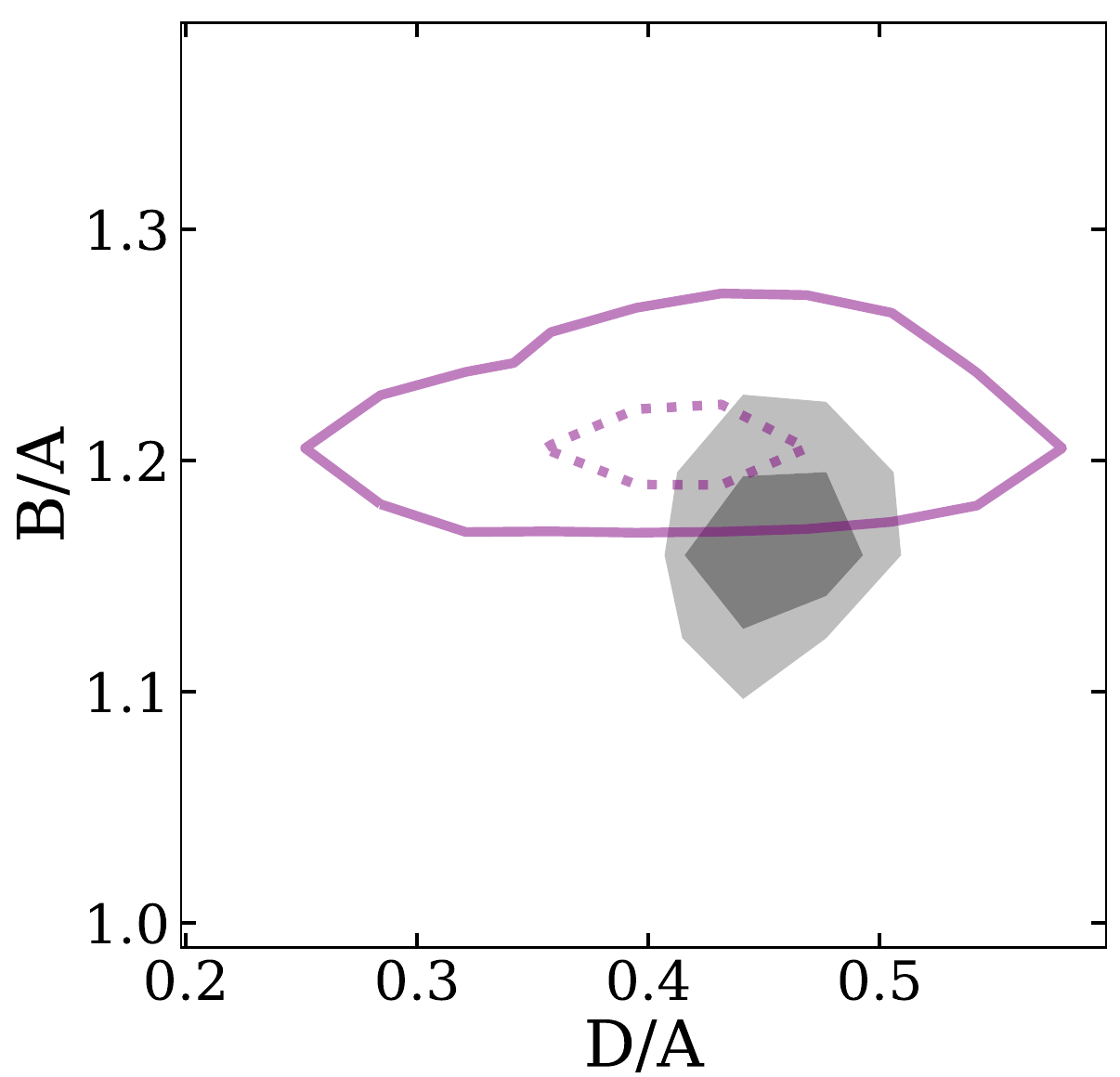}

\caption{Spectral fitting results for WGD J2038. \emph{Top Row}: Model fit with 68\% confidence interval to image spectra showing contributions from all spectral components used in the fit. \emph{Bottom Row}: Comparison between model and grism traces in the 2D grism image computed using a PSF-weighted sum along the y axis. Traces are computed after spectra from all images have been added to the 2D image and are thus affected by blending between neighbouring images.  \emph{Lower Row}: Comparison between observed narrow-line flux ratios, and predicted flux ratios based on model fit to observed image positions. Dotted lines and dark contour represent one sigma confidence intervals for the model and data respectively, while solid lines and light contours represent two sigma confidence intervals.}
}
\end{figure*}

\label{lastpage}
\bsp
\end{document}